\documentclass[a4paper,fleqn,usenatbib]{mnras}

\usepackage{newtxtext,newtxmath}

\usepackage{newtxtext,newtxmath}

\usepackage[T1]{fontenc}
\usepackage{ae,aecompl}

\usepackage{graphicx}	
\usepackage{amsmath}	
\usepackage{amssymb}	

\usepackage{cancel}
\usepackage{amsfonts}
\usepackage{algorithm}
\usepackage{algpseudocode}

\makeatletter
\def\BState{\State\hskip-\ALG@thistlm}
\makeatother

\usepackage{listings}
\usepackage{enumitem}
\usepackage{calc}
\usepackage{bm}
\usepackage{times}
\usepackage{cases}
\usepackage{textgreek}

\graphicspath{{./images/}}

\usepackage[english]{babel}
\usepackage{dashrule}
\usepackage{multirow}

\usepackage{tabularx}
\usepackage[usenames, dvipsnames]{color}
\definecolor{myblue}{rgb}{0,0,1}
\definecolor{mygray}{rgb}{0.5, 0.5, 0.5}
\definecolor{mypink1}{rgb}{1, 0, 1}

\title[Joint Stokes imaging for RI]
{Sparse interferometric Stokes imaging under polarization constraint (Polarized SARA)}

\author[J. Birdi et al.]{
Jasleen Birdi,\thanks{E-mail: jb36@hw.ac.uk}
Audrey Repetti,
and Yves Wiaux
\\
Institute of Sensors, Signals and Systems, Heriot-Watt University, Edinburgh EH14 4AS, UK}

\date{Accepted XXX. Received YYY; in original form ZZZ}

\pubyear{2017}

\DeclareMathSymbol{\phi}{\mathalpha}{operators}{8}

\begin{document}
\label{firstpage}
\pagerange{\pageref{firstpage}--\pageref{lastpage}}
\maketitle

\begin{abstract}
We develop a novel algorithm for sparse Stokes parameters imaging in radio interferometry under the polarization constraint. The latter is a physical non-linear relation between the Stokes parameters, imposing that the polarization intensity is a lower bound on the total intensity. To solve the joint inverse Stokes imaging problem including this bound, we leverage epigraphical projection techniques in convex optimization and design a primal-dual method offering a highly flexible and parallelizable structure. In addition, we propose to regularize each Stokes parameter map through an average sparsity prior in the context of a reweighted analysis approach (SARA). The resulting approach is dubbed Polarized SARA. We demonstrate on simulated observations of M87 with the Event Horizon Telescope that imposing the polarization constraint leads to superior image quality. The results also confirm that the performance of the average sparsity prior surpasses the alternative state-of-the-art priors for polarimetric imaging.

\end{abstract}




\section{Introduction} \label{sec:intro}

The study of the polarized emissions from various astrophysical sources in the universe provides invaluable information about the origin as well as the medium of propagation of these emissions. In most of the cases, these sources generate appreciable linearly polarized radiations and only negligible circularly polarized radiations. Thus, of particular interest is the study of linearly polarized emissions. These radiations can be generated, for instance, due to the synchrotron emission from the electrons in high-energy objects \citep{Ginzburg1965}. Analysis of these polarized emissions gives insights into the strength and orientation of the magnetic field in the sources. Moreover, while traversing, the interaction with the magnetized plasma along the line of sight to the source can modify the polarization state of these radiations via processes like Faraday rotation \citep{Pacholczyk1970, Simard-Normandin1981}. As a result, the polarized emissions also characterize the magnetic field distributions of these plasmas \citep{Dreher1987, Brentjens2005}. This all indicates the importance of imaging these polarized emissions, which is referred to as the polarimetric imaging.

In the context of polarimetric imaging for radio interferometry (RI), the intensity distribution of the sky image of interest is characterized by the Stokes parameters- $I, Q, U$ and $V$, which are all real valued. While $I$ represents the total intensity of the radio emissions, $Q,\, U$ and $V$ describe the polarization state of the electromagnetic radiations coming from the target area of the sky. In particular,  $Q$ and $U$ refer to the linear polarization, and $V$ denotes the circular polarization. 
Furthermore, the linear polarization image $P$ is given by $P = Q + \bold{i} \,U$. The magnitude of this complex valued quantity provides the linear polarization intensity, while the electric vector polarization angle (EVPA) can be obtained from its phase. 
Importantly, the Stokes parameters are not completely independent but rather constrained by a physical non-linear relation imposing that the polarization intensity is a lower bound on the total intensity: $\sqrt{Q^2 + U^2 + V^2} \leq I$. One can also see this constraint, namely the polarization constraint, as the generalization of the more simple positivity constraint on the intensity image in the context of unpolarized imaging.

In order to produce the linear polarimetric images at very high angular resolutions, one of the possibilities is to leverage the technique of linear very long baseline interferometry (VLBI) \citep{Roberts1994}. VLBI basically consists of a collection of radio antennas, spread all across the Earth or even in space (Space VLBI), with the aim of producing images of the target sources in the sky at unprecedented angular resolutions. More recently, the Event Horizon Telescope (EHT)\footnote{http://eventhorizontelescope.org} has been designed to observe the immediate environment around a black hole at angular resolutions comparable to the event horizon. The EHT is essentially a ground-based VLBI array. Its primary observing targets are the super-massive black hole Sgr $\text{A}^*$ at the center of the Milky Way galaxy \citep{Doeleman2008}, and the nucleus of M87, a giant elliptical galaxy in the constellation Virgo \citep{Doeleman2012, Akiyama2015}. Note that in view of the discussion above, polarimetric imaging with the EHT data can yield extremely valuable information about the magnetic field distribution and magnetized plasma in the regions around these targets.
However, the radio interferometers do not directly provide an image as an output. Indeed, the interferometric measurements consist of complex visibilities which are related to the Fourier transform of the brightness distribution of the sky image of interest \citep{Thompson2001}. Since each such visibility is acquired by a pair of antennas in the interferometer and the number of antennas are limited, only an incomplete sampling of the Fourier plane is observed.
This leads to a highly under-determined problem of RI image reconstruction. There exists a number of methods in the literature to solve the corresponding problem. All these methods were initially developed for the Stokes $I$ imaging only. Later on, some of these methods have been extended to polarimetric imaging. In this context, the most widely used algorithm is \textsc{clean} \citep{Hogbom1974}.
It implements a greedy, non-linear iterative deconvolution approach. With regards to polarimetric imaging, \textsc{clean} solves for each of the Stokes images in the same manner, although totally independently. Basically, given the Stokes visibilities corresponding to any of the Stokes parameters, each iteration involves the computation of the residual dirty image wherein the maximum intensity pixel is searched for. This is followed by a beam removal step where, based on a \emph{loop gain factor}, a fraction of this pixel's value convolved with the dirty beam is removed. The process is continued until the maximum intensity value in the dirty image becomes lower than some threshold value. Working pixel by pixel, \textsc{clean} implicitly considers the sought image to be sparse. Many variants of \textsc{clean} have also been proposed these last years, notably its multi-scale \citep{Cornwell2008} and adaptive scale version \citep{Bhatnagar2004}. It is worth mentioning that in essence, \textsc{clean} has been shown to be very similar to some of the existing optimization algorithms in literature. In particular, it shares many attributes with the matching pursuit algorithm \citep{Mallat1993}. Lately, the analogies of \textsc{clean} with a sparsity regularized gradient descent method have been shown in \cite{Rau2009, Carrillo2014, Onose2016}. 
More recently, another technique called Generalized Complex \textsc{clean} has been proposed in \cite{Pratley2016} for polarimetric imaging. This technique is basically a modification of the \textsc{clean} algorithm. Unlike CLEANing independently for the real valued Stokes $Q$ and $U$ images, as done in the traditional \textsc{clean} methods, the authors in \cite{Pratley2016} propose to CLEAN the complex valued linear polarization image $P$. It offers the advantage of rotational invariance and detection of more true components in sources near the noise level.

In the past years, a great interest raised for compressive sensing (CS) based techniques. In the context of RI imaging, these techniques were pioneered by \citet{Wiaux2009}, followed by other works including \citet{Wiaux2010, Li2011, Carrillo2012, Carrillo2014, Garsden2015, Onose2016}, to name a few. In particular, these techniques reconstruct the image of interest by leveraging the sparsity of the sought image, either in the image domain or in a transformed domain. Though applied only for Stokes $I$ image reconstruction, the quality of reconstruction obtained by these techniques have shown to outperform that obtained by \textsc{clean}. Very recently, the first application of these sparsity regularized methods for polarimetric imaging has been developed in \cite{Akiyama2017b}. In this case, the authors promote the sparsity of the underlying images using the $\ell_1$ norm along with the total variation (TV) regularization \citep{Rudin1992, Chambolle1997}, and solve the resultant problem using a monotonic version of fast iterative shrinkage/thresholding algorithm (FISTA) \citep{Beck2009b, Beck2009a}. The authors validate their technique on simulated EHT data and obtain super-resolved Stokes images. The resolution of the reconstructed images is much higher than that obtained by $\textsc{clean}$. However, similar to \textsc{clean}, this sparsity based approach also solves independently for the Stokes images.

In practice, the Stokes images are correlated. As previously discussed, due to the polarization constraint, the intensity in each pixel of the total intensity image cannot be smaller than the corresponding polarization intensity. Nevertheless, up to the best of our knowledge, none of the previously mentioned methods take this constraint explicitly into account. It is worth mentioning then that in the absence of this constraint, non-physical reconstructions may be produced. One way to reconstruct the images with physical meaning is to make use of maximum entropy methods (MEM). MEM based techniques have already proven their worth in RI imaging \citep{Cornwell1985}. These methods aim to find an image that maximizes the entropy function while being consistent with the acquired data. As an extension to polarimetric MEM, a special entropy function incorporating this polarization constraint is used \citep{Narayan1986, Holdaway1990, Coughlan2016, Chael2016}. Note that these works rely on gradient-descent based techniques which are restricted to solve minimization problems with differentiable objective functions. In particular, non-smooth sparsity priors, e.g., $\ell_1$ norm, cannot be incorporated using these approaches. For an overview on gradient-descent based algorithms, we refer the reader to \citet{Bertsekas1999}, and references therein.

In this article, we propose a new method for joint estimation of Stokes images. More specifically, we develop a sparse imaging method that jointly solves for the Stokes parameter maps under the polarization constraint. Our contribution is twofold. Firstly, within the proposed sparse modelling framework, the novelty of our method lies in taking into account the polarization constraint explicitly in the image reconstruction problem. Since this constraint cannot be handled by the classical optimization approaches, we propose to enforce it by employing the technique of epigraphical projections, which consists in splitting the associated constraint set into easily manageable sets \citep{Chierchia2015}. Secondly, we generalize to polarimetric imaging the sparsity averaging reweighted analysis (SARA) approach introduced for Stokes $I$ imaging in \cite{Carrillo2013, Carrillo2014}. The resultant approach, referred to as Polarized SARA, now promotes average sparsity of each of the Stokes images $I$, $Q$, $U$ and $V$. While the original SARA for intensity imaging also imposes positivity of the Stokes $I$ map, Polarized SARA accounts for the polarization constraint for joint Stokes imaging.
In order to solve the corresponding image reconstruction problem, we develop an iterative algorithm based on a primal-dual method \citep{Combettes2011, Condat2013, Vu2013, Pesquet2015}. It is important to emphasize that, even if the method is described in the context of sparsity averaging regularization, the proposed algorithm can be used for any other prior.

Note that although the sparse modelling techniques have already been applied for polarimetric imaging in \cite{Akiyama2017b}, as pointed out earlier, this work was restricted to $\ell_1$ regularization and TV-based regularizations. Moreover, the authors in \cite{Akiyama2017b} do not consider a joint imaging problem, instead solve independently for each of the Stokes parameters, without imposing the polarization constraint. It is worth noting that their imaging problem can be solved using a similar primal-dual based approach, as the one developed in this article. For the sake of completeness, we also propose to introduce the polarization constraint in their image reconstruction problem, and solve the resulting problem using epigraphical projections in a primal-dual framework.

The outline of the article is as follows. In Section~\ref{sec:RI_imaging}, we lay the foundations for our work. In particular, we describe the RI imaging model for full polarization and pose the inverse problem for polarimetric imaging. In Section~\ref{sec:CS_opt}, we discuss about the CS and the optimization frameworks, which are then used to design our polarimetric method. We give a detailed description of the proposed approach in Section~\ref{sec:prop_app}. We then investigate the performance of the proposed algorithm on simulated EHT data in Section~\ref{sec:sim_app}. We conclude in Section~\ref{sec:conc} and discuss the future work.


\vspace*{-0.6cm}
\section{Full polarization Observation model}\label{sec:RI_imaging}
\subsection{Problem description}
A radio interferometer comprises an array of antennas, designed to image the radio sources in a given sky area. In order to image the Stokes parameters, each antenna pair in the interferometer provides a set of measurements, called the visibilities, at time instants $t \in \{1, \ldots, T\}$. 
For a better understanding, let us consider an interferometer consisting of $N_a$ antennas, such that each antenna pair is labelled by $(\alpha, \beta) \in \{1,\ldots, N_a\}^2$, with $\alpha < \beta$. Furthermore, referring the separation between each antenna pair by the term baseline, let the projected baseline components corresponding to an antenna pair $(\alpha, \beta)$ at time $t$ be denoted by $(u_{\alpha, \beta, t}, v_{\alpha, \beta, t}, w_{\alpha, \beta, t})$, expressed in units of the observation wavelength. While $\bm{u}_{\alpha, \beta, t} = (u_{\alpha, \beta,t},v_{\alpha, \beta, t})$ describes the components in the plane perpendicular to the line of sight, $w_{\alpha, \beta, t}$ refers to the component in the line of sight. Within the same coordinate system, the Stokes parameters of the target sky image can be expressed by the corresponding components $(l, m, n)$. Here $\bm{l} = (l,m)$ and $n(\bm{l}) = \sqrt{1- l^2 - m^2 }$, with $l^2 + m^2 \leq 1$. Remark that the target sky intensity distribution can vary with the observation time and frequency. In this context, for each time instant $t$ and observation frequency $\nu$, the radio interferometric measurement equation (RIME) is given by \citep{Smirnov2011}
\vspace*{-0.4cm}
\begin{equation}
\label{eq:RIME}
\bm{{y}}^\top_{\alpha, \beta, t, \nu} =  \int  \, {\bm{{S}}}^\top_{t, \nu} (\bm{l}) \, \bm{\mathsf{L}} \,  \bm{\mathsf{D}}^\top_{\alpha, \beta, t, \nu}(\bm{l}) \, e^{-i2\pi \bm{u}_{\alpha, \beta, t} \cdot \bm{l}} $\text{d}$^2\bm{l},
\end{equation}
where $(.)^\top$ denotes the transpose operation of its argument. In equation~\eqref{eq:RIME}, $\bm{{y}}_{\alpha, \beta, t, \nu} \in \mathbb{C}^{4}$ is the full polarization visibility vector corresponding to the measurements made by the antenna pair $(\alpha, \beta)$ at time instant $t$ and observation frequency $\nu$. Essentially, it can be seen as the Fourier transform of the product of brightness distribution and the Mueller matrix, at the spatial frequency $\bm{u}_{\alpha,\beta, t}$ \citep{Hamaker1996, Thompson2001, Rau2009}. More specifically, $\bm{\mathsf{L}}$ denotes the linear operator which acts on the Stokes vector ${\bm{S}}^\top_{t, \nu}(\bm{l}) = [I, \, Q, \, U, \, V]_{t, \nu}(\bm{l})$ to give the brightness vector as follows
\begin{equation} 
{\bm{{S}}}^\top_{t, \nu}(\bm{l}) \bm{\mathsf{L}} = [I + Q, \, U + $\textbf{i}$ V, \, U - $\textbf{i}$ V , \, I - Q ]_{t, \nu}(\bm{l}).
\end{equation} 
The $4 \times 4$ Mueller matrix $\bm{\mathsf{D}}_{\alpha, \beta, t, \nu}(\bm{l})$ is obtained from the Kronecker product of $2 \times 2$ Jones matrices defined for antennas $\alpha$ and $\beta$ at time $t$ and frequency $\nu$ \citep{Hamaker1996}. 
The Mueller matrix for each antenna pair is generally dependent on the angular position $\bm{l}$ on the sky to incorporate not only the direction-independent effects (DIEs) but also the direction-dependent effects (DDEs). These effects are either known (e.g. the $w$ component \citep{Cornwell2008b, Offringa2014, Dabbech2017}), or need to be calibrated \citep{Smirnov2011, Salvini2014, Smirnov2015, Weeren2016, Repetti2017, Sokolowski2017}. The Mueller matrices can also have non-zero off-diagonal components, corresponding to the cross-polarization leakage. Therefore, each component of the visibility vector $\bm{{y}}_{\alpha, \beta, t, \nu}$ gathers contributions from all the Stokes parameters.

For the sake of our considerations, we restrict our settings to the brightness distributions with neither time nor frequency dependency, with a single observation frequency. This implies that $\bm{S}_{t,\nu} (\bm{l})  = \bm{S} (\bm{l})$, and the frequency index is dropped for all other variables for ease of notation.

In order to recover the Stokes images from the given measurements, we consider a discretized version of the inverse problem in equation~\eqref{eq:RIME}. To this aim, we denote by $\overline{\bm{\mathsf{S}}} =[\overline{\bm{s}}_1, \overline{\bm{s}}_2, \overline{\bm{s}}_3, \overline{\bm{s}}_4] \in \mathbb{R}^{N \times 4}$ the Stokes matrix, corresponding to the concatenation of the discretized Stokes maps. In this context, $\overline{\bm{s}}_1,\overline{\bm{s}}_2, \overline{\bm{s}}_3$ and $\overline{\bm{s}}_4$, each belonging to $\mathbb{R}^N$, denote the discretization of the Stokes images $I$, $Q$, $U$ and $V$, respectively. 
In accordance with equation~\eqref{eq:RIME}, we denote the measurement matrix by $\bm{\mathsf{Y}} \in \mathbb{C}^{M \times 4}$, where each row $m \in \{1, \ldots, M\}$ corresponds to the full polarization measurements acquired by a given antenna pair $(\alpha, \beta)$ at time $t$. With these definitions in mind, the discretized measurement model is given by
\vspace*{-0.25cm}
\begin{equation} \label{eq:RIME_disc}
\bm{\mathsf{Y}} = {{{{\phi}}}}(\overline{\bm{\mathsf{S}}}) + {\bm{\mathsf{E}}},
\end{equation}
where the measurements are corrupted by a random additive noise, represented by the matrix $\bm{\mathsf{E}} \in \mathbb{C}^{M \times 4}$. Each column of the matrix $\bm{\mathsf{E}}$ is a realization of an independent and identically distributed (i.i.d.) Gaussian noise. 
The measurement operator ${\phi}: \mathbb{C}^{N \times 4} \to \mathbb{C}^{M \times 4}$ in equation~\eqref{eq:RIME_disc} acts on the Stokes matrix to give the measurements $\bm{\mathsf{Y}}$. 
In particular, at each observation instant $t$ and within the considered discrete setting, it consists in computing the Fourier transform of the multiplication of the Mueller matrix $\bm{\mathsf{D}}_{\alpha, \beta, t}$ with the brightness matrix $\overline{\bm{\mathsf{S}}} \bm{\mathsf{L}}$, at the sampled frequencies $\bm{u}_{\alpha, \beta, t}$ for all the antenna pairs. 
It is to be mentioned here that calibrating for the DIEs and DDEs is out of the scope of the current article wherein the operator $\phi$ is assumed to be known beforehand.

\subsection{Polarization constraint}
\label{ssec:pol_cons}
One of the key contributions of the current work is exploiting the polarization constraint for the estimation of the Stokes images by solving problem~\eqref{eq:RIME_disc}. This constraint, as discussed in Section~\ref{sec:intro}, physically links the Stokes images, and is defined as follows$\colon$
\begin{equation} \label{eq:flux_bound}
\forall  n \in \{ 1, \ldots, N \}, \quad - {{\mathsf{S}}}_{n,1} + \|\bm{\mathsf{S}}_{n,2:4} \|_2 \leq 0,
\end{equation}
where, for every $n \in \{1, \ldots, N\}$, $\mathsf{S}_{n,1}$ denotes the $n^{\text{th}}$ coefficient of the first column of the matrix $\bm{\mathsf{S}}$, and the notation ${\bm{\mathsf{S}}_{n,2:4}}$ signifies the $n^{\text{th}}$ coefficients of the columns 2 to 4 of the matrix $\bm{\mathsf{S}}$.
However, the underlying constraint set associated with~\eqref{eq:flux_bound} is very complex, and difficult to handle. In the past years, few methods have been developed to take this constraint into account within the MEM framework \citep{Narayan1986, Chael2016}. Nonetheless, concerning the sparsity based techniques, the only existing method for polarimetric imaging (proposed in \citet{Akiyama2017b}) does not impose~\eqref{eq:flux_bound} explicitly on the target Stokes images. As described later, we develop a method to enforce this constraint within the sparse modelling framework.
%

%

%
%
%

\vspace*{-0.5cm}
\section{Compressive sensing \& optimization}\label{sec:CS_opt}
\subsection{Compressive sensing framework}
\label{ssec:CS}
In the recent years, CS framework has gained a lot of attention in the research community \citep{Candes2006, Donoho2006}. In particular for RI imaging, it has been introduced by \cite{Wiaux2009}. These CS techniques often employ convex optimization methods to solve ill-posed inverse problems, exploiting the sparsity of the sought image. They have been shown to provide better reconstruction quality than that obtained by the standard \textsc{clean} algorithm for RI imaging. However, it is worth mentioning that reducing the computational cost of these optimization based methods is still an area of ongoing investigation. 

Generally, inverse problems of the form of equation~\eqref{eq:RIME_disc} are ill-posed or ill-conditioned. Indeed, as described in Section~\ref{sec:RI_imaging}, the measurements only contain a partial information of the image of interest, often corrupted by an additive noise. Then, solving problem~\eqref{eq:RIME_disc} is a challenging task, and a suitable approach needs to be adapted. To overcome this difficulty, the CS theory assumes that the target signal has a sparse representation in a dictionary. More specifically, given a sparsifying dictionary $\mathsf{\bm{\Psi}} \in \mathbb{C}^{N \times J}$, the sparse representation of the signal $\overline{\bm{\mathsf{S}}}$ in this dictionary is given by $\mathsf{\bm{\Psi}}^\dagger \overline{\bm{\mathsf{S}}}$, where $\mathsf{\bm{\Psi}}^\dagger$ is the adjoint operator of $\mathsf{\bm{\Psi}}$. The first task is to choose an appropriate dictionary for the image under consideration. In this context, many studies have been performed in the past years \citep{Rubinstein2010, Starck2010}. 
These studies indicate that for the simplest case of considering an already sparse image (i.e. an image consisting of point sources), $\mathsf{\bm{\Psi}}$ can be chosen to be the dirac basis, i.e. identity matrix with $J=N$, promoting sparsity in the image domain itself. However, when the underlying images are smooth and have more complex structures, the possibilities of this sparsifying dictionary $\mathsf{\bm{\Psi}}$ include the wavelet domain \citep{Mallat2009}, or a collection of wavelet bases \citep{Carrillo2012}. Furthermore, in the case of piece-wise constant images, gradient domain is usually the best option for promoting sparsity, using total variation based regularizations \citep{Rudin1992, Wiaux2010}. Once a suitable dictionary is chosen, within the CS framework, the next task is to solve the underlying inverse problem by promoting the sparsity of the sought image in the domain determined by the chosen dictionary.
 
The best way to promote sparsity of a variable is to use $\ell_0$ pseudo norm \citep{Donoho1995}. Essentially, the $\ell_0$ pseudo norm of any signal is defined to be the number of its non-zero components. Therefore, by definition, minimizing this norm eventually leads to reduction in the number of non-zero elements, thereby promoting sparsity of the given signal. However, owing to its non-convexity, solving the problem for $\ell_0$ norm is often intractable, especially for large dimensional problems. Instead, its convex relaxation, i.e. the $\ell_1$ norm, is often considered \citep{Chen2001}. A probable drawback of the $\ell_1$ norm as a sparsity inducing term is its dependency on the magnitude of the signal coefficients. To mimic the behaviour of the $\ell_0$ pseudo norm more appropriately, this dependency can be alleviated by the use of weighted $\ell_1$ norm. For any matrix $\bm{\mathsf{S}} \in \mathbb{R}^{N \times 4}$, its weighted $\ell_1$ norm is defined as 
\vspace*{-0.15cm}
\begin{equation}
\| \mathsf{\bm{\Psi}}^\dagger {{\bm{\mathsf{S}}}} \|_{\bm{\mathsf{W}},1} = \sum_{i=1}^4 \sum_{j=1}^J \mathsf{W}_{j,i} \left| [\mathsf{\bm{\Psi}}^\dagger {{\bm{\mathsf{S}}}} ]_{j,i} \right|, 
\end{equation}
where the notation $[.]_{j,i}$ stands for the coefficient in the $j^{\text{th}}$ row and $i^{\text{th}}$ column of the argument matrix. Additionally, $\bm{\mathsf{W}} \in \mathbb{R}_+^{J \times 4}$ is the weighting matrix such that its each column comprises of the weights for the signal in the corresponding column of the matrix $\bm{\mathsf{S}}$.
With this approach, the estimate of the signal $\overline{\bm{\mathsf{S}}}$ from the degraded measurements $\bm{\mathsf{Y}}$ is defined to be a solution of the following minimization problem
\begin{equation} \label{eq:const_eq}
\underset{\bm{\mathsf{S}} \in \mathbb{R}^{N \times 4}}{\operatorname{minimize}} \, \| \mathsf{\bm{\Psi}}^\dagger {{\bm{\mathsf{S}}}} \|_{\bm{\mathsf{W}},1} \, \, \, \text{subject to} \,\, \, \| \bm{\mathsf{Y}} - {{\phi}} (\bm{\mathsf{S}}) \|_2 \leq \epsilon, 
\end{equation}
where $\epsilon > 0$ is an upper bound on the norm of the additive noise. Moreover, the Lagrangian function associated with problem~\eqref{eq:const_eq} provides its unconstrained formulation, given by
\begin{equation} \label{eq:unconst_eq}
\underset{\bm{\mathsf{S}} \in \mathbb{R}^{N \times 4}}{\operatorname{minimize}} \, {\mu} \| \mathsf{\bm{\Psi}}^\dagger {{\bm{\mathsf{S}}}} \|_{\bm{\mathsf{W}},1} + \frac{1}{2} \| \bm{\mathsf{Y}} - {{\phi}} (\bm{\mathsf{S}}) \|_2^2,
\end{equation}
where $\mu > 0$. Many efficient convex optimization techniques, discussed in the next section, can be used to solve the resultant minimization problem, whether it is the constrained~\eqref{eq:const_eq} or unconstrained~\eqref{eq:unconst_eq} formulation. Note that while in problem~\eqref{eq:unconst_eq} the parameter $\mu$ needs to be tuned, problem~\eqref{eq:const_eq} requires the value of $\epsilon$ to be specified, which can be theoretically determined. Therefore, problem~\eqref{eq:const_eq} is often preferred over problem~\eqref{eq:unconst_eq}.

Regarding the choice of the weights, these should be such that the small valued coefficients are penalized. To determine these weights, \cite{Candes2008} proposed to solve iteratively a sequence of the weighted $\ell_1$ minimization problems (either problem~\eqref{eq:const_eq} or~\eqref{eq:unconst_eq}) - referred to as the reweighting scheme. By doing so, the weights at each iteration are computed by taking the inverse of the solution from the previous iteration. In the context of radio interferometric imaging, the effectiveness of this scheme has been demonstrated in \cite{Carrillo2012, Onose2017}.

%
%
%
%
\subsection{Optimization framework} \label{ssec:opt_frame}
In order to solve the inverse problem~\eqref{eq:RIME_disc} using the CS framework described above, we resort to convex optimization techniques. In particular, in this work we develop an iterative algorithm based on proximal splitting methods. The main advantages of these methods are their flexibility to deal with very sophisticated minimization problems, and their scalability offering the possibility to handle large dimensional variables. An overview of these methods has been provided in \cite{Combettes2011, Komodakis2015}. They can be employed to solve a wide class of problems which can be expressed in the following form
\begin{equation}
\underset{\bm{\mathsf{S}} \in \mathbb{R}^{N \times 4}}{\operatorname{minimize}}  \, \sum_{k=1}^K f_k(\bm{\mathsf{S}}), 
\label{eq:prox_split}
\end{equation} 
where for $ k \in \{1, \ldots, K\}, f_k$ is a proper, lower-semicontinuous convex function from $\mathbb{R}^{N \times 4}$ to $]-\infty, +\infty]$. It is important to emphasize that problem~\eqref{eq:prox_split} is very general and that most of the problems encountered in signal and image processing applications can be written using this formulation. Indeed any constrained problem can be reformulated as~\eqref{eq:prox_split}. This can be achieved by casting one of the functions $f_k$ as an indicator function of the constraint set of interest. Let $\mathcal{C} \subset \mathbb{R}^{N \times 4}$ represent such a non-empty closed convex set. Then, the indicator function of this set, at a given point $\bm{\mathsf{S}} \in \mathbb{R}^{N \times 4}$, is defined as 
\[
\iota_{\mathcal{C}}(\bm{\mathsf{S}}) = \begin{cases}
                                0,  & \text{if} \, \, \bm{\mathsf{S}} \in \mathcal{C}, \\
                                +\infty, & \text{otherwise.} 
                                \end{cases}
                                \]
Another interesting point of problem~\eqref{eq:prox_split} is that it can take into account both smooth and non-smooth functions. In practice, to handle these functions, proximal splitting methods will use the gradients of the smooth functions and the proximity operators of the non-smooth functions. Formally, the proximity operator of a function $f \colon \mathbb{R}^{N \times 4} \to ]-\infty, +\infty]$ at the point $\bm{\mathsf{S}}$ is defined as
\begin{equation}
\operatorname{prox}_{f} (\bm{\mathsf{S}}) 
= \underset{\bm{\mathsf{U}} \in \mathbb{R}^{N \times 4}}{\operatorname{argmin}} \,\, f(\bm{\mathsf{U}}) + \frac{1}{2}\|\bm{\mathsf{U}}-\bm{\mathsf{S}}\|_2^2 .
\label{eq:prox}
\end{equation}
This operator can be seen as a generalization of the projection operator $\bm{\mathcal{P}}_\mathcal{C}$ onto the set $\mathcal{C} \subset \mathbb{R}^{N \times 4}$, i.e., 
\begin{equation}
\bm{\mathcal{P}}_\mathcal{C} (\bm{\mathsf{S}}) = \underset{\bm{\mathsf{U}} \in \mathcal{C}}{\operatorname{argmin}} \|\bm{\mathsf{U}}-\bm{\mathsf{S}}\|_2^2\ .
\end{equation} 
Based on the proximal splitting methods, problems of the form~\eqref{eq:prox_split} can be efficiently solved by several existing algorithms, e.g. forward-backward algorithm (also known as iterative soft thresholding algorithm- ISTA) \citep{Chen1997, Tseng2000}, Douglas-Rachford algorithm \citep{Eckstein1992}, to name a few.

In the particular yet common case of composite problems where the non-smooth functions are composed with a linear operator, adapted methods need to be designed. For instance, consider the following problem 
\begin{equation}
\underset{\bm{\mathsf{S}} \in \mathbb{R}^{N \times 4}}{\operatorname{minimize}}\,  f_1(\bm{\mathsf{S}}) + f_2(\bm{\mathsf{T}}\bm{\mathsf{S}}), 
\label{eq:primal_def}
\end{equation}
where $f_1 \colon \mathbb{R}^{N \times 4} \to ]-\infty, +\infty]$ is a convex differentiable function, $f_2 \colon \mathbb{R}^{Q \times 4} \to ]-\infty, +\infty]$ is a non-smooth, proper, lower-semicontinuous convex function, and $\bm{\mathsf{T}} \in \mathbb{R}^{Q \times N}$ is a linear operator. Note that this problem can be solved using the forward-backward algorithm, alternating between a gradient step on $f_1$ and a proximity step on $f_2 \circ \bm{\mathsf{T}}$. However, using this approach may require the inversion of the operator $\bm{\mathsf{T}}$ or performing sub-iterations to compute the proximity step. This can be problematic especially when the dimension of the underlying problem increases. To overcome this issue, recently several primal-dual methods have been proposed \citep{Chambolle2010, Combettes2011, Condat2013, Vu2013, Combettes2014, Komodakis2015}. Basically, they provide \emph{full splitting} and solve simultaneously for primal and dual problems. More formally, the dual problem associated with the primal problem~\eqref{eq:primal_def} is given by
\begin{equation}
\underset{\bm{\mathsf{V}} \in \mathbb{R}^{Q \times 4}} {\operatorname{minimize}} \, f_1^* (-\bm{\mathsf{T}}^\dagger \bm{\mathsf{V}}) + f_2^*(\bm{\mathsf{V}}),
\label{eq:dual_def}
\end{equation}
where $f_1^*$ (resp. $f_2^*$) is the Legendre-Fenchel conjugate function of $f_1$ (resp. $f_2$) \citep{Bauschke2011}, and $\bm{\mathsf{T}}^\dagger$ is the adjoint operator of $\bm{\mathsf{T}}$. Concerning the splitting in the primal-dual methods, it is achieved over all the functions involved in the minimization problem, including the gradient and proximity operators as well as the involved linear operator, thereby preventing the need to invert the linear operator. These methods, thus, offer computational advantages over other splitting methods. 


\section{Polarized SARA}
\label{sec:prop_app}
\subsection{Objective function for polarimetric imaging}
\label{ssec:obj_fun}

The estimation of the Stokes images from the degraded measurements consists in solving the inverse problem~\eqref{eq:RIME_disc}. Given its ill-posedness, ensuring the data fidelity is not sufficient to recover the sought images. Indeed, following the CS theory described in Section~\ref{ssec:CS}, the problem needs to be regularized and \emph{a priori} information needs to be injected in the reconstruction process. We thus aim to solve a minimization problem consisting of a data fidelity and a regularization term. The data fidelity term given by
\begin{multline}  \label{eq:f_diff}
f \big( \phi(\bm{\mathsf{S}}) \big) = \iota_{\mathbb{B} (\bm{\mathsf{Y}}, \epsilon)} \big( \phi (\bm{\mathsf{S}}) \big) \\ \text{with} \, \,  \mathbb{B}(\bm{\mathsf{Y}}, \epsilon) = \{ \phi (\bm{\mathsf{S}}) \in \mathbb{C}^{M \times 4} \colon \| \phi (\bm{\mathsf{S}}) - \bm{\mathsf{Y}} \|_2 \leq \epsilon \},
\end{multline}
constraints the residual to lie within an $\ell_2$ ball centred in $\bm{\mathsf{Y}}$ and whose radius is determined by the noise level $\epsilon$.

Concerning the regularization term, we will use a hybrid function taking into account the following prior information. 

\vspace*{-0.2cm}
\paragraph*{Real-valuedness.} The Stokes images should be real-valued. This condition can be enforced by the use of an indicator function of a set $\mathbb{U} = \mathbb{R}^{N \times 4}$.

\paragraph*{Sparsifying regularization.} Leveraging the CS theory, we promote sparsity of the Stokes images in a sparsifying dictionary $\mathsf{\bm{\Psi}}$. 
In this context, as discussed in Section~\ref{ssec:CS}, adopting the reweighting scheme which consists in iteratively solving the problems with weighted $\ell_1$ norm provides a better estimation of the sought images in $\ell_0$ sense as compared to $\ell_1$ norm. Therefore, we propose to use the weighted $\ell_1$ norm as the sparsifying regularization term
\begin{equation} \label{eq:g_l1}
g ( \mathsf{\bm{\Psi}}^\dagger \bm{\mathsf{S}} ) = 
\| \mathsf{\bm{\Psi}}^\dagger {{\bm{\mathsf{S}}}} \|_{\bm{\mathsf{W}},1} .
\end{equation}
It is to be emphasized here that using the weighted $\ell_1$ norm also offers the advantage of no tuning of regularization parameters, unlike the case of $\ell_1$ norm. More precisely, since Stokes $Q$, $U$ and $V$ images are lower in intensity than Stokes $I$, the latter dominates in the $\ell_1$ norm sparsity inducing term. To overcome this unequal contribution of the Stokes images, different regularization parameters need to be chosen for each image. On the contrary, thanks to the weights in~\eqref{eq:g_l1}, all the Stokes images are normalized, thereby having equal importance in this sparsity term, avoiding the need to use any additional parameters to enhance the contribution of the Stokes $Q$, $U$ and $V$ images. 

Note that any linear sparsifying operator could be used in the proposed method. However, inspired by the sparsity averaging proposed in \cite{Carrillo2013, Carrillo2014} for Stokes $I$ imaging, we extend it to polarimetric imaging and choose to promote sparsity averaging for each of the Stokes parameter maps $I$, $Q$, $U$ and $V$. 
It consists in choosing $\mathsf{\bm{\Psi}}$ as the concatenation of the first eight Daubechies wavelets and Dirac basis (see e.g. \citet{Mallat2009, Onose2016}). Using this dictionary coupled with the reweighting scheme corresponds to the SARA regularization.

\paragraph*{Polarization constraint.} We exploit the correlation between the Stokes images by enforcing the polarization constraint, described earlier in Section~\ref{ssec:pol_cons}. 
Let $\mathbb{P}$ be the associated polarization constraint set, defined as
\begin{equation} \label{eq:flux_cons}
\mathbb{P}  = \left\{ \bm{\mathsf{S}} \in \mathbb{R}^{N \times 4} \, \middle\vert  \big( \forall  n \in \{ 1, \ldots, N \} \big) - {{\mathsf{S}}}_{n,1} + \|\bm{\mathsf{S}}_{n,2:4} \|_2 \leq 0 \right\}.
\end{equation}
Then, the polarization constraint can be imposed by the use of an indicator function of the set $\mathbb{P}$.
Additionally, this constraint can be expressed as ${{\mathsf{S}}}_{n,1} \geq \|\bm{\mathsf{S}}_{n,2:4} \|_2$, where $n \in \{1,\ldots,N\}$. It can then be noticed that it imposes the polarization intensity as a lower bound on the total intensity image, and thus implicitly enforces the positivity of the total intensity image (Stokes $I$).

With the above mentioned prior information at hand, our resultant minimization problem to be solved for the Stokes images is given by
\begin{equation} \label{eq:min_gen}
\underset{\bm{\mathsf{S}} \in \mathbb{R}^{N \times 4}} {\operatorname{minimize}} \, f \big( {{\phi}} (\bm{\mathsf{S}}) \big) + \iota_{\mathbb{U}} (\bm{\mathsf{S}}) + g(\mathsf{\bm{\Psi}}^\dagger \bm{\mathsf{S}}) + \iota_{\mathbb{P}} (\bm{\mathsf{S}}).
\end{equation}
It can be observed that enforcing the polarization constraint in problem~\eqref{eq:min_gen} involves projecting the variable $\bm{\mathsf{S}}$ onto the set $\mathbb{P}$. However, the associated projection does not have a closed form. Therefore, to impose this constraint, we employ a splitting technique based on epigraphical projection \citep{Chierchia2015}. Remark that this recently proposed technique is used to handle minimization problems involving sophisticated constraints (see e.g. 
\citet{Harizanov2013, Chierchia2014, Moerkotte2015, Gheche2016}).

Note that, in the case when the polarization constraint is not taken into account, the positivity of the total intensity image is no longer ensured and it needs to be imposed explicitly. 
This can be done by modifying the set $\mathbb{U}$ to a set $\mathbb{U}^\prime$ given by
\begin{equation} \label{eq:pos_set}
\mathbb{U}^\prime = \left\{\bm{\mathsf{S}} \,  \middle\vert \, \bm{\mathsf{S}}_{\colon,1} \in \mathbb{R}_+^N, \, \bm{\mathsf{S}}_{\colon,2:4} \in \mathbb{R}^{N \times 3} \right\}.
\end{equation}
In such a case, problem~\eqref{eq:min_gen} simplifies to
\begin{align} \label{eq:min_gen_matrix_no_flux}
\underset{\substack{\bm{\mathsf{S}} \in \mathbb{R}^{N \times 4}}} {\operatorname{minimize}} \, &  f \big({{\phi}} (\bm{\mathsf{S}}) \big) + \iota_{\mathbb{U}^\prime} (\bm{\mathsf{S}})  +  g(\mathsf{\bm{\Psi}}^\dagger \bm{\mathsf{S}}).
\end{align}

\subsection{Epigraphical projection}

The requirement to satisfy the polarization constraint is that the Stokes matrix belong to the set $\mathbb{P}$. However, as pointed out earlier, this constraint is difficult to manage. To circumvent this difficulty, we utilize the epigraphical projection techniques recently developed by \citet{Chierchia2015}. Leveraging these techniques, we propose to introduce an auxiliary variable $\bm{\mathsf{Z}} \in \mathbb{R}^{N \times 2}$ in the minimization problem~\eqref{eq:min_gen} and thereby, splitting the polarization constraint set into simpler constraint sets, such that the projection onto these sets can be efficiently computed. Doing so, problem~\eqref{eq:min_gen} can be equivalently rewritten as
\begin{subequations}\label{eq:min_gen_g}
\begin{equation} 
\underset{\substack{\bm{\mathsf{S}} \in \mathbb{R}^{N \times 4},  \bm{\mathsf{Z}} \in \mathbb{R}^{N \times 2}}} {\operatorname{minimize}} \, f \big( \phi (\bm{\mathsf{S}}) \big) + \iota_{\mathbb{U}} (\bm{\mathsf{S}}) + g(\mathsf{\bm{\Psi}}^\dagger \bm{\mathsf{S}})
\tag{\ref{eq:min_gen_g}}
\end{equation}
subject to ($\forall  n \in \{ 1, \ldots, N \}$) 
\begin{numcases}{}
h_1({\mathsf{S}}_{n,1}) \leq {{\mathsf{Z}}_{n,1}}, \label{eq:h1_cons} \\
h_2({\bm{\mathsf{S}}}_{n,2:4}) \leq {{\mathsf{Z}}_{n,2}},\label{eq:h2_cons} \\
{{\mathsf{Z}}}_{n,1} + {{\mathsf{Z}}}_{n,2} \leq 0,\label{eq:zeta_cons} 
\end{numcases}
\end{subequations}
where the functions $h_1$ and $h_2$ are defined as 
\begin{equation} \label{eq:def_h1}
( \forall \zeta \in \mathbb{R}) \quad h_1(\zeta) = -\zeta, 
\end{equation}
\begin{align} \label{eq:def_h2}
\big(\forall ({\bm{\zeta}}) \in \mathbb{R}^{3} \big) \quad h_2({\bm{\zeta}}) & = \| {\bm{\zeta}} \|_2.
\end{align}
To understand the above mentioned modified minimization problem, one can observe that the polarization constraint set $\mathbb{P}$, defined in equation~\eqref{eq:flux_cons}, can be equivalently rewritten as
\begin{multline} \label{eq:flux_recast}
\mathbb{P}  = \left\{ \bm{\mathsf{S}} \in \mathbb{R}^{N \times 4} \, \middle\vert \,    \big( \forall  n \in \{ 1, \ldots, N \} \big)\right. \\
 \left. h_1({\mathsf{S}}_{n,1}) + h_2 (\bm{\mathsf{S}}_{n,2:4}) \leq 0  \right\}.
\end{multline}
Therefore, the Stokes matrix $\bm{\mathsf{S}}$ satisfying the constraint defined by set $\mathbb{P}$ is equivalent to have the variables $(\bm{\mathsf{S}}, \bm{\mathsf{Z}})$ satisfying the constraints defined by the equations~\eqref{eq:h1_cons} - \eqref{eq:zeta_cons}.

To simplify the notation of the minimization problem~\eqref{eq:min_gen_g}, we need to introduce the definition of the epigraph of a proper, lower semi-continuous function $\psi \colon \mathbb{R}^N \to ]-\infty, +\infty]$. Essentially, it corresponds to the set of points lying on or above the graph of $\psi$, and formally, it is given by \citep{Rockafellar1997}$\colon$
\begin{equation} \label{eq:epi_def}
\operatorname{epi} \psi = \left\{ (\bm{z}, \gamma) \in \mathbb{R}^N \times \mathbb{R} \, \middle\vert \, \psi(\bm{z}) \leq \gamma \right\}.
\end{equation}
%
%
%
%
\begin{algorithm}
\caption{Primal-dual algorithm for joint Stokes imaging}\label{algo_polar}
\begin{algorithmic}[1]

\vspace*{0.1cm}
\State {\textbf{given}}
$\, \bm{\mathsf{S}}^{(0)} \in \mathbb{R}^{N \times 4}, \, \bm{\mathsf{Z}}^{(0)} \in \mathbb{R}^{N \times 2}, \, \bm{\mathsf{A}}^{(0)} \in \mathbb{R}^{J \times 4}, \bm{\mathsf{B}}^{(0)} \in \mathbb{R}^{M \times 4}, \, \bm{\mathsf{C}}^{(0)} \in \mathbb{R}^{N \times 4}, \, \bm{\mathsf{D}}^{(0)} \in \mathbb{R}^{N \times 2}, \gamma \bm{\mathsf{W}} \in \mathbb{R}_+^{J \times 4}$

\vspace*{0.1cm}

\State \textbf{repeat for} $k = 0, 1, \ldots$
\vspace*{0.1cm}
\Statex \quad \, \fbox{\textbf{Primal updates}}

\vspace{0.1cm}

\State  \label{alg:primal_update}
\quad \, $\bm{\mathsf{S}}^{(k+1)} = \bm{\mathcal{P}}_{\mathbb{U}} \bigg(\bm{\mathsf{S}}^{(k)} - \tau \, \big(\mathsf{\bm{\Psi}} \bm{\mathsf{A}}^{(k)} + \phi^\dagger (\bm{\mathsf{B}}^{(k)}) + \bm{\mathsf{C}}^{(k)} \big) \bigg)$

\vspace*{0.2cm}

\State \label{alg:primal_update_zeta}
\quad \, $\bm{\mathsf{Z}}^{(k+1)} = \bm{\mathcal{P}}_{\mathbb{V}} \bigg(\bm{\mathsf{Z}}^{(k)} - \tau \bm{\mathsf{D}}^{(k)} \bigg)$

\vspace*{0.2cm}
\Statex \quad \, \fbox{\textbf{Dual updates}}

\vspace*{0.1cm}

\Statex \, \quad \, \underline{Promoting sparsity:}

\vspace*{0.1cm}
\State \label{alg:a_update}
\quad \, $\widetilde{\bm{\mathsf{A}}}^{(k)} = \bm{\mathsf{A}}^{(k)} + \rho_1 \mathsf{\bm{\Psi}}^\dagger \bigg( 2 \, \bm{\mathsf{S}}^{(k+1)} - \bm{\mathsf{S}}^{(k)} \bigg)$

\vspace*{0.2cm}

\State \label{alg:prox_l1}
\quad \, $\bm{\mathsf{A}}^{(k+1)} = \widetilde{\bm{\mathsf{A}}}^{(k)} - \rho_1 \bm{\mathcal{T}}_{\gamma \bm{\mathsf{W}}/\rho_1} \bigg(  \widetilde{\bm{\mathsf{A}}}^{(k)}/\rho_1 \bigg)$

\vspace*{0.1cm}

\Statex \, \quad \, \underline{Projecting onto $\ell_2$ ball:}

\vspace*{0.1cm}

\State \label{alg:b_update}
\quad \, $\widetilde{\bm{\mathsf{B}}}^{(k)} = \bm{\mathsf{B}}^{(k)} + \rho_2 \, \phi \bigg( 2 \, \bm{\mathsf{S}}^{(k+1)} - \bm{\mathsf{S}}^{(k)} \bigg)$

\vspace*{0.2cm}

\State \label{alg:prox_l2}
\quad \, $\bm{\mathsf{B}}^{(k+1)} = \widetilde{\bm{\mathsf{B}}}^{(k)} - \rho_2 \, \bm{\mathcal{P}}_{\mathbb{B}(\bm{\mathsf{Y}}, \epsilon/\rho_2)} \bigg(  \widetilde{\bm{\mathsf{B}}}^{(k)}/\rho_2 \bigg)$

\vspace*{0.1cm}

\Statex \, \quad \, \underline{Performing epigraphical projection:}

\vspace{0.1cm}
\State \label{alg:c_update}
\quad \, $ \widetilde{\bm{\mathsf{C}}}^{(k)} = \bm{\mathsf{C}}^{(k)} + \rho_3 \bigg( 2 \, \bm{\mathsf{S}}^{(k+1)} - \bm{\mathsf{S}}^{(k)} \bigg)$

\vspace*{0.2cm}

\State \label{alg:c_zeta1_update}
\quad \, $\widetilde{\bm{\mathsf{D}}}^{(k)} = \bm{\mathsf{D}}^{(k)} + \rho_3 \bigg( 2 \, \bm{\mathsf{Z}}^{(k+1)} - \bm{\mathsf{Z}}^{(k)} \bigg)$

\vspace*{0.2cm}

\State \label{alg:proj_epi_h1}
\quad \, $\left[\begin{matrix}
\bm{\mathsf{C}}_{:,1}^{(k+1)} \\
\vspace*{-0.2cm} \\
 \bm{\mathsf{D}}_{:,1}^{(k+1)}
\end{matrix}\right]
= 
\left[\begin{matrix}
\widetilde{\bm{\mathsf{C}}}_{:,1}^{(k)} \\
\vspace*{-0.2cm} \\
\widetilde{\bm{\mathsf{D}}}_{:,1}^{(k)}
\end{matrix}\right]
- \rho_3 \, \bm{\mathcal{P}}_{\mathbb{E}_1}
\left(\frac{1}{\rho_3} \left[\begin{matrix}
\widetilde{\bm{\mathsf{C}}}_{:,1}^{(k)} \\
\vspace*{-0.2cm} \\
\widetilde{\bm{\mathsf{D}}}_{:,1}^{(k)}
\end{matrix}\right]
\right)$

\vspace*{0.2cm}

\State \label{alg:proj_epi_h2}
\quad \, $\left[\begin{matrix}
\bm{\mathsf{C}}_{:,2:4}^{(k+1)} \\
\vspace*{-0.2cm} \\
\bm{\mathsf{D}}_{:,2}^{(k+1)}
\end{matrix}\right]
= 
\left[\begin{matrix}
\widetilde{\bm{\mathsf{C}}}_{:,2:4}^{(k)} \\
\vspace*{-0.2cm} \\
\widetilde{\bm{\mathsf{D}}}_{:,2}^{(k)}
\end{matrix}\right]
- \rho_3 \, \bm{\mathcal{P}}_{\mathbb{E}_2}
\left(\frac{1}{\rho_3} \left[\begin{matrix}
 \widetilde{\bm{\mathsf{C}}}_{:,2:4}^{(k)} \\
  \vspace*{-0.2cm} \\
 \widetilde{\bm{\mathsf{D}}}_{:,2}^{(k)} 
\end{matrix}\right]
\right)
$

\vspace*{0.2cm}

\State \textbf{until convergence} \label{alg:d2_update_end}

\end{algorithmic}
\end{algorithm}
%
%
Using this definition, conditions~\eqref{eq:h1_cons} and ~\eqref{eq:h2_cons} respectively represent the epigraph of the functions $h_1$ and $h_2$. More precisely, condition~\eqref{eq:h1_cons} implies that for every $n \in \{1,\dots,N\}$, $({\mathsf{S}}_{n,1}, {\mathsf{Z}}_{n,1}) \in \operatorname{epi} h_1$. For a compact notation, we define $\mathbb{E}_1 = (\operatorname{epi} h_1)^N $ to be the product space such that
\begin{equation}
(\bm{\mathsf{S}}_{:,1}, \bm{\mathsf{Z}}_{:,1}) \in \mathbb{E}_1
\Leftrightarrow 
(\forall n \in \{1,\dots,N\}) \, \, ({\mathsf{S}}_{n,1}, {\mathsf{Z}}_{n,1}) \in \operatorname{epi} h_1.
\end{equation}
Similarly, defining $\mathbb{E}_2 = (\operatorname{epi} h_2)^N $, condition~\eqref{eq:h2_cons} is equivalent to
\begin{equation}
(\bm{\mathsf{S}}_{:,2:4},\bm{\mathsf{Z}}_{:,2} \big) \in \mathbb{E}_2
\Leftrightarrow 
(\forall n \in \{1,\dots,N\}) \, \,
 (\bm{\mathsf{S}}_{n,2:4}, {\mathsf{Z}}_{n,2}) \in \operatorname{epi} h_2.
\end{equation}
Thus, the constraints~\eqref{eq:h1_cons} and~\eqref{eq:h2_cons} can be imposed as $\iota_{\mathbb{E}_1}(\bm{\mathsf{S}}_{:,1}, \bm{\mathsf{Z}}_{:,1})$ and $\iota_{\mathbb{E}_2}  (\bm{\mathsf{S}}_{:,2:4}, \bm{\mathsf{Z}}_{:,2} )$, respectively. Furthermore, to impose condition~\eqref{eq:zeta_cons}, we introduce 
\begin{equation} \label{eq:set_V}
\mathbb{V} = \left\{ \bm{\mathsf{Z}} \in \mathbb{R}^{N \times 2} \, \middle\vert \, \big(\forall  n \in \{ 1, \ldots, N \} \big)  \; {{\mathsf{Z}}}_{n,1} + {{\mathsf{Z}}}_{n,2} \leq 0  \right\}.
\end{equation}
Then, condition~\eqref{eq:zeta_cons} can be represented as an indicator function of the set $\mathbb{V}$.
Finally, imposing the constraints~\eqref{eq:h1_cons} - \eqref{eq:zeta_cons} using their respective indicator functions leads to the following minimization problem
\begin{multline} \label{eq:min_gen2}
\underset{\substack{\bm{\mathsf{S}} \in \mathbb{R}^{N \times 4}, \\ \bm{\mathsf{Z}} \in \mathbb{R}^{N \times 2}}} {\operatorname{minimize}}  \, \,   f \big( \phi (\bm{\mathsf{S}}) \big) + \iota_{\mathbb{U}} (\bm{\mathsf{S}})  +  \gamma \, g(\mathsf{\bm{\Psi}}^\dagger \bm{\mathsf{S}})
 + \iota_{\mathbb{V}}(\bm{\mathsf{Z}}) 
  \\  
 + \iota_{\mathbb{E}_1}(\bm{\mathsf{S}}_{:,1}, \bm{\mathsf{Z}}_{:,1}) +  \iota_{\mathbb{E}_2}  (\bm{\mathsf{S}}_{:,2:4}, \bm{\mathsf{Z}}_{:,2} ),
\end{multline}
where $\gamma > 0$ is a free parameter only affecting the convergence speed. 
%

%
%
%
%

Note that the minimization problem~\eqref{eq:min_gen2} considers the SARA regularization and imposes the polarization constraint explicitly. We refer to this proposed method of joint Stokes imaging as Polarized SARA. In the same line of thought, solving the problem of Stokes imaging with SARA regularization but without polarization constraint, i.e. problem~\eqref{eq:min_gen_matrix_no_flux}, is termed as Polarized SARA without constraint.

\subsection{Algorithm formulation}
In order to solve the resultant problem~\eqref{eq:min_gen2}, we develop a method based on a primal-dual forward-backward algorithm, which offers a highly flexible and parallelizable structure \citep{Condat2013, Vu2013, Pesquet2015}. The proposed algorithm is given in Algorithm~\ref{algo_polar}. As described in Section~\ref{ssec:opt_frame}, primal-dual methods consist in solving jointly the primal and the dual problem. In our case, the primal problem to be solved is given in equation~\eqref{eq:min_gen2}. This problem can be written in a compact form as follows
\begin{align} \label{eq:min_gen_simp}
\underset{\substack{\bm{\mathsf{S}} \in \mathbb{R}^{N \times 4}, \\ \bm{\mathsf{Z}} \in \mathbb{R}^{N \times 2}}} {\operatorname{minimize}} \, \, &  {q}(\bm{\mathsf{S}}, \bm{\mathsf{Z}}) + \gamma \, \widetilde{g}(\mathsf{\bm{\Psi}}^\dagger \bm{\mathsf{S}}, \bm{\mathsf{Z}}) + \widetilde{f} \big( {{\phi}} (\bm{\mathsf{S}} ), \bm{\mathsf{Z}} \big) + {l}(\bm{\mathsf{S}}, \bm{\mathsf{Z}}),
\end{align}
where,
\begin{alignat*}{2}
& {q}(\bm{\mathsf{S}}, \bm{\mathsf{Z}}) && = \iota_{\mathbb{U}} (\bm{\mathsf{S}}) + \iota_{\mathbb{V}} (\bm{\mathsf{Z}}), \\
& \widetilde{g}(\mathsf{\bm{\Psi}}^\dagger \bm{\mathsf{S}}, \bm{\mathsf{Z}}) && = g(\mathsf{\bm{\Psi}}^\dagger \bm{\mathsf{S}}), \\
& \widetilde{f} \big({{\phi}} (\bm{\mathsf{S}} ), \bm{\mathsf{Z}} \big) && = f \big( {{\phi}} (\bm{\mathsf{S}}) \big), \\
& {l}(\bm{\mathsf{S}}, \bm{\mathsf{Z}}) && = \iota_{\mathbb{E}_1}(\bm{\mathsf{S}}_{:,1}, \bm{\mathsf{Z}}_{:,1}) +  \iota_{\mathbb{E}_2} (\bm{\mathsf{S}}_{:,2:4},  \bm{\mathsf{Z}}_{:,2} ).
\end{alignat*}
Then, according to Section~\ref{ssec:opt_frame} (see e.g. \citet{Komodakis2015} for further detail), the dual problem associated with~\eqref{eq:min_gen_simp} is given by
%
\begin{multline} \label{eq:min_gen2_dual}
\underset{\substack{\bm{\mathsf{A}} \in \mathbb{R}^{J \times 4},  \bm{\mathsf{B}} \in \mathbb{C}^{M \times 4}, \\ \bm{\mathsf{C}} \in \mathbb{R}^{N \times 4},  \bm{\mathsf{D}} \in \mathbb{R}^{N \times 2}}} {\operatorname{minimize}} \, \,  {q}^* (- \mathsf{\bm{\Psi}} \bm{\mathsf{A}} - \phi^\dagger(\bm{\mathsf{B}}) - \bm{\mathsf{C}}, - \bm{\mathsf{D}}) + \frac{1}{{\gamma}} \, \widetilde{g}^* (\bm{\mathsf{A}}, \bm{\mathsf{D}})   \\ 
+  \widetilde{f}^*(\bm{\mathsf{B}}, \bm{\mathsf{D}}) + {l}^* (\bm{\mathsf{C}}, \bm{\mathsf{D}}) .
\end{multline}
Note that in problem~\eqref{eq:min_gen2_dual}, $\bm{\mathsf{A}} \in \mathbb{R}^{J \times 4} $ is the dual variable corresponding to the non-smooth $\ell_1$ term (the function $g$ in problem~\eqref{eq:min_gen2}), $ \bm{\mathsf{B}} \in \mathbb{C}^{M \times 4} $ is the dual variable associated with the data fidelity term (the function $f$ in problem~\eqref{eq:min_gen2}), whereas $ \bm{\mathsf{C}} \in \mathbb{R}^{N \times 4} $ and $\bm{\mathsf{D}} \in \mathbb{R}^{N \times 2} $ are associated with the indicator functions of the epigraphs of $h_1$ and $h_2$ in problem~\eqref{eq:min_gen2}.

Using this primal-dual formulation, Algorithm~\ref{algo_polar} solves alternately for the above mentioned primal and dual problems. In this regard, the algorithm can be seen as consisting of two major steps, denoted by: Primal updates and Dual updates. Essentially, each iteration $k \in \mathbb{N}$ involves updating the primal variables $\bm{\mathsf{S}}$ and $\bm{\mathsf{Z}}$, followed by the update of the dual variables $\bm{\mathsf{A}}, \bm{\mathsf{B}}$, $\bm{\mathsf{C}}$ and $\bm{\mathsf{D}}$.   
For each such iteration $k$, these two major steps are detailed in the following.

\subsubsection{Primal updates}
In Algorithm~\ref{algo_polar}, the primal variables are incremented using steps~\ref{alg:primal_update} and~\ref{alg:primal_update_zeta}. These updates have a structure reminiscent of the forward-backward steps. More precisely, forward-backward step consists of alternating between a gradient step and a proximity (or projection) step, whereas in the absence of any smooth term, only the proximity step is performed. This structure can be observed in the update of $\bm{\mathsf{S}}$ (i.e. step~\ref{alg:primal_update}), where a projection onto the set $\mathbb{U}$ is carried out. One can notice that an additive term appears in this update. This additional term allows to take into account the dual variables $\bm{\mathsf{A}}$, $\bm{\mathsf{B}}$ and $\bm{\mathsf{C}}$ associated with the sparsity prior, the data fidelity term and the epigraphical constraints, respectively. 
The same analogy can be observed for the updating step~\ref{alg:primal_update_zeta} of the variable $\bm{\mathsf{Z}}$. This update takes into account the dual variable $\bm{\mathsf{D}}$ associated with the epigraphical constraints, followed by a projection onto the set $\mathbb{V}$. 

Note that in steps~\ref{alg:primal_update} and~\ref{alg:primal_update_zeta}, the projections onto the sets $\mathbb{U}$ and $\mathbb{V}$, respectively, need to be performed. Firstly, for any matrix $\bm{\mathsf{X}}$ of size $N \times 4$, the projection onto the set $\mathbb{U}$ is simply given by
\begin{equation} \label{eq:real_proj}
\bm{\mathcal{P}}_{\mathbb{U}} \big( \bm{\mathsf{X}} \big) = \operatorname{Re} \big({\bm{\mathsf{X}}} \big),
\end{equation}
where the operator $\operatorname{Re} (.)$ provides the real part of its argument.

Secondly, using Proposition 2.1 from \citet{Chierchia2015}, the projection onto the set $\mathbb{V}$ is performed as
\begin{equation}
(\forall \, \bm{\mathsf{U}} \in \mathbb{R}^{N \times 2}) \quad \bm{\mathcal{P}}_{\mathbb{V}} (\bm{\mathsf{U}}) = \widetilde{\bm{\mathsf{U}}}, 
\end{equation}
where $\widetilde{\bm{\mathsf{U}}} \in \mathbb{R}^{N \times 2}$ is defined such that, for every $n \in \{1,\dots,N \}$, 
\begin{equation}
\big(\widetilde{{\mathsf{U}}}_{n,1}, \widetilde{{\mathsf{U}}}_{n,2}\big) = \begin{cases}
       ({{\mathsf{U}}}_{n,1}, {{\mathsf{U}}}_{n,2}),& \text{if} \,\, {{\mathsf{U}}}_{n,1} + {{\mathsf{U}}}_{n,2} \leq 0, \\
                                         \frac{1}{2} ({{\mathsf{U}}}_{n,1} -{{\mathsf{U}}}_{n,2}, {{\mathsf{U}}}_{n,2} - {{\mathsf{U}}}_{n,1}),&\text{otherwise}.
                                         \end{cases} 
\end{equation}

\subsubsection{Dual updates} 
The dual variables are updated in steps~\ref{alg:a_update} - \ref{alg:proj_epi_h2} of Algorithm~\ref{algo_polar}.
At each iteration $k$, it requires the evaluation of the proximity operators of the associated functions.

\paragraph*{Sparsity prior.} Steps~\ref{alg:a_update} and~\ref{alg:prox_l1} consist in updating the dual variable $\bm{\mathsf{A}}$ associated with the sparsity prior defined in equation~\eqref{eq:g_l1}. In particular, step~\ref{alg:prox_l1} requires to perform the proximity operator of the $\ell_1$ norm, which corresponds to the soft-thresholding operation \citep{Chaux2007}, using the soft-threshold sizes given by $\gamma \bm{\mathsf{W}} \in \mathbb{R}_+^{J \times 4}$. Furthermore, $\bm{\mathcal{T}}$ stands for the soft-thresholding operator and is defined as
\begin{equation}
(\forall \, {\bm{\mathsf{A}}} \in \mathbb{R}^{J \times 4}) \quad \bm{\mathcal{T}}_{\gamma \bm{\mathsf{W}}/\rho_1} \bigg( {\bm{\mathsf{A}}/\rho_1} \bigg) = {\bm{\mathsf{H}}}\,,
\end{equation}
where ${\bm{\mathsf{H}}} \in \mathbb{R}^{J \times 4}$ is defined as follows
\begin{equation} \label{eq:soft_thresh}
(\forall i \in \{1,2,3,4\}) \quad {\bm{\mathsf{H}}}_{:,i} = \operatorname{prox}_{\rho_1^{-1} \gamma \bm{\mathsf{W}}_{\colon,i} \|.\|_1}\, \big({\bm{\mathsf{A}}}_{:,i} \big).
\end{equation}
Note that the soft-thresholding operation is performed component-wise. Therefore, in equation~\eqref{eq:soft_thresh}, with $ j \in \{1, \ldots, J\}$, the $j^{\text{th}}$ component of each vector ${\bm{\mathsf{H}}}_{:,i}$ is given by
\begin{equation}
 {{\mathsf{H}}}_{j,i} = \begin{cases}
               - {{\mathsf{A}}}_{j,i}  + \gamma {\mathsf{W}}_{j,i} /\rho_1 \, , & \text{if} \,\, {{\mathsf{A}}}_{j,i} < -\gamma {\mathsf{W}}_{j,i}/\rho_1\, , \\
               0, & \text{if} \,\, - \gamma {\mathsf{W}}_{j,i}/\rho_1 \leq {{\mathsf{A}}}_{j,i} \leq \gamma {\mathsf{W}}_{j,i}/\rho_1\, , \\
               {{\mathsf{A}}}_{j,i}  -  \gamma {\mathsf{W}}_{j,i}/\rho_1\, , & \text{otherwise}.
               \end{cases} 
\end{equation} 
Essentially, this operation forces the elements, which are smaller than some threshold, to zero, while the rest of the elements are reduced by this threshold value. Therefore, iteration-by-iteration, it removes the smaller values and finally, only the elements having significant values are left, hence promoting sparsity.

\paragraph*{Data fidelity.} Steps~\ref{alg:b_update} and~\ref{alg:prox_l2} are involved in the update of the dual variable $\bm{\mathsf{B}}$ corresponding to the data fidelity term given in equation~\eqref{eq:f_diff}. This update consists in performing the projection onto the $\ell_2$ ball ${\mathbb{B}} (\bm{\mathsf{Y}}, \theta)$ of radius $\theta = \epsilon/ \rho_2$ and centred in $\bm{\mathsf{Y}}$. For any matrix ${\bm{\mathsf{B}}} \in \mathbb{C}^{M \times 4}$, the corresponding projection is given by
\begin{equation}
\bm{\mathcal{P}}_{\mathbb{B}(\bm{\mathsf{Y}}, \theta)} (\bm{\mathsf{B}}/\rho_2) = 
\begin{cases}
 \displaystyle \theta \frac{(\bm{\mathsf{B}}/\rho_2) - \bm{\mathsf{Y}}}{\| (\bm{\mathsf{B}}/\rho_2)  - \bm{\mathsf{Y}}\|_2 } + \bm{\mathsf{Y}}, & \text{if} \, \|(\bm{\mathsf{B}}/\rho_2)  - \bm{\mathsf{Y}}\|_2 > \theta, \\
\bm{\mathsf{B}}/\rho_2, & \text{otherwise}.
               \end{cases} 
\end{equation}

\paragraph*{Epigraphical projections and polarization constraint.}
Steps~\ref{alg:c_update} -~\ref{alg:proj_epi_h2} consist in updating the dual variables $\bm{\mathsf{C}}$ and $\bm{\mathsf{D}}$, which are associated with the epigraphical projections on the functions $h_1$ and $h_2$. Recall that these projections are required to enforce the polarization constraint. In particular, in step~\ref{alg:proj_epi_h1}, the vectors $(\widetilde{\bm{\mathsf{C}}}_{:,1}, \widetilde{\bm{\mathsf{D}}}_{:,1})$ need to be projected onto the epigraph of $h_1$, while in step~\ref{alg:proj_epi_h2}, the projection of $(\widetilde{\bm{\mathsf{C}}}_{:,2:4}, \widetilde{\bm{\mathsf{D}}}_{:,2} \big)$ onto the epigraph of $h_2$ is required. These projections can be performed in the following manner. 
Firstly, for any two variables $(\bm{c}, \bm{d}) \in (\mathbb{R}^N)^2$, following \citet{Chierchia2015}, the projection onto the epigraph of the function $h_1$ boils down to
\begin{equation} \label{eq:proj_e1}
\bm{\mathcal{P}}_{\mathbb{E}_1} (\bm{c}, \bm{d}) = (\widetilde{{c}}_{n}, \widetilde{{d}}_{n})_{1 \leq n \leq N}, 
\end{equation}
where
\begin{equation}
(\widetilde{{c}}_{n}, \widetilde{{d}}_{n}) = \begin{cases}                                        
          ({c}_{n}, {d}_{n}), & \text{if} \, \, {c}_{n} + {d}_{n} \geq 0, \\
          \bigg( \frac{1}{2} ({c}_{n} - {d}_{n}), \frac{{d}_{n} - {c}_{n}}{2} \bigg), &  \text{otherwise}.
          \end{cases} 
\end{equation}   
Secondly, for every $({\bm{\mathsf{R}}} \in \mathbb{R}^{N \times 3}, \bm{d} \in \mathbb{R}^N)$, the projection onto the epigraph of the function $h_2$ is given by
\begin{equation} \label{eq:proj_e2}
\bm{\mathcal{P}}_{\mathbb{E}_2} \big( {\bm{\mathsf{R}}}, \bm{d} \big) = (\widetilde{{\bm{\mathsf{R}}}}_{n,\colon}, \widetilde{{d}}_{n})_{1 \leq n \leq N},
\end{equation}
such that
\begin{align} \label{eq:proj_e2_def}
(& \widetilde{{\bm{\mathsf{R}}}}_{n,:}, \, \widetilde{{d}}_{n}) =  \nonumber \\
& \begin{cases}
   \big( \bm{0}, 0 \big), & \text{if} \, \, \| {\bm{\mathsf{R}}}_{n,\colon}
    \|_2 < - {d}_{n}, \\
 \big( {\bm{\mathsf{R}}}_{n,\colon},  {d}_{n} \big),     &  \text{if} \, \, \| {\bm{\mathsf{R}}}_{n,\colon} \|_2 <  {d}_{n}, \\
  \alpha_n \big( {\bm{\mathsf{R}}}_{n,\colon}, \| {\bm{\mathsf{R}}}_{n,\colon} \|_2 \big) , & \text{otherwise},
 \end{cases}
\end{align}  
 where $\alpha_n = \frac{1}{2} \bigg( 1 + \frac{{d}_{n}}{\| {\bm{\mathsf{R}}}_{n,\colon} \|_2} \bigg)$ \citep{Chierchia2015}.                                    

\subsubsection{Convergence properties}

The choice of the step sizes $\tau, \rho_1, \rho_2$ and $\rho_3$ governs the convergence of Algorithm~\ref{algo_polar} to the solution of the minimization problem~\eqref{eq:min_gen2}. More precisely, these parameters should be chosen in a manner that the following holds \citep{Pesquet2015}
\begin{equation}
\frac{1}{\tau} - \rho_1 \, \| {\mathsf{\bm{\Psi}}}\|_{sp}^2 - \rho_2 \| \phi \|_{sp}^2 - \rho_3 \geq 0,
\end{equation}
where $\|.\|_{sp}$ denotes the spectral norm of its argument. 
Then, under this condition, the sequence of iterates $(\bm{\mathsf{S}}^{(k)})_{k \in \mathbb{N}}$ generated by Algorithm~\ref{algo_polar} converges to a solution to problem~\eqref{eq:min_gen2}. 

Additional remarks can be made regarding the proposed approach. The whole procedure of updating primal variables followed by the dual variables in Algorithm~\ref{algo_polar} is repeated until the required convergence criterion, defined by the user, is met. Furthermore, it is worth mentioning that the different steps involved in Algorithm~\ref{algo_polar} can be implemented in parallel for each vector within the defined matrices. It therefore presents a highly parallelizable algorithm.

\subsubsection{Reweighting scheme}
As discussed earlier, we aim to solve the weighted $\ell_1$ minimization problem iteratively, and hence approaching towards the solution in the $\ell_0$ sense. 
More precisely, each reweighting iteration, indexed by $l \in \mathbb{N}$, consists in solving the weighted $\ell_1$ minimization problem~\eqref{eq:min_gen2} using Algorithm~\ref{algo_polar}. 
The weights for each iteration are computed from the previous solution of the Stokes images as follows:
\vspace*{-0.15cm}
\begin{equation} \label{eq:weights}
(\forall j \in \{1, \ldots, J\}) \quad \mathsf{W}_{j,i}^{(l+1)} = \frac{\delta^{(l+1)}}{\delta^{(l+1)} +  \left| \left[\mathsf{\bm{\Psi}}^\dagger \bm{\mathsf{S}}^{(l)} \right]_{j,i} \right| } \, ,
\end{equation}
where $i \in \{1, \ldots, 4\}$ and $\delta^{(l+1)} > 0$ is decreased iteration-by-iteration such that $\delta^{(l+1)} \to 0$ when $l \to +\infty$, and hence the weighted $\ell_1$ norm approaches to the $\ell_0$ norm. The newly computed weights are used to update the soft-threshold sizes as $\gamma \bm{\mathsf{W}}^{(l+1)}$.

To initialize the weights, recall that one of the benefits of using this reweighting scheme is to avoid the tuning of any additional regularization parameters. Keeping this in mind, Algorithm~\ref{algo_polar} is used to solve the problem consisting of only data fidelity term and the positivity constraint, without imposing sparsity and polarization constraint, i.e. problem~\eqref{eq:min_gen_matrix_no_flux} with function $g = 0$. Note that this corresponds to solving a constrained version of the non-negative least squares problem. Formally, in the algorithm, it consists of updating the primal variable $\bm{\mathsf{S}}$ and the dual variable $\bm{\mathsf{B}}$, with the rest of the primal and dual variables appearing in Algorithm~\ref{algo_polar} taken to be zero, and the set $\mathbb{U}$ modified to set $\mathbb{U}^\prime$ to take into account positivity of the total intensity image. 
The solutions obtained are then used to compute the weights (as defined in~\eqref{eq:weights}) for $l = 1$ and hence the soft-thresholding sizes for the first reweighting iteration. Subsequent reweighting iterations solve for the problem~\eqref{eq:min_gen2} by passing on the updated sizes alongwith the solutions of the primal and dual variables from the previous iteration to Algorithm~\ref{algo_polar}. The resultant reweighting scheme is described in Algorithm~\ref{algo_rw}. This process is repeated until convergence. 
%
\begin{algorithm}
\caption{Reweighting scheme}\label{algo_rw}
\begin{algorithmic}[1]
\State {\textbf{given}} 
$\, \bm{\mathsf{S}}^{(0)}, \bm{\mathsf{Z}}^{(0)}, \bm{\mathsf{A}}^{(0)}, \bm{\mathsf{B}}^{(0)}, \bm{\mathsf{C}}^{(0)}, \bm{\mathsf{D}}^{(0)}, \gamma \bm{\mathsf{W}}^{(0)}$

\vspace*{0.1cm}

\State \textbf{repeat for} $l = 0, 1, \ldots$
\vspace*{0.1cm}

\State \hspace*{0.2cm} $[\bm{\mathsf{S}}^{(l+1)}, \bm{\mathsf{Z}}^{(l+1)}, \bm{\mathsf{A}}^{(l+1)}, \bm{\mathsf{B}}^{(l+1)}, \bm{\mathsf{C}}^{(l+1)}, \bm{\mathsf{D}}^{(l+1)}]$ = Algorithm~\ref{algo_polar}

\vspace*{0.1cm}

\State \hspace*{0.2cm} $\forall i \in \{1,2,3,4\}$, \textbf{compute} $\bm{\mathsf{W}}_{\colon,i}^{(l+1)}$ \text{as per equation~\eqref{eq:weights}}

\vspace*{0.1cm}

\State \textbf{until convergence}

\end{algorithmic}
\end{algorithm}
%

\subsection{Polarization constraint for TV based problems}
\label{ssec:gen_TV}

In the context of EHT imaging for full polarization, the authors in \cite{Akiyama2017b} consider independent problems for each of the Stokes parameters. This is achieved by post-multiplying equation~\eqref{eq:RIME_disc} with ${\bm{\mathsf{L}}^{-1}}$, the inverse of the operator $\bm{\mathsf{L}}$, thereby converting brightness matrix back to Stokes matrix and giving $\bm{\mathsf{Y}} {\bm{\mathsf{L}}^{-1}}  =  \widetilde{\phi} ( \overline{\bm{\mathsf{S}}}) + \bm{\mathsf{E}} {\bm{\mathsf{L}}^{-1}}$. 
Here $\widetilde{\phi}$ is a modified operator incorporating ${\bm{\mathsf{L}}^{-1}}$ in the operator $\phi$. In this manner, in the absence of DDEs, each column of $\widetilde{\bm{\mathsf{Y}}} = \bm{\mathsf{Y}} {\bm{\mathsf{L}}^{-1}}$ corresponds to the measured visibilities related to the respective Stokes parameter in the matrix $\overline{\bm{\mathsf{S}}}$. 
The authors then propose to solve the following minimization problem
\begin{equation} \label{eq:Akiyama}
\underset{\bm{\mathsf{S}}\in \mathbb{R}^{N \times 4}} {\operatorname{minimize}}  \, \,   \frac{1}{2} \| \widetilde{\bm{\mathsf{Y}}} - \widetilde{\phi} ( {\bm{\mathsf{S}}}) \|_2^2 +      \iota_{\mathbb{U}^\prime}(\bm{\mathsf{S}}) + \check{g}(\mathsf{\bm{\Psi}}^\dagger \bm{\mathsf{S}}), 
\end{equation}
where the unconstrained problem formulation is used and the data fidelity term is given by the squared $\ell_2$ term (first term in~\eqref{eq:Akiyama}). The polarization constraint is not imposed, justifying the use of the set ${\mathbb{U}^\prime}$ to impose positivity of the total intensity image. Furthermore, the third term, $\check{g}(\mathsf{\bm{\Psi}}^\dagger \bm{\mathsf{S}})$ in~\eqref{eq:Akiyama} is the sparsity prior imposing the sparsity of the sought images in some sparsifying dictionary. In this regard, \cite{Akiyama2017b} shows that the TV and $\ell_1 + $TV sparsifying regularizations are effective in producing super-resolved images and lead to better reconstruction quality than using the standard \textsc{clean} method. Note that for these regularizations, the authors have considered the isotropic TV norm. Basically, the TV norm is defined for a 2D image $\bm{\mathsf{U}} \in \mathbb{R}^{N_1 \times N_2}$ as the $\ell_{2,1}$ norm of the horizontal and the vertical gradients of this image \citep{Rudin1992}\footnote{In the current work, the images are represented in vectorized form of dimension $N= N_1 \times N_2$. However, the 2D images can easily be obtained by reshaping these vectors.}. 
More formally, it is given by 
\begin{align} \label{eq:TV}
\| \bm{\mathsf{U}} \|_{\operatorname{TV}} & = \| \bm{\nabla} \bm{\mathsf{U}} \|_{2,1} \nonumber \\ 
& = \sum_{n_1 =1}^{N_1} \sum_{n_2 =1}^{N_2} \sqrt{| [\nabla_{x} \bm{\mathsf{U}}]_{n_1,n_2} |^2 + | [\nabla_{y} \bm{\mathsf{U}}]_{n_1,n_2} |^2} ,
\end{align}
where $\bm{\nabla} = [\nabla_x, \nabla_y] $ is the concatenation of the horizontal gradient operator $\nabla_x \colon \mathbb{R}^{N_1 \times N_2} \to \mathbb{R}^{N_1 \times N_2}$ and the vertical gradient operator $\nabla_y \colon \mathbb{R}^{N_1 \times N_2} \to \mathbb{R}^{N_1 \times N_2}$. 

We hereby propose to generalize the minimization problem~\eqref{eq:Akiyama} solved by \cite{Akiyama2017b} to incorporate explicitly the polarization constraint. It amounts to 
\begin{multline} \label{eq:Akiyama_2}
\underset{\substack{\bm{\mathsf{S}} \in \mathbb{R}^{N \times 4}, \\ \bm{\mathsf{Z}} \in \mathbb{R}^{N \times 2}}} {\operatorname{minimize}}  \, \,   \frac{1}{2} \| \widetilde{\bm{\mathsf{Y}}} - \widetilde{\phi}(\bm{\mathsf{S}}) \|_2^2  + \iota_{\mathbb{U}} (\bm{\mathsf{S}})  +   \check{g}(\mathsf{\bm{\Psi}}^\dagger \bm{\mathsf{S}})
 + \iota_{\mathbb{V}}(\bm{\mathsf{Z}}) 
  \\  
 + \iota_{\mathbb{E}_1}(\bm{\mathsf{S}}_{:,1}, \bm{\mathsf{Z}}_{:,1}) +  \iota_{\mathbb{E}_2}  (\bm{\mathsf{S}}_{:,2:4}, \bm{\mathsf{Z}}_{:,2} ).
\end{multline}
This problem can be solved by using a modified version of the primal-dual method proposed in Algorithm~\ref{algo_polar}. In particular, Algorithm~\ref{algo_polar} can incorporate any convex sparsity regularization function, and can be adapted for the unconstrained problem of interest. The resultant algorithm is provided in Algorithm~\ref{algo_polar_TV}, consisting of the following amendments made in Algorithm~\ref{algo_polar}.

\noindent (i). While Algorithm~\ref{algo_polar} has been provided for the constrained formulation with the data fidelity term defined in~\eqref{eq:f_diff}, problem~\eqref{eq:Akiyama_2} where the data consistency is instead ensured by a differentiable squared $\ell_2$ term, can still be solved. More precisely, the update of variable $\bm{\mathsf{B}}$ in Algorithm~\ref{algo_polar} is no longer required, instead the gradient term $(\widetilde{\phi}^\dagger (\widetilde{\phi}(\bm{\mathsf{S}}) -\widetilde{\bm{\mathsf{Y}}}))$ is added in the update of the variable $\bm{\mathsf{S}}$, as shown in Step~\ref{alg:primal_update_TV} of Algorithm~\ref{algo_polar_TV}. 

\vspace*{0.1cm}
\noindent (ii). Regarding the sparsity prior, depending on the chosen regularization, the corresponding thresholding operator $\bm{\mathcal{T}}$ and the dictionary $\mathsf{\bm{\Psi}}$ needs to be modified in step~\ref{alg:prox_l1}. With this in mind, either of the TV and $\ell_1 + $TV sparsifying regularizations can be taken into account in Algorithm~\ref{algo_polar_TV} as follows:
\vspace*{-0.1cm}
\paragraph*{TV regularization:} The sparsity prior is given by $\check{g}(\mathsf{\bm{\Psi}}^\dagger \bm{\mathsf{S}})= \sum_{i=1}^4 \mu_i \| \bm{\nabla} \bm{\mathsf{S}}^\prime_i \|_{2,1}$, where $\bm{\mathsf{S}}^\prime_i$ 
is the reshaped matrix form of the vector $\bm{\mathsf{S}}_{\colon,i}$ and for every $i \in \{1,2,3,4\}$, $\mu_i > 0$ is the regularization parameter. Thus, in Algorithm~\ref{algo_polar_TV}, $\mathsf{\bm{\Psi}} = \bm{\nabla}$ and the operator $\bm{\mathcal{T}} = \bm{\mathcal{T}}_{\text{TV}}$ is the proximity operator for the TV norm \citep{Chambolle1997, Beck2009a}, using the threshold-size $\bm{\Lambda} = [\mu_1, \mu_2, \mu_3, \mu_4]$ in step~\ref{alg:prox_l1_TV}.

\begin{algorithm}
\caption{Primal-dual algorithm to solve problem~\eqref{eq:Akiyama_2}}\label{algo_polar_TV}
\begin{algorithmic}[1]

\vspace*{0.1cm}
\State {\textbf{given}}
$\, \bm{\mathsf{S}}^{(0)} \in \mathbb{R}^{N \times 4}, \, \bm{\mathsf{Z}}^{(0)} \in \mathbb{R}^{N \times 2}, \, \bm{\mathsf{A}}^{(0)} \in \mathbb{R}^{J \times 4},  \bm{\mathsf{C}}^{(0)} \in \mathbb{R}^{N \times 4}, \, \bm{\mathsf{D}}^{(0)} \in \mathbb{R}^{N \times 2}$

\vspace*{0.1cm}

\State \textbf{repeat for} $k = 0, 1, \ldots$
\vspace*{0.1cm}
\Statex \quad \, \fbox{\textbf{Primal updates}}

\vspace{0.1cm}

\State  \label{alg:primal_update_TV}
\quad \, $\bm{\mathsf{S}}^{(k+1)} = \bm{\mathcal{P}}_{\mathbb{U}} \bigg(\bm{\mathsf{S}}^{(k)} - \tau \, \widetilde{\phi}^\dagger \big(\widetilde{\phi} (\bm{\mathsf{S}}^{(k)}) - \widetilde{\bm{\mathsf{Y}}} \big) - \tau \big(\mathsf{\bm{\Psi}} \bm{\mathsf{A}}^{(k)}+ \bm{\mathsf{C}}^{(k)} \big) \bigg)$

\vspace*{0.2cm}

\State \label{alg:primal_update_zeta_TV}
\quad \, $\bm{\mathsf{Z}}^{(k+1)} = \bm{\mathcal{P}}_{\mathbb{V}} \bigg(\bm{\mathsf{Z}}^{(k)} - \tau \bm{\mathsf{D}}^{(k)} \bigg)$

\vspace*{0.2cm}
\Statex \quad \, \fbox{\textbf{Dual updates}}

\vspace*{0.1cm}

\Statex \, \quad \, \underline{Promoting sparsity:}

\vspace*{0.1cm}
\State \label{alg:a_update_TV}
\quad \, $\widetilde{\bm{\mathsf{A}}}^{(k)} = \bm{\mathsf{A}}^{(k)} + \rho_1 \mathsf{\bm{\Psi}}^\dagger \bigg( 2 \, \bm{\mathsf{S}}^{(k+1)} - \bm{\mathsf{S}}^{(k)} \bigg)$

\vspace*{0.2cm}

\State \label{alg:prox_l1_TV}
\quad \, $\bm{\mathsf{A}}^{(k+1)} = \widetilde{\bm{\mathsf{A}}}^{(k)} - \rho_1 \bm{\mathcal{T}}_{\bm{\Lambda}/\rho_1} \bigg(  \widetilde{\bm{\mathsf{A}}}^{(k)}/\rho_1 \bigg)$

\vspace*{0.1cm}

\Statex \, \quad \, \underline{Performing epigraphical projection:}

\vspace{0.1cm}
\State \label{alg:c_update_TV}
\quad \, $ \widetilde{\bm{\mathsf{C}}}^{(k)} = \bm{\mathsf{C}}^{(k)} + \rho_3 \bigg( 2 \, \bm{\mathsf{S}}^{(k+1)} - \bm{\mathsf{S}}^{(k)} \bigg)$

\vspace*{0.2cm}

\State \label{alg:c_zeta1_update_TV}
\quad \, $\widetilde{\bm{\mathsf{D}}}^{(k)} = \bm{\mathsf{D}}^{(k)} + \rho_3 \bigg( 2 \, \bm{\mathsf{Z}}^{(k+1)} - \bm{\mathsf{Z}}^{(k)} \bigg)$

\vspace*{0.2cm}

\State \label{alg:proj_epi_h1_TV}
\quad \, $\left[\begin{matrix}
\bm{\mathsf{C}}_{:,1}^{(k+1)} \\
\vspace*{-0.2cm} \\
 \bm{\mathsf{D}}_{:,1}^{(k+1)}
\end{matrix}\right]
= 
\left[\begin{matrix}
\widetilde{\bm{\mathsf{C}}}_{:,1}^{(k)} \\
\vspace*{-0.2cm} \\
\widetilde{\bm{\mathsf{D}}}_{:,1}^{(k)}
\end{matrix}\right]
- \rho_3 \, \bm{\mathcal{P}}_{\mathbb{E}_1}
\left(\frac{1}{\rho_3} \left[\begin{matrix}
\widetilde{\bm{\mathsf{C}}}_{:,1}^{(k)} \\
\vspace*{-0.2cm} \\
\widetilde{\bm{\mathsf{D}}}_{:,1}^{(k)}
\end{matrix}\right]
\right)$

\vspace*{0.2cm}

\State \label{alg:proj_epi_h2_TV}
\quad \, $\left[\begin{matrix}
\bm{\mathsf{C}}_{:,2:4}^{(k+1)} \\
\vspace*{-0.2cm} \\
\bm{\mathsf{D}}_{:,2}^{(k+1)}
\end{matrix}\right]
= 
\left[\begin{matrix}
\widetilde{\bm{\mathsf{C}}}_{:,2:4}^{(k)} \\
\vspace*{-0.2cm} \\
\widetilde{\bm{\mathsf{D}}}_{:,2}^{(k)}
\end{matrix}\right]
- \rho_3 \, \bm{\mathcal{P}}_{\mathbb{E}_2}
\left(\frac{1}{\rho_3} \left[\begin{matrix}
 \widetilde{\bm{\mathsf{C}}}_{:,2:4}^{(k)} \\
  \vspace*{-0.2cm} \\
 \widetilde{\bm{\mathsf{D}}}_{:,2}^{(k)} 
\end{matrix}\right]
\right)
$

\vspace*{0.2cm}

\State \textbf{until convergence} \label{alg:d2_update_end_TV}

\end{algorithmic}
\end{algorithm}

\paragraph*{$\ell_1 + $TV regularization:} It consists of two terms, $\check{g}(\mathsf{\bm{\Psi}}^\dagger \bm{\mathsf{S}})= \sum_{i=1}^4 \upsilon_{1,i} \|\bm{\mathsf{S}}_{\colon,i} \|_1 + \sum_{i=1}^4 \upsilon_{2,i} \| \bm{\nabla} \bm{\mathsf{S}}^\prime_i \|_{2,1}$, with $\upsilon_{1,i}$ and $\upsilon_{2,i} > 0$. The first term, i.e. $\ell_1$ norm, imposes sparsity of the underlying images in the Dirac basis. The second term, i.e. TV term, promotes the sparsity in the gradient domain~\eqref{eq:TV}. As a result, in this case,
the sparsifying dictionary $\mathsf{\bm{\Psi}} = [{\mathbf{1}}, \bm{\nabla} ]$ is the concatenation of the identity matrix (Dirac basis) and the gradient basis, respectively. Similarly, the operator $\bm{\mathcal{T}} = [\bm{\mathcal{T}}_{\bm{1}}, \bm{\mathcal{T}}_{\text{TV}}]$ is the concatenation of the proximity operators corresponding to the $\ell_1$ and the TV norms, associated with the thresholding sizes $\bm{\upsilon}_1 = [\upsilon_{1,1}, \upsilon_{1,2}, \upsilon_{1,3}, \upsilon_{1,4}]$ and $\bm{\upsilon}_2 = [\upsilon_{2,1}, \upsilon_{2,2}, \upsilon_{2,3}, \upsilon_{2,4}]$, respectively, i.e. $\bm{\Lambda} = [\bm{\upsilon}_1, \bm{\upsilon}_2]$ in step~\ref{alg:prox_l1_TV}.

Note that the same algorithm can be used to solve problem~\eqref{eq:Akiyama} as well wherein the polarization constraint is not imposed and only steps~\ref{alg:primal_update_TV},~\ref{alg:a_update_TV} and \ref{alg:prox_l1_TV} in Algorithm~\ref{algo_polar_TV} need to be executed.

Lastly, regarding the terminology, solving problem~\eqref{eq:Akiyama_2} considering TV regularization (resp. $\ell_1$~+~TV) is referred to as TV (resp. $\ell_1$~+~TV) problem with constraint. Similarly, solving problem~\eqref{eq:Akiyama} with TV regularization (resp. $\ell_1$~+~TV) is termed as TV (resp. $\ell_1$~+~TV) problem without constraint.


\vspace*{-0.5cm}
\section{Simulations and Results} \label{sec:sim_app} 

%
In this section, we discuss the considered simulation settings and describe the different cases for simulations. We then investigate the performance of the proposed Polarized SARA method, implemented in \textsc{matlab}, on simulated EHT datasets. 

\vspace*{-0.2cm}
\subsection{Simulation Setup}
\label{ssec:setup}

Without any loss of generality for the proposed algorithm, we consider the idealistic case and work in the absence of DDEs. 
In such a scenario, the Mueller matrix is essentially the identity matrix, and the measurement operator takes the form
\begin{equation} \label{eq:meas_op}
{{\phi}}(\overline{\bm{\mathsf{S}}}) = \mathsf{\bm{\Phi}} \overline{\bm{\mathsf{{S}}}} \, \bm{\mathsf{L}} \, \,  \text{with} \, \, \mathsf{\bm{\Phi}} = \bm{\mathsf{G F Z}}.
\end{equation}
Essentially, the operator $\bm{\mathsf{L}} \in \mathbb{C}^{4 \times 4}$ acts on the Stokes matrix $\overline{\bm{\mathsf{{S}}}}$ to give the brightness matrix as described in Section~\ref{sec:RI_imaging}. Once the brightness matrix is obtained, the zero padding operator $\bm{\mathsf{Z}} \in \mathbb{C}^{ \alpha N \times N}$ is used to oversample the image contained in each column of the brightness matrix, by a factor of $\alpha$ in each dimension. Then, the task is to compute the Fourier transform of the oversampled images at the spatial frequencies sampled by the interferometer. However, evaluating the Fourier transform directly at these sampled points incur a high computational cost. Thus, we use a fast implementation of the Fourier transform in terms of a fast Fourier transform (FFT), performed by a matrix $\bm{\mathsf{F}} \in \mathbb{C}^{ \alpha N \times \alpha N}$. Since FFT provides discrete Fourier coefficients, the operation of degridding these discrete Fourier points to the continuous samples is performed by a matrix $\bm{\mathsf{G}} \in \mathbb{C}^{M \times \alpha N}$. More precisely, each row of this matrix $\bm{\mathsf{G}}$ consists of convolution kernels, which are modelled with compact support in the Fourier domain \citep{Fessler2003}. Note that the matrix $\bm{\mathsf{Z}}$ consists of oversampling and scaling of the sought images in order to compensate for the interpolation errors introduced by the matrix $\bm{\mathsf{G}}$.

\begin{figure}
\vspace*{-0.15cm}
\centering
\includegraphics[width=0.8\columnwidth]{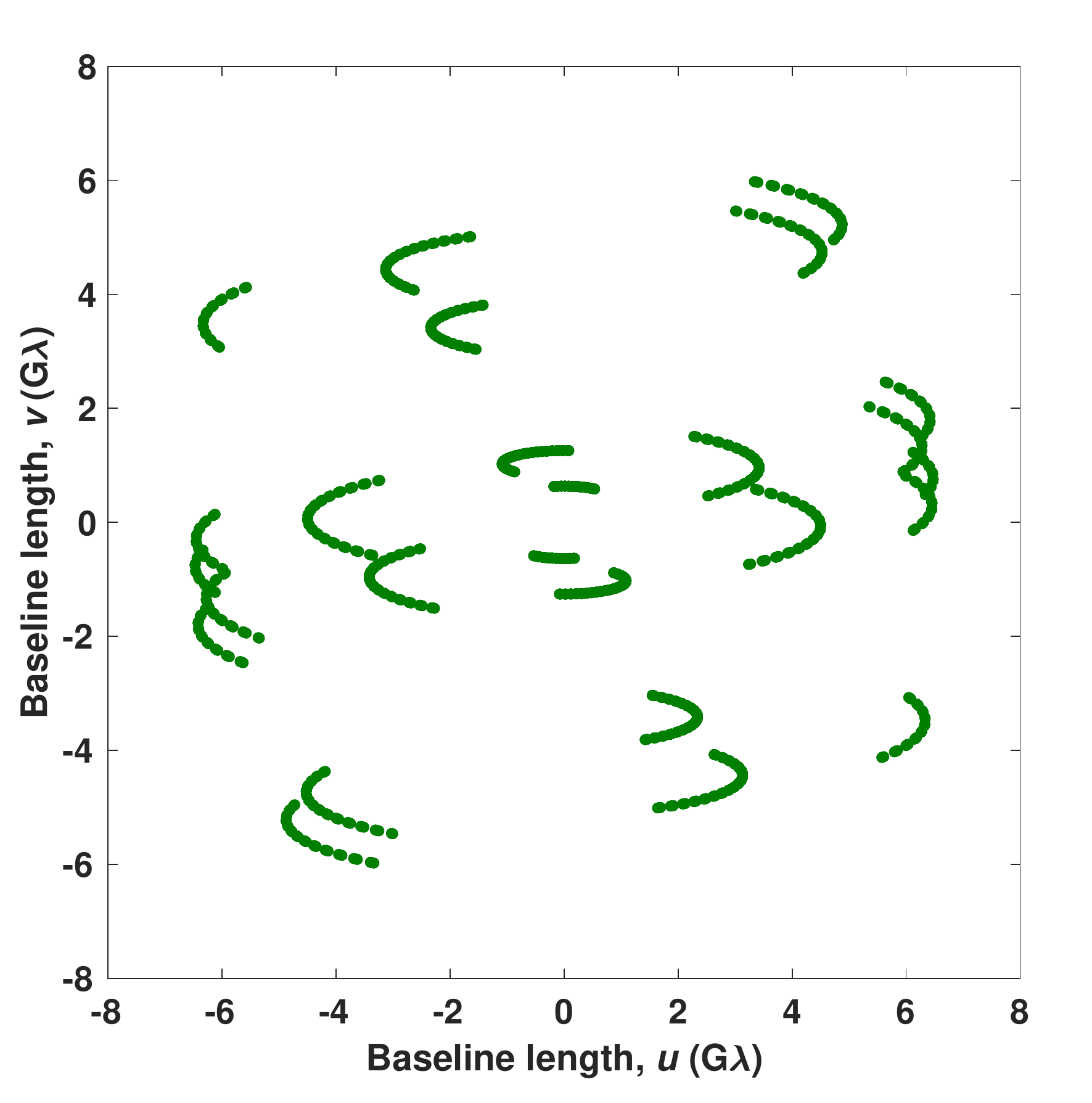}
\vspace*{-0.1cm}
\caption{The EHT $u-v$ coverage used for simulations, taken from \citet{Akiyama2017b, Akiyama2017a}. It corresponds to the meaurements made at 1.3~mm  (230 GHz) using six stations of the VLBI array.}
\label{fig:uv_cov}
\end{figure}

It can be noticed that using equation~\eqref{eq:meas_op}, in the absence of DDEs, the measurement model given in equation~\eqref{eq:RIME_disc} can be seen as the Fourier transform of the brightness matrix $\overline{\bm{\mathsf{S}}} \bm{\mathsf{L}}$ computed at the sampled frequencies. With these measurement settings, we perform tests on the EHT $u-v$ coverage, as shown in Fig.~\ref{fig:uv_cov}. This realistic coverage, adopted from \cite{Akiyama2017b, Akiyama2017a}, corresponds to the measurements made at wavelength \textlambda ~= 1.3 mm (i.e. observation frequency of 230 GHz), using a VLBI array consisting of six stations. In this case, the maximum observation baseline, $\text{B}_{\text{max}}$ = 7.2 G\textlambda. 

\begin{figure}
\centering
\begin{tabular}{@{}c@{}c@{}}
\hspace*{-0.6cm}\includegraphics[width = 4.42cm]{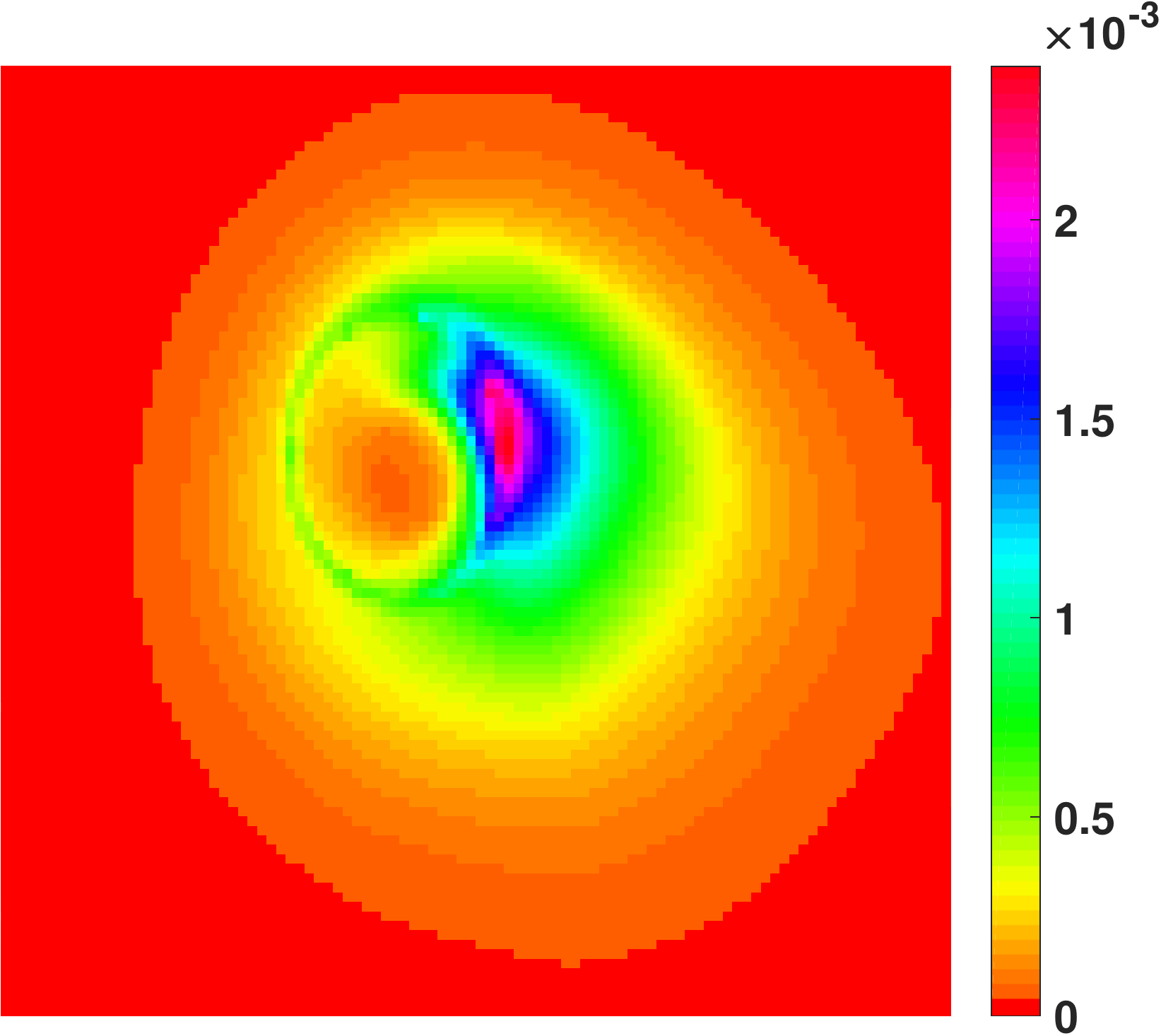} &
\hspace*{-1.1cm}\includegraphics[width = 4.41cm]{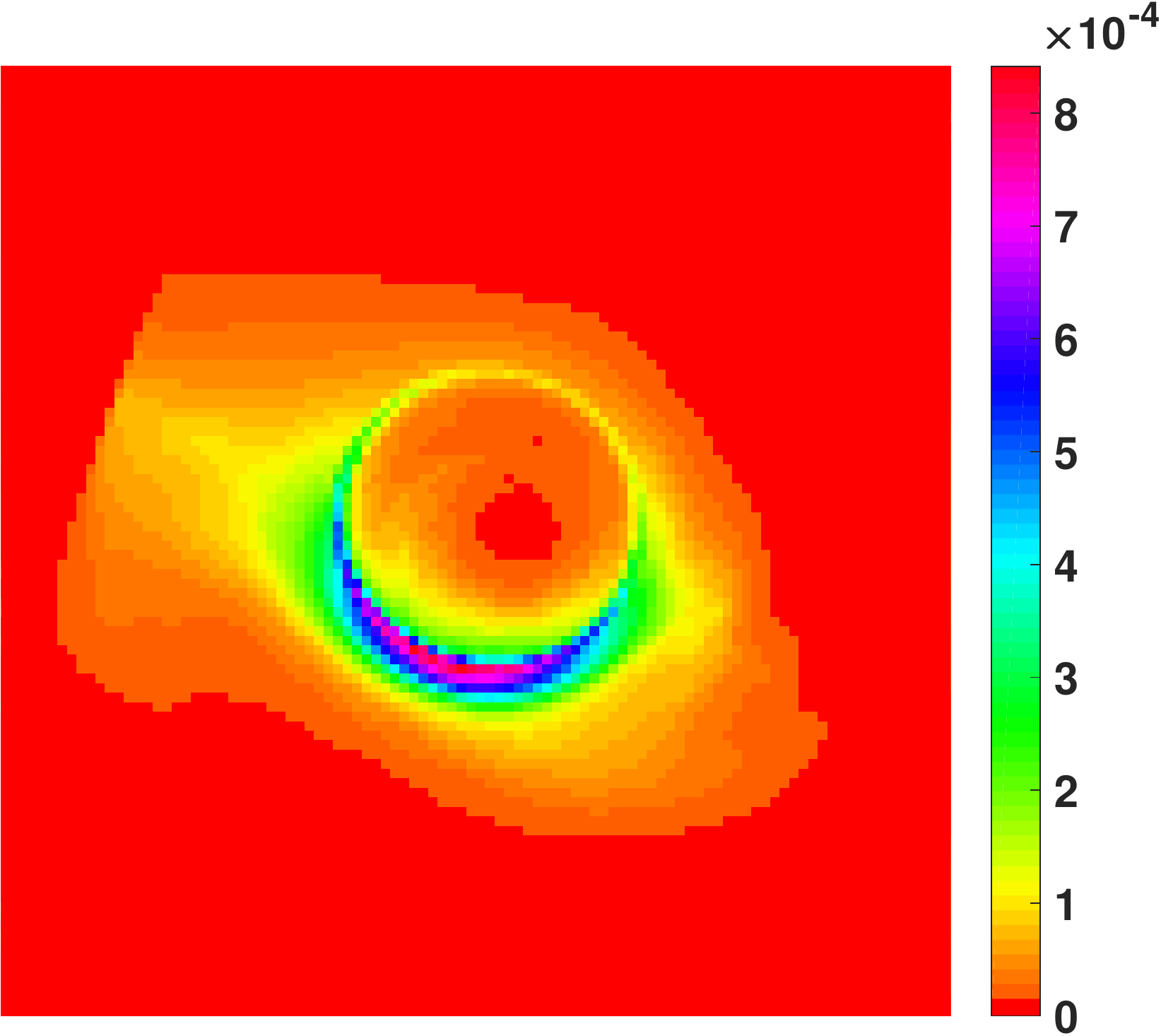} \\
\includegraphics[trim ={2cm 0 0 0.2cm},clip,width=5cm]{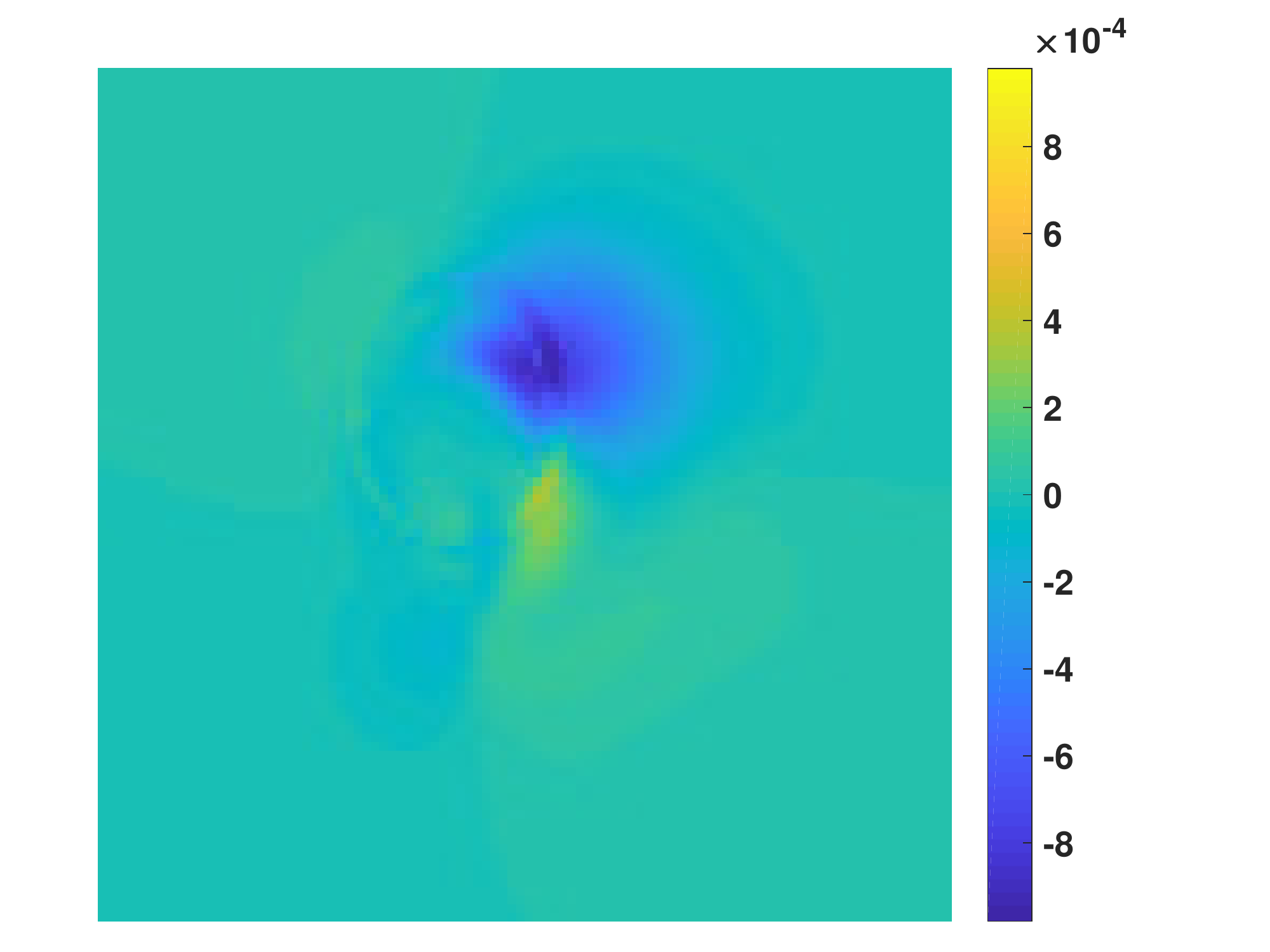} &
\hspace*{-0.5cm}\includegraphics[trim ={2cm 0 0 0.2cm},clip,width=5cm]{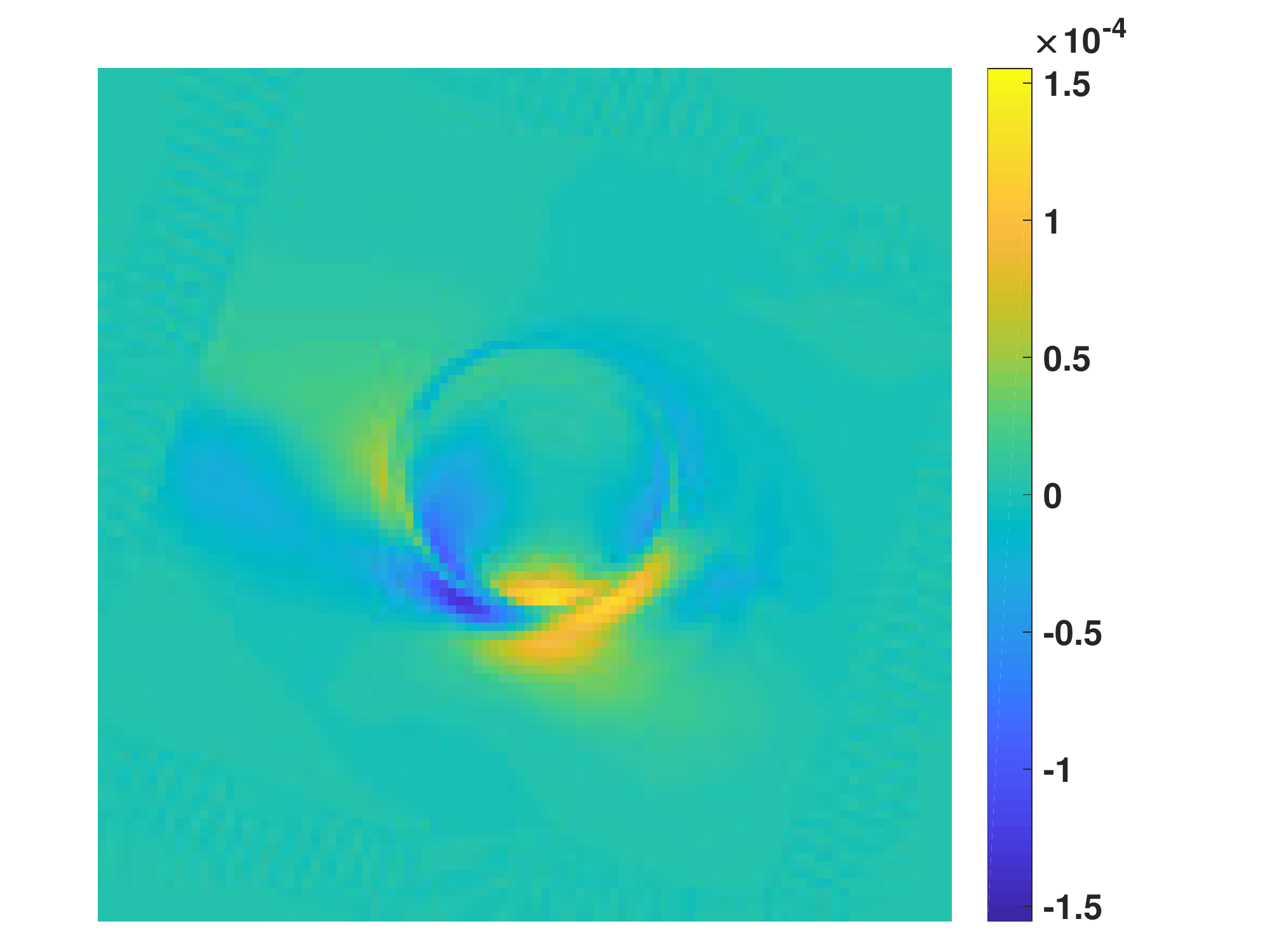} \\
\includegraphics[trim ={2cm 0 0 0.2cm},clip,width=5cm]{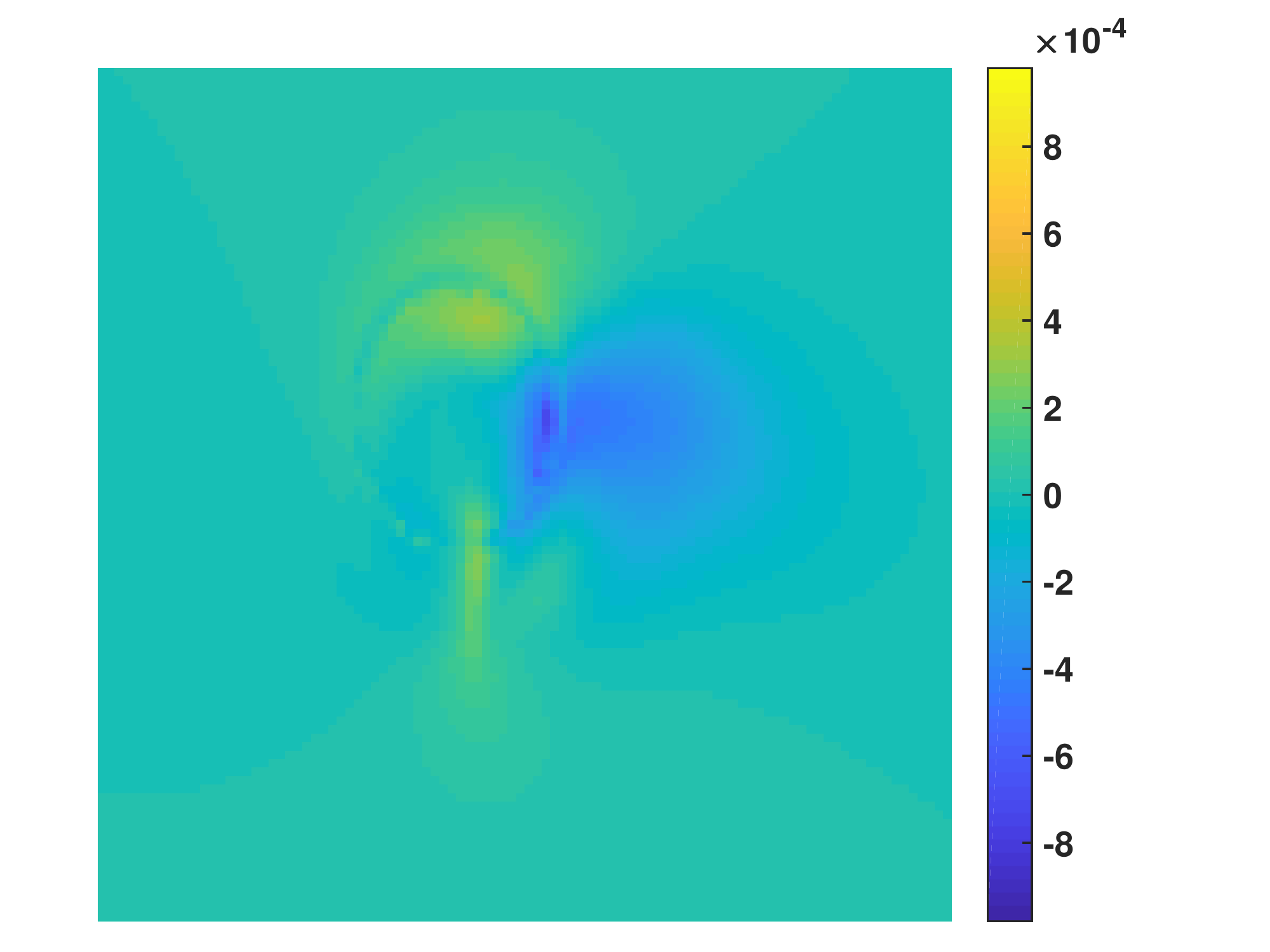} &
\hspace*{-0.5cm}\includegraphics[trim ={2cm 0 0 0.2cm},clip,width=5cm]{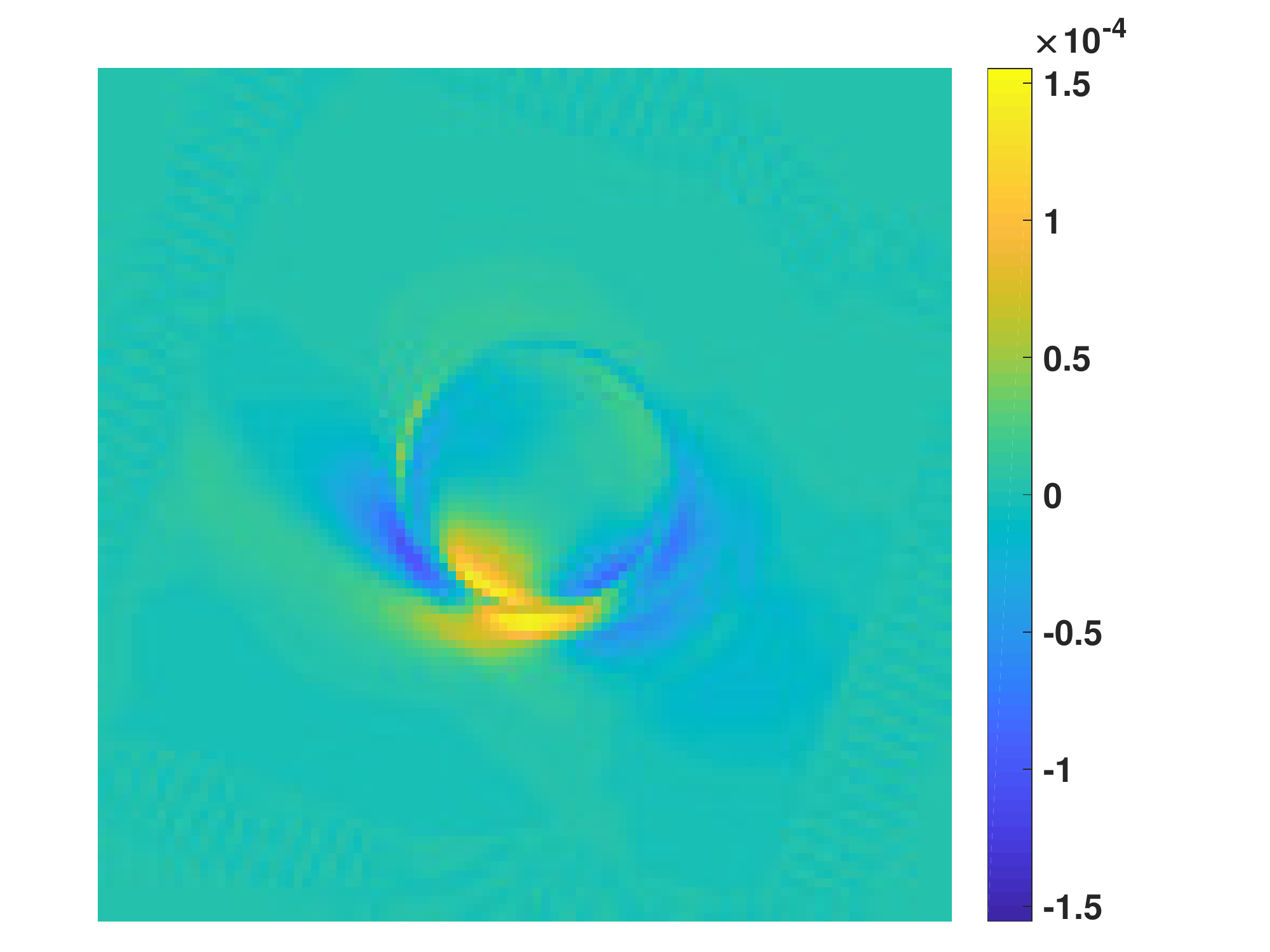} \\
\includegraphics[trim ={2cm 0 0 0.2cm},clip,width=5cm]{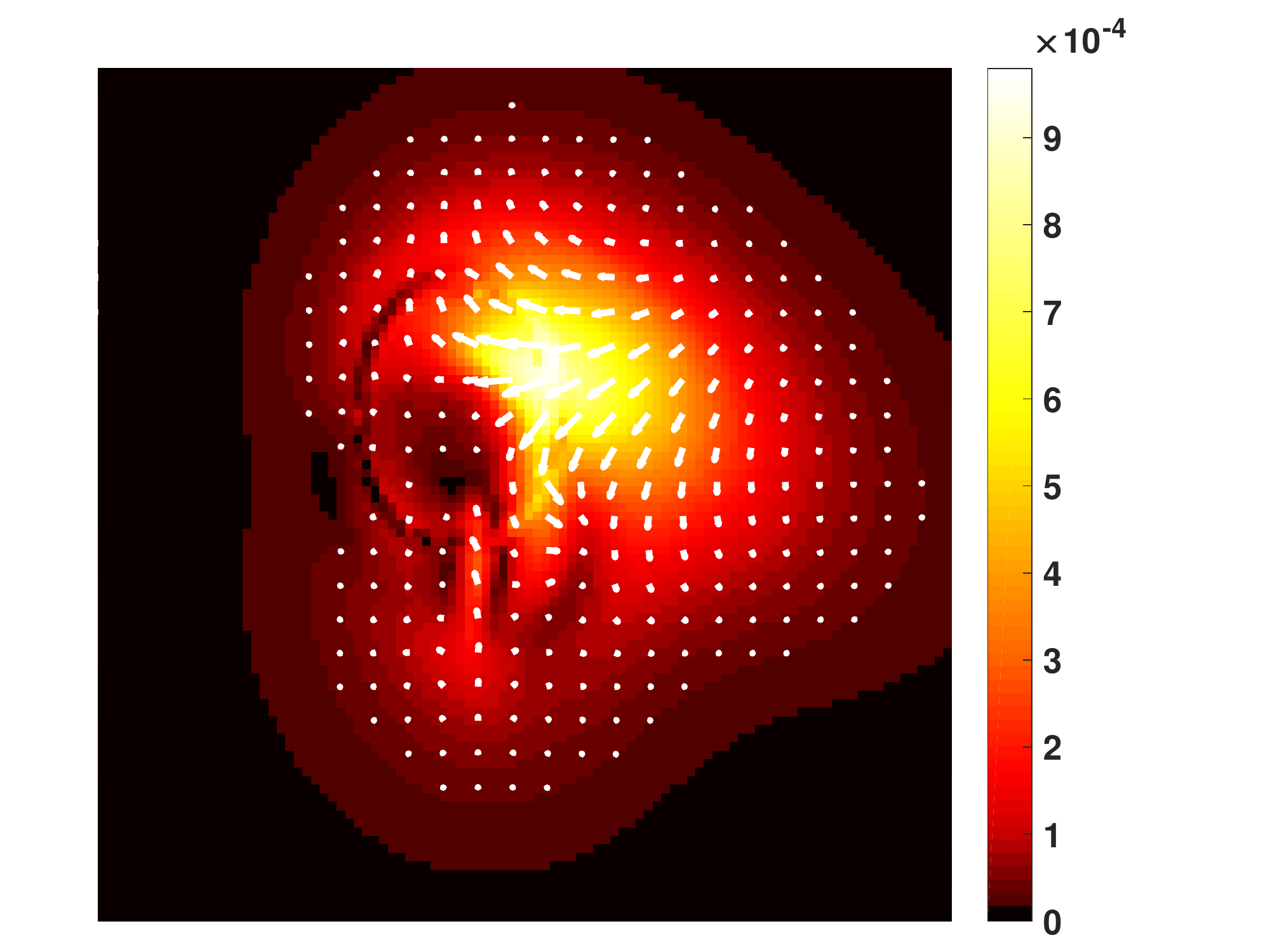} &
\hspace*{-0.5cm}\includegraphics[trim ={2cm 0 0 0.2cm},clip,width=5cm]{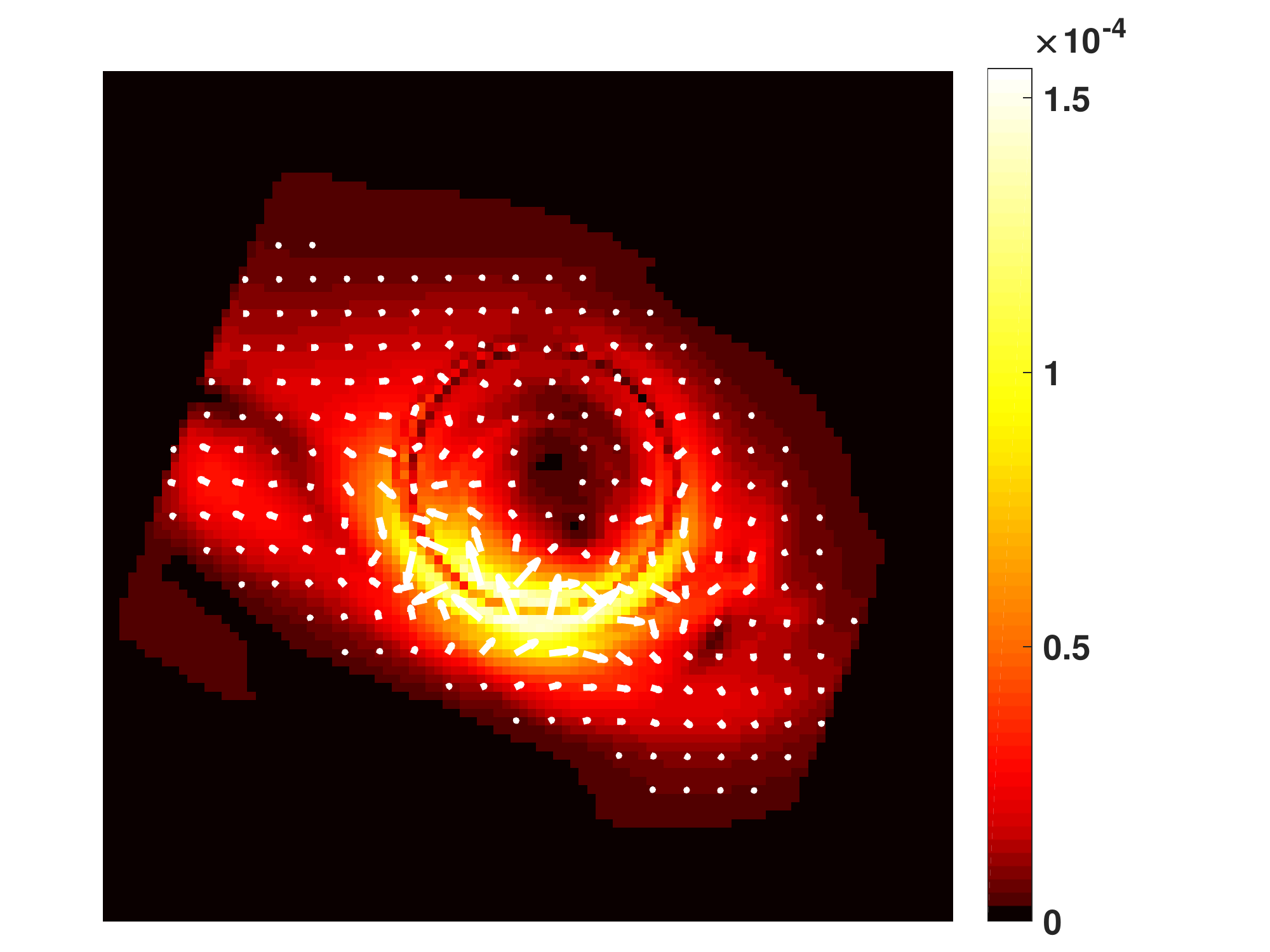} \\
\end{tabular}
\caption{The two sets of ground truth images used for performing simulations. The first column corresponds to the forward-jet model images \citep{Broderick2009, Lu2014}. The second column corresponds to the counter-jet model images \citep{Dexter2012}. For both the columns, each row shows the following images: Stokes $I$ (first row), Stokes $Q$ (second row), Stokes $U$ (third row), and the linear polarization image $P$ (fourth row). For the latter, the electric vector polarization angle (EVPA) distribution is shown in white bars, plotted over the linear polarization intensity ($|P|$). All the images are shown in linear scale.}
\label{fig:true_images}
\end{figure}

Concerning the images used for simulations, as mentioned in Section~\ref{sec:intro}, only the linearly polarized emissions from most of the radio sources are detectable, while the circularly polarized emissions are negligible. Thus, in our simulations, we aim to solve only for the Stokes $I, Q$ and $U$ images, considering $\bm{s}_4 = 0$ and the Stokes matrix as $\bm{\mathsf{S}} = [\bm{s}_1, \bm{s}_2, \bm{s}_3]$. One can observe that the application of the proposed approach to the considered case is straightforward. In practice, we consider two sets of images, based on physically motivated models of M87 radio emissions at 1.3 mm wavelength. The first set of images consist of a forward-jet model, which was initially developed in \cite{Broderick2009}. We use the version of this model presented in \cite{Lu2014}, coherent with the EHT observations at the considered wavelength. The second set of images involves a counter-jet model. It is based on general relativistic magnetohydrodynamic (GRMHD) simulation results \citep{Dexter2012} and polarimetric radiative transfer calculations \citep{Dexter2016}. These two sets of images are displayed in Fig.~\ref{fig:true_images} in first and second column, respectively. 
In both the cases, the true Stokes $I$, $Q$ and $U$ images are presented along with the linear polarization image $P$, respectively in the rows one to four. Note that for both these sets, we consider the image size $N = 100 \times 100$ with the field of view of 200 $\mu$as. Then, the resultant pixel size of $2 \mu$as corresponds to a scale of the event horizon radius.  
For both the sets of model images, we simulate the noisy measurements as per equation~\eqref{eq:RIME_disc} with the measurement operator given by~\eqref{eq:meas_op}. We consider the measurements related to each column of the brightness matrix are corrupted by a Gaussian noise with the same variance $\sigma^{2}$, where $\sigma = 5 \times 10^{-3}$, consistent with the realistic EHT settings. Furthermore, in the considered settings, the residual norm resembles the $\chi^{2}$ distribution with $8 M$ degrees of freedom, and the bound $\epsilon$ for the $\ell_2$ ball $\mathbb{B}$ defined in~\eqref{eq:f_diff} can be determined from the noise variance $\sigma^{2}/2$ of the real and imaginary parts of the noise. We thus set this bound as $\epsilon^{2} = \big(8 M + 2 \sqrt{2(8 M)} \big) \sigma^{2}/2$, where the bound $\epsilon^{2}$ is taken to be 2 standard deviations above the mean of the $\chi^{2}$ distribution \citep{Carrillo2012}.


\vspace*{-0.2cm}

\subsection{Effect of polarization constraint}
\label{ssec:pol_const_effect}

As previously discussed in Section~\ref{ssec:obj_fun}, the polarization constraint needs to be satisfied by the Stokes images to avoid unphysical reconstructions. To validate the importance of imposing this constraint explicitly in the reconstruction process, we perform tests with and without this constraint. The case of imposing this constraint, i.e. Polarized SARA, consists of solving the minimization problem~\eqref{eq:min_gen2} using Algorithm~\ref{algo_rw} with each reweighting iteration consisting of implementing Algorithm~\ref{algo_polar}. On the contrary, Polarized SARA without  constraint implies solving the minimization problem~\eqref{eq:min_gen_matrix_no_flux}. Note that in this context, Algorithm~\ref{algo_polar} can still be used to solve the weighted $\ell_1$ minimization problems within the reweighting scheme. However, Algorithm~\ref{algo_polar} is employed to solve only for the Stokes matrix $\bm{\mathsf{S}}$ (step~\ref{alg:primal_update}), taking into account only the sparsity prior (steps~\ref{alg:a_update} and \ref{alg:prox_l1}) and the data fidelity term (steps~\ref{alg:b_update} and \ref{alg:prox_l2}). 
Additionally, recall that in the absence of the polarization constraint, the positivity of the Stokes $I$ image is taken into account by the use of the modified set $\mathbb{U}^\prime$~\eqref{eq:pos_set}. In step~\ref{alg:primal_update} of Algorithm~\ref{algo_polar}, the projection needs to be performed on this set. This projection also  consists in taking the real part of its arguments as described in~\eqref{eq:real_proj}, with an extra step of considering only the positive values for Stokes $I$, i.e. $\bm{\mathsf{X}}_{\colon,1} = \operatorname{max} \bigg\{ \operatorname{Re} \big({\bm{\mathsf{X}}}_{\colon,1} \big), \bm{0} \bigg\}$.

In order to compare between the tests performed with and without imposing polarization constraint, we keep track of the pixels not satisfying this constraint~\eqref{eq:flux_bound}. In particular, it consists in analyzing the polarization error image, $\bm{p} \in \mathbb{R}_+^N$, where, for every $n \in \{1, \ldots, N\}$, 
\[
p_n = \begin{cases}
                 -\mathsf{S}_{n,1} + \| \bm{\mathsf{S}}_{n,2:3} \|_2 ,  & \text{if} \, \, -\mathsf{S}_{n,1} + \| \bm{\mathsf{S}}_{n,2:3} \|_2 > \zeta , \\
                                0, & \text{otherwise.} 
                                \end{cases}
                                \]
Basically, this image is generated by taking the difference between the linear polarization intensity image and the total intensity image. Note that only the pixels with values larger than some threshold $\zeta$ are retained, while the others are put to zero. In essence, this image is a representation of the pixels not satisfying the polarization constraint~\eqref{eq:flux_bound}, and having values greater than $\zeta$. The value of $\zeta$ is taken to be 3 times the rms noise, which is estimated from the residual image. Thus, by considering this threshold, the pixels with values smaller than the noise level are discarded. Finally, we denote the percentage of the non-zero pixels in the image $\bm{p}$ by $N_p$, where $N_p \in [0, 100]$.

\subsection{Comparison with the other methods}
\label{ssec:comp_methods}
In the context of EHT imaging for full polarization, as mentioned earlier, the work in \cite{Akiyama2017b} represents the only existing method within the sparse modelling framework, aiming to solve the problem~\eqref{eq:Akiyama}. In Section~\ref{ssec:gen_TV}, we have proposed to generalize this problem by taking into account the polarization constraint and hence solving for problem~\eqref{eq:Akiyama_2}. 
Keeping these in mind, we therefore compare the results obtained by the following: Polarized SARA, Polarized SARA without constraint, TV problem with and without constraint, $\ell_1$~+~TV problem with and without constraint.

It is important to emphasize that all these problems are solved using primal-dual approaches. More specifically, while the first two problems are solved by Algorithm~\ref{algo_rw} which incorporates Algorithm~\ref{algo_polar} in each iteration, Algorithm~\ref{algo_polar_TV} is used to solve the last four problems.  
Remark that these comparisons essentially correspond to analyzing the performance of different sparsifying regularizations for EHT imaging. The performed comparisons are also beneficial in analyzing the importance of polarization constraint for different sparsifying regularizations.

In order to be coherent with previous studies \citep{Chael2016, Akiyama2017b, Akiyama2017a} for EHT imaging, we also perform comparison with the widely used Cotton-Schwab \textsc{clean} (\textsc{cs-clean}) algorithm \citep{Schwab1984}. To this purpose, for each considered dataset in this article, we implemented \textsc{cs-clean} in the Common Astronomy Software Applications (CASA) package\footnote{https://casa.nrao.edu/}.

\subsection{Comparison in the super-resolution regime}
\label{ssec:super_res}

Another comparison which can be made between the results obtained by different sparsifying regularizations is regarding the optimal resolution achieved by the respective reconstructed images, especially in the super-resolution regime, i.e. when one goes beyond the nominal interferometric resolution (\textlambda/$\text{B}_{\text{max}}$), also referred to as the diffraction limit. In this context, we adopt the comparison scheme proposed by \cite{Akiyama2017b, Akiyama2017a}. It consists in convolving the reconstructed images with circular Gaussian beams of varying full width half maximum (FWHM) sizes. We then compute the NRMSE between these convolved images and the corresponding ground truth images. Note that such a convolution varies the resolution of the underlying images. Therefore, not only the reconstruction errors, but also the errors due to loss of resolution will contribute to the computed NRMSEs. 
We compare the curves obtained from the TV problem with constraint, $\ell_1$~+~TV problem with constraint, Polarized SARA and \textsc{cs-clean}. For the TV and $\ell_1$~+~TV problems with constraint, we consider the implementation of Algorithm~\ref{algo_polar_TV}. 
This already provides a better scenario for these TV based regularizations, and hence accounts for a fair comparison of the performance of these methods with the Polarized SARA.

\vspace*{-0.1cm}
\subsection{Simulation settings}
For each of the cases discussed earlier, we perform 5 simulations varying the noise realizations. In order to stop the computation of the algorithm at convergence, we consider a stopping criterion.

Firstly, we ensure that the at convergence the residual norm is in the vicinity of the $\ell_2$ upper bound $\epsilon$ defined in Section~\ref{ssec:setup}, i.e. $\| \phi (\bm{\mathsf{S}}) - \bm{\mathsf{Y}} \|_2 \leq (1 + \vartheta) \epsilon$, where $\vartheta > 0$ is a tolerance parameter. We set it to be equal to $5 \times 10^{-3}$. 
In addition, as second stopping criterion, we impose the relative variation between two consecutive iterates to be very small, i.e.:
\begin{equation} \label{eq:stop_crit}
\underset{i \in \{1,2,3\}} {\operatorname{max}} \bigg( \| \bm{\mathsf{S}}_{\colon,i}^{(k+1)} - \bm{\mathsf{S}}_{\colon,i}^{(k)} \|_2 / \|\bm{\mathsf{S}}_{\colon,i}^{(k)}\|_2 \bigg) \leq \varepsilon,
\end{equation}
where $\varepsilon > 0$.
Concerning the case with the polarization constraint, not only the above two mentioned criteria are taken into account, but we also verify that the constraint is satisfied, up to a small error, i.e., $N_p \leq \varrho$, where $\varrho > 0$. 

As described previously, the proposed Polarized SARA method as well as the Polarized SARA without constraint method incorporate the reweighting scheme (Algorithm~\ref{algo_rw}), wherein we perform 10 reweighting iterations. For each iteration and for both the methods, we choose $\varepsilon = 10^{-5}$ in~\eqref{eq:stop_crit}. In addition to this, we choose $\varrho = 0.5$ to stop Algorithm~\ref{algo_polar} for Polarized SARA. Note that choosing $\varrho = 0.5$ stops the algorithm when only 0.5$\%$ of the pixels in the polarization error image, generated from the reconstructed Stokes images, are not satisfying the constraint~\eqref{eq:flux_bound}.

Regarding the implementation of Algorithm~\ref{algo_polar_TV} to solve for the TV and $\ell_1$~+~TV problems without constraint, we choose $\varepsilon = 10^{-5}$ for the forward-jet model, and $7\times 10^{-6}$ for the counter-jet model for the stopping criterion. While solving for the TV and $\ell_1$~+~TV problems with constraint, we also choose $\varrho = 0.5$. For the threshold parameters $\bm{\Lambda}$ in Algorithm~\ref{algo_polar_TV}, we tune these values to minimize the normalized root mean square error (NRMSE). For any true image $\overline{\bm{s}}$ and the corresponding reconstructed image $\bm{s}$, NRMSE is defined as~
\begin{equation}
\text{NRMSE} = \sqrt{\frac{\sum_n |s_n - \overline{s}_n|^2}{\sum_n |\overline{s}_n|^2}}.
\end{equation}
Therefore, with this definition, lower the NRMSE, better is the reconstruction. 
\begin{table}
\centering
\begin{tabular}{@{}c@{}|c|c|c}
\hline
Polarization 
& \multirow{2}{*}{TV} 
& \multirow{2}{*}{$\ell_1$~+~TV} 
& \multirow{2}{*}{SARA} \\
constraint &&&	\\
\hline\hline
\multicolumn{4}{c}{Stokes $I$ image - SNR\big/ NRMSE} \\
\hline
without constraint & 27.53 \big/ 0.2549 & 27.64 \big/ 0.2542 & 33.09 \big/ 0.1912 \\
with constraint &  28.19 \big/ 0.2442 & 28.72 \big/ 0.2378 & 33.15 \big/ 0.1906  \\
\hline \hline
\multicolumn{4}{c}{Linear polarization image - SNR\big/ NRMSE} \\
\hline
without constraint & 23.66 \big/ 0.3063 & 24.46 \big/ 0.2944 & 27.54 \big/ 0.2527 \\
with constraint & 24.92 \big/ 0.2876 & 24.91 \big/ 0.2878 & 28.96 \big/  0.2350 \\
\hline
\multicolumn{4}{c}{(a)} \\
\multicolumn{4}{c}{}	\\ 
\hline
Polarization 
& \multirow{2}{*}{TV} 
& \multirow{2}{*}{$\ell_1$~+~TV} 
& \multirow{2}{*}{SARA} \\
constraint &&& \\
\hline \hline
\multicolumn{4}{c}{Stokes $I$ image - SNR\big/ NRMSE} \\
\hline
without constraint & 12.81 \big/ 0.5269 & 12.82 \big/ 0.5268 & 15.97 \big/ 0.4502 \\
with constraint & 13.51 \big/ 0.5089 & 13.51 \big/ 0.5090 & 16.71 \big/ 0.4337 \\
\hline \hline
\multicolumn{4}{c}{Linear polarization image - SNR\big/ NRMSE} \\
\hline
without constraint &  5.03 \big/ 0.7781 & 5.85 \big/ 0.7469 & 9.01 \big/ 0.6374 \\
with constraint & 8.62 \big/ 0.6500 & 8.77 \big/ 0.6449 & 9.51 \big/ 0.6215\\
\hline
\multicolumn{4}{c}{(b)} \\
\multicolumn{4}{c}{} \\
\end{tabular}
\caption{SNR and NRMSE values for the reconstructed images corresponding to the (a) forward-jet model, and (b) counter-jet model, obtained by different sparsifying regularizations. For each case, the mean values (computed over 5 simulations) are shown for the Stokes $I$ image and the linear polarization image reconstructed with and without imposing the polarization constraint. }
\label{tab:nrmse}
\end{table}


\begin{table}
\centering
\begin{tabular}{c|c|c|c}
\hline
Model & TV & $\ell_1$~+~TV & SARA \\
\hline
Forward-jet & 20.47 & 16.14 & 15.02 \\
Counter-jet & 62.97 & 59.96 & 41.97 \\
\hline
\end{tabular}
\caption{Percentage of pixels not satisfying the polarization constraint in the reconstructed images obtained by without imposing the polarization constraint in the reconstruction process. The percentage is listed for the reconstructed images corresponding to the forward-jet model (first row), and the counter-jet model (second row). Each column represents the values corresponding to the reconstructed images obtained by different sparsifying regularizations. For each case, the shown values correspond to the mean values computed over 5 simulations.}
\label{tab:flux}
\end{table}

\begin{figure*}
\centering
\begin{tabular}{@{}c@{}c@{}}
\multicolumn{2}{c}{{\color{mygray} \rule[0.05cm]{0.5cm}{2pt}} Model \quad
{\color{red} \hdashrule[0.06cm]{0.5cm}{1.5pt}{1.5pt}} TV \quad
{\color{green} \hdashrule[0.4ex]{1cm}{0.5mm}{%
  1.5mm 1.5pt 0.6mm 1pt}} $\ell_1$~+~TV \quad
{\color{myblue} \hdashrule[0.5ex]{0.8cm}{0.5mm}{1.5mm 2pt}} Polarized SARA \quad
{\color{mypink1} \rule[0.08cm]{0.5cm}{1pt}} {CS-CLEAN}} \\
\hspace{-0.1cm}\includegraphics[width=1\columnwidth]{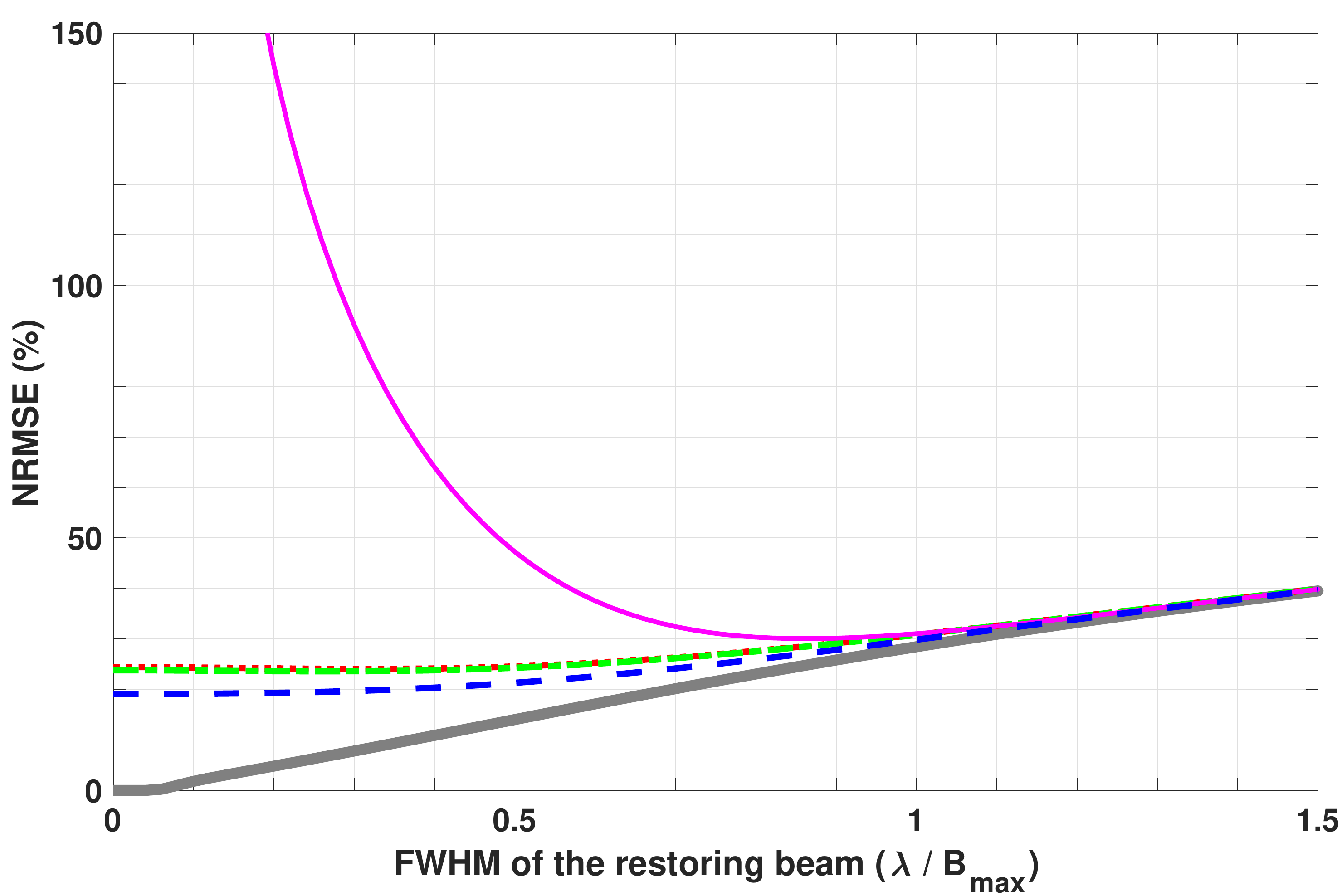} &
\hspace{0.15cm}\includegraphics[width=1\columnwidth]{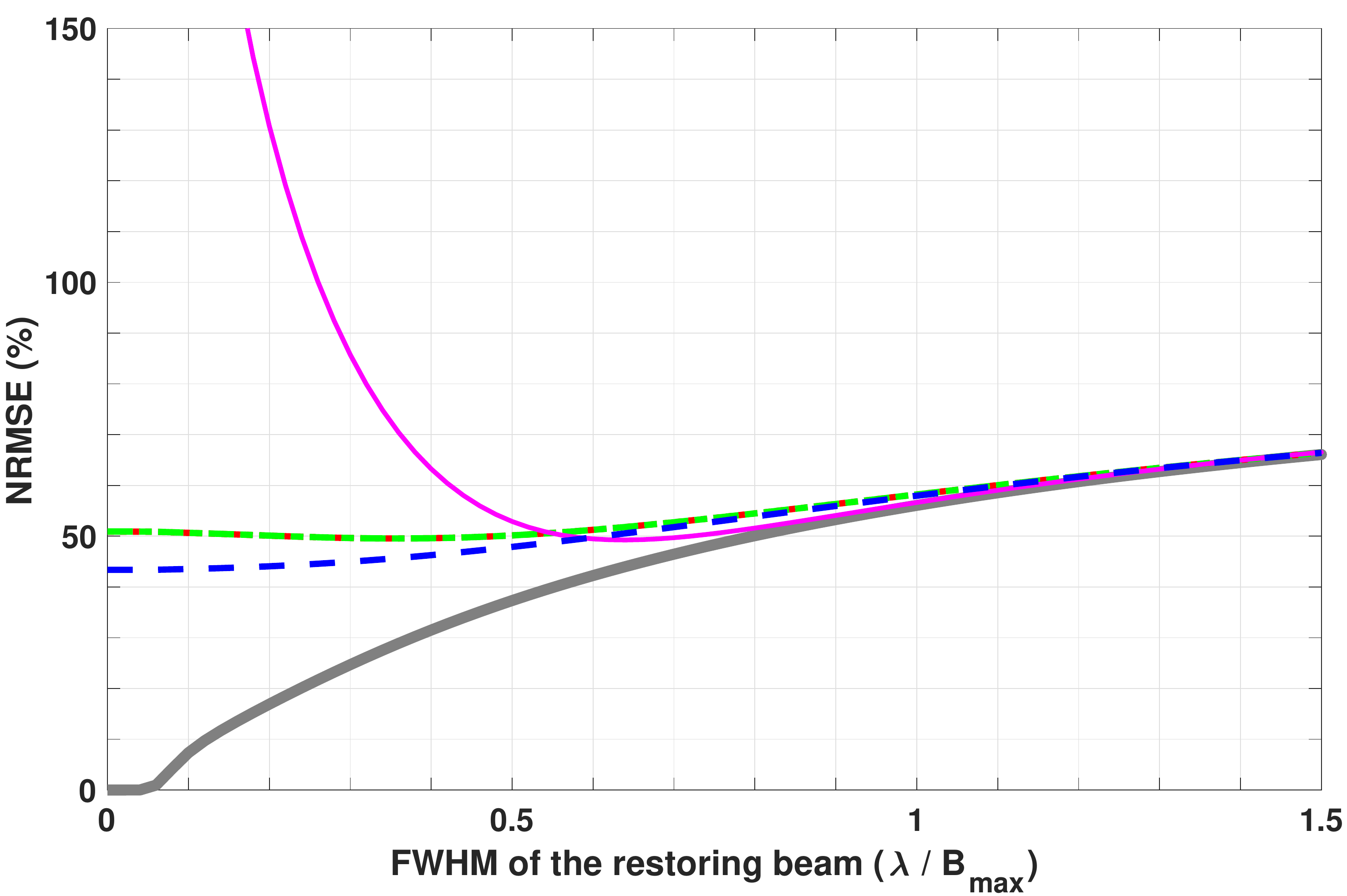} \\
\hspace{-0.1cm}\includegraphics[width=1\columnwidth]{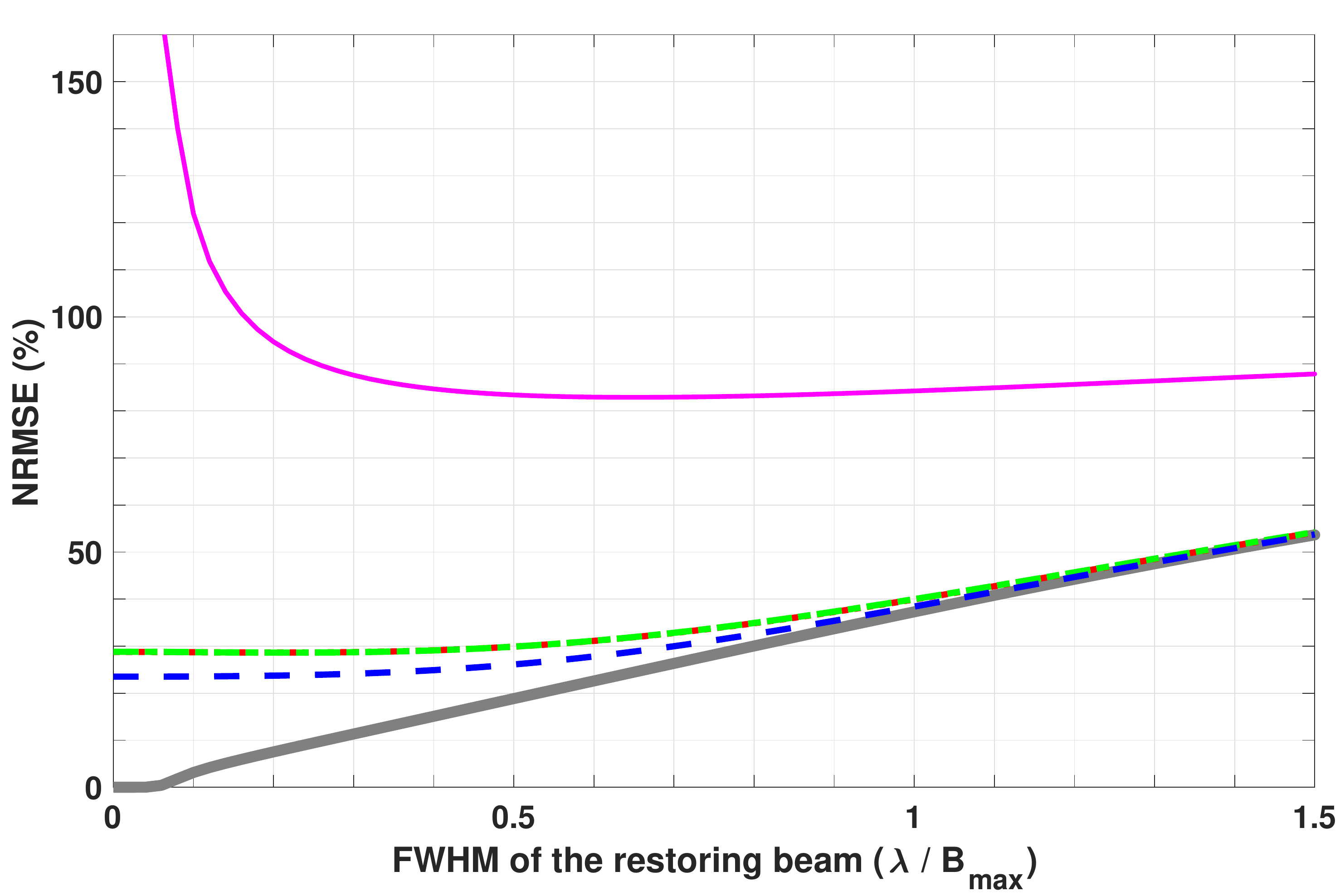} &
\hspace{0.15cm}\includegraphics[width=1\columnwidth]{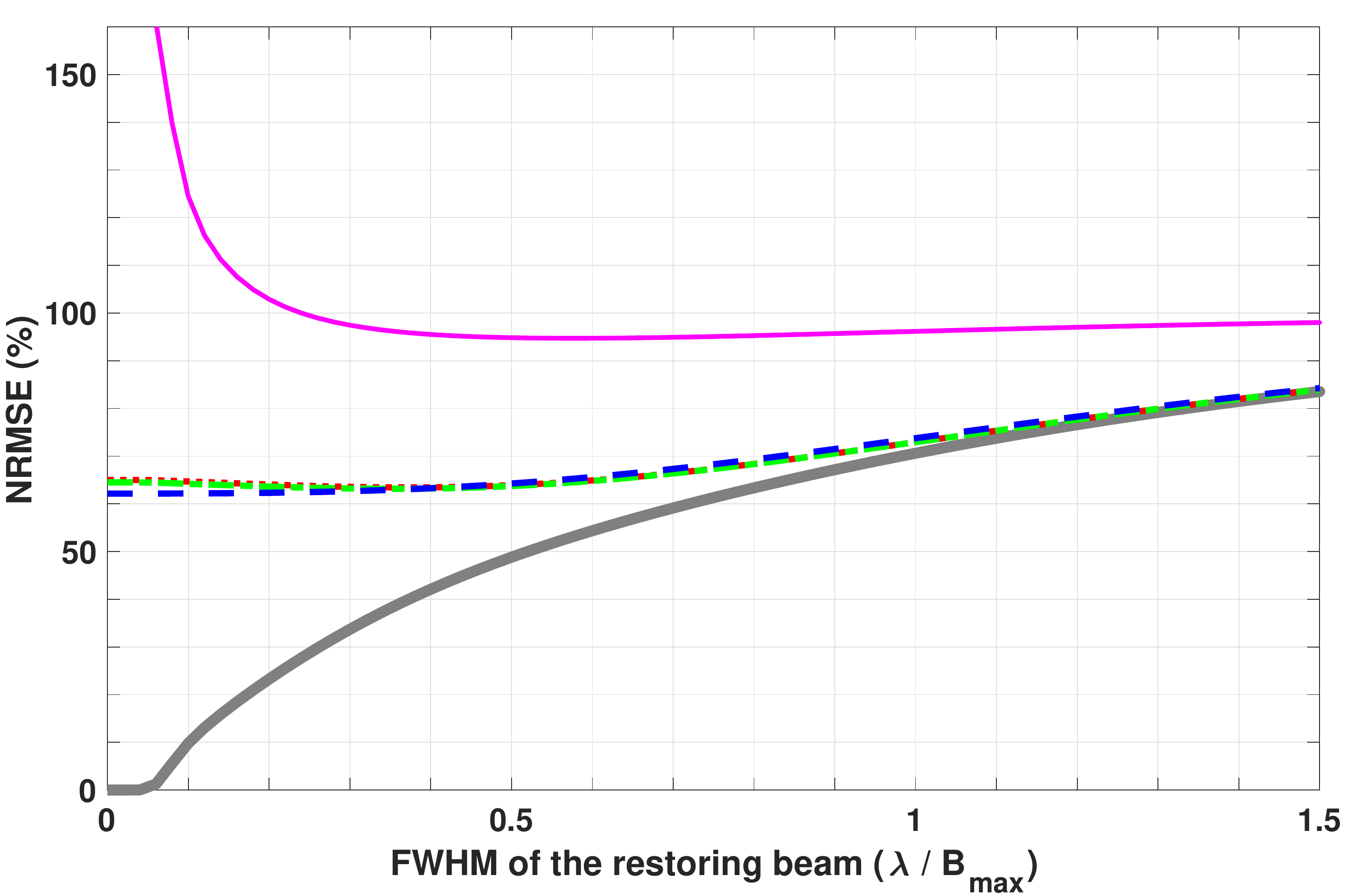} \\
(a) & \hspace{0.15cm} (b)
\end{tabular}
\caption{The NRMSE plots for the Stokes $I$ image (first row) and the linear polarization image (second row) corresponding to (a) forward-jet model, and (b) counter-jet model, as a function of the FWHM size of the restoring beam. The NRMSE is evaluated between the original ground truth image and the beam-convolved reconstructed images. Different curves represent the errors for different regularizations while incorporating the polarization constraint: TV problem with constraint (dotted red curve), $\ell_1$~+~TV problem with constraint (dash-dotted green curve), Polarized SARA (dashed blue curve). 
While the pink curve (continuous thin curve) corresponds to the reconstructions obtained by \textsc{\textsc{cs-clean}}, the grey curve (continuous thick curve) shows the errors for the Model, i.e., error between the original ground truth image and the beam-convolved ground truth image.}
\label{fig:I_nrmse}
\end{figure*}

\subsection{Results and discussion}
\label{sec:results}
For a quantitative comparison between the reconstructed images from different cases, the reconstruction quality is assessed in terms of NRMSE as well as signal-to-noise ratio (SNR). It is defined as
\begin{equation}
\text{SNR} = 20 \log_{10} \left(\frac{||\overline{\bm{s}}||_2}{\|\bm{s} - \overline{\bm{s}}\|_2}\right), 
\end{equation}
implying that higher SNR corresponds to better reconstruction quality. These NRMSE and SNR values for the reconstructed Stokes $I$ image and the linear polarization image $P$, generated from the reconstructed Stokes $Q$ and $U$ images, are listed in Table~\ref{tab:nrmse} for both set of models: (a) forward-jet model, and (b) counter-jet model. In each case, the shown value corresponds to the mean value computed over the performed 5 simulations. It can be observed from Table~\ref{tab:nrmse} that, on the one hand, for a given regularization, imposing the polarization constraint yields lesser error (and thus higher SNR) in the reconstructions than that obtained by without imposing it. More precisely, irrespective of the chosen regularization, enforcement of this constraint leads to improvement in the reconstruction quality. On the other hand, comparison between different regularizations shows that the SARA regularization performs significantly better than the other two regularizations, having~$\sim$~1-5 dBs higher SNR. This holds true not only for Polarized SARA but also for Polarized SARA without constraint. This indicates the importance of choosing a suitable dictionary for better reconstruction. Concerning the importance of the polarization constraint, we quantify it by giving the percentage ($N_p$) of the pixels not fulfilling this constraint in Table~\ref{tab:flux}. In particular, this table provides the values of $N_p$ in the cases with absence of enforcement of the polarization constraint, whereas in its presence $N_p \leq 0.5 \%$ as specified in the stopping criterion. Table~\ref{tab:flux} then demonstrates that without imposing this constraint, an appreciable percentage of pixels have non-physical values. Additionally, in terms of the sparsifying regularizations, it again indicates the better performance of the SARA regularization in comparison with the others. 

For the comparison in the super-resolution regime and as mentioned in Section~\ref{ssec:super_res}, the NRMSE plots for both the forward-jet (first column) and counter-jet (second column) model are shown in Fig.~\ref{fig:I_nrmse}. The first and second row respectively display the plots for the Stokes $I$ and the linear polarization image $P$. 
In all these plots, the curve (continuous thick, grey curve) labelled Model, depicts the NRMSE values between the ground truth images convolved with circular Gaussian beams of varying FWHM sizes and the original ground truth images. Relating it to the previous discussion (Section~\ref{ssec:super_res}), this curve basically represents the minimum attainable errors at any given resolution, arising purely because of the loss of resolution. The other curves correspond to the following: TV problem with constraint (dotted red curve), $\ell_1$~+~TV problem with constraint (dash-dotted green curve), Polarized SARA (dashed blue curve). 
We also give the curve (continuous thin, pink curve) obtained by the reconstructions from the widely used Cotton-Schwab \textsc{clean} (\textsc{cs-clean}) algorithm \citep{Schwab1984} using uniform weighting.

It can be seen that for the Stokes $I$ images, \textsc{cs-clean} NRMSE values start to increase rapidly in the super-resolution regime, where the diffraction limit is specified by the FWHM of size 1. This indicates the inability of \textsc{cs-clean} to produce super-resolved images. Moreover, in this case, the minimum errors are obtained at a resolution of $\sim$ 50 - 80~$\%$ of the diffraction limit. On the contrary, for the other considered sparsifying regularizations, the NRMSE values vary gradually even in the super-resolution regime. Infact the error tends to decrease. It can be noticed that the values for the TV and $\ell_1$~+~TV problems with constraint are quite close, whereas the errors from the Polarized SARA are lesser than that obtained by the former two. Another interesting observation is related to the resolution where the minimum error is achieved by these regularizations. While for the TV and $\ell_1$~+~TV problems, it is at $\sim$ 25 - 35~$\%$ of the diffraction limit, the corresponding value for the Polarized SARA is 0~$\%$. This highlights that the reconstructions obtained by the latter do not need to be convolved with a restoring beam. This is in contrast to the results obtained by other curves, where convolution with a restoring beam is required to get the minimum error. The same features are noticed from the plots of the linear polarization images. 
Note that in this case, the errors obtained by \textsc{cs-clean} are quite large, with the minimum being at around 60~$\%$ of the interferometric resolution. These large errors indicate that \textsc{cs-clean} is not particularly suitable for recovering the linearly polarized emission images.

\begin{figure*}
\centering
\begin{tabular}{@{}c@{}c@{}c@{}c@{}}
\hspace*{-0.0cm}\includegraphics[width = 4.5cm]{true_I_new_avery-eps-converted-to.pdf} \\
\hspace*{-0cm}\includegraphics[width = 4.5cm]{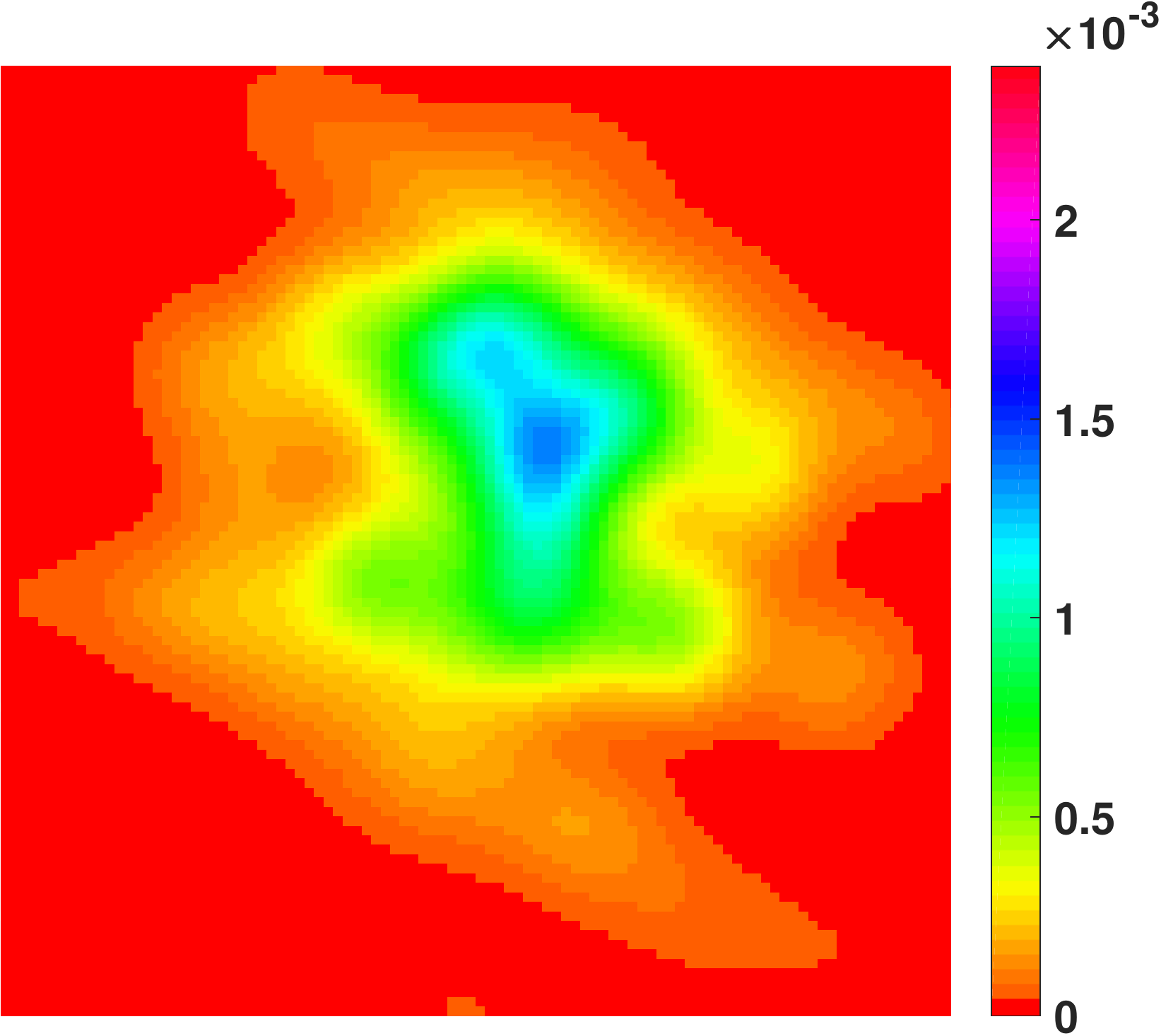} &
\hspace*{-0cm}\includegraphics[width = 4.5cm]{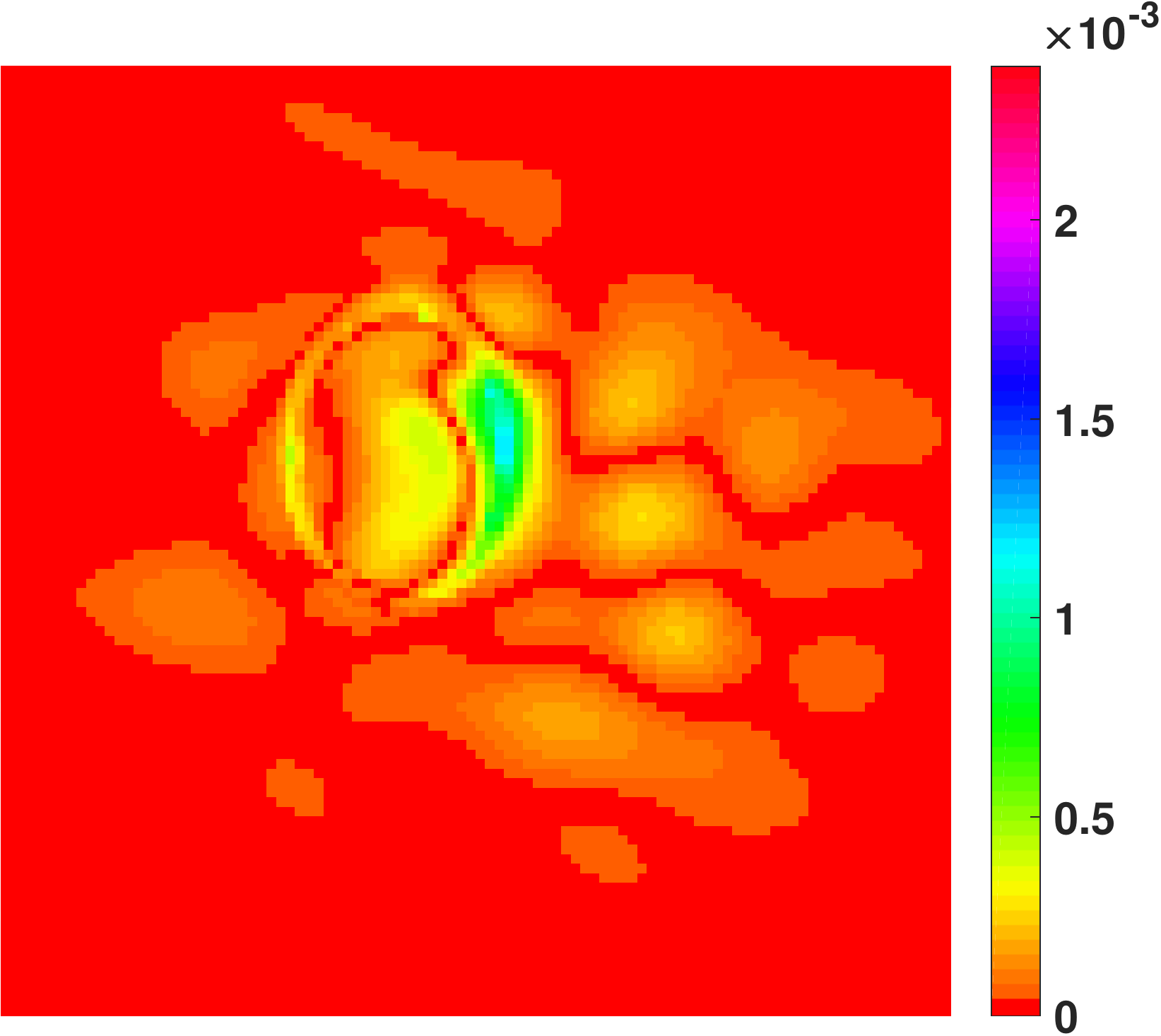} \\
\hspace*{-0.0cm}\includegraphics[width = 4.5cm]{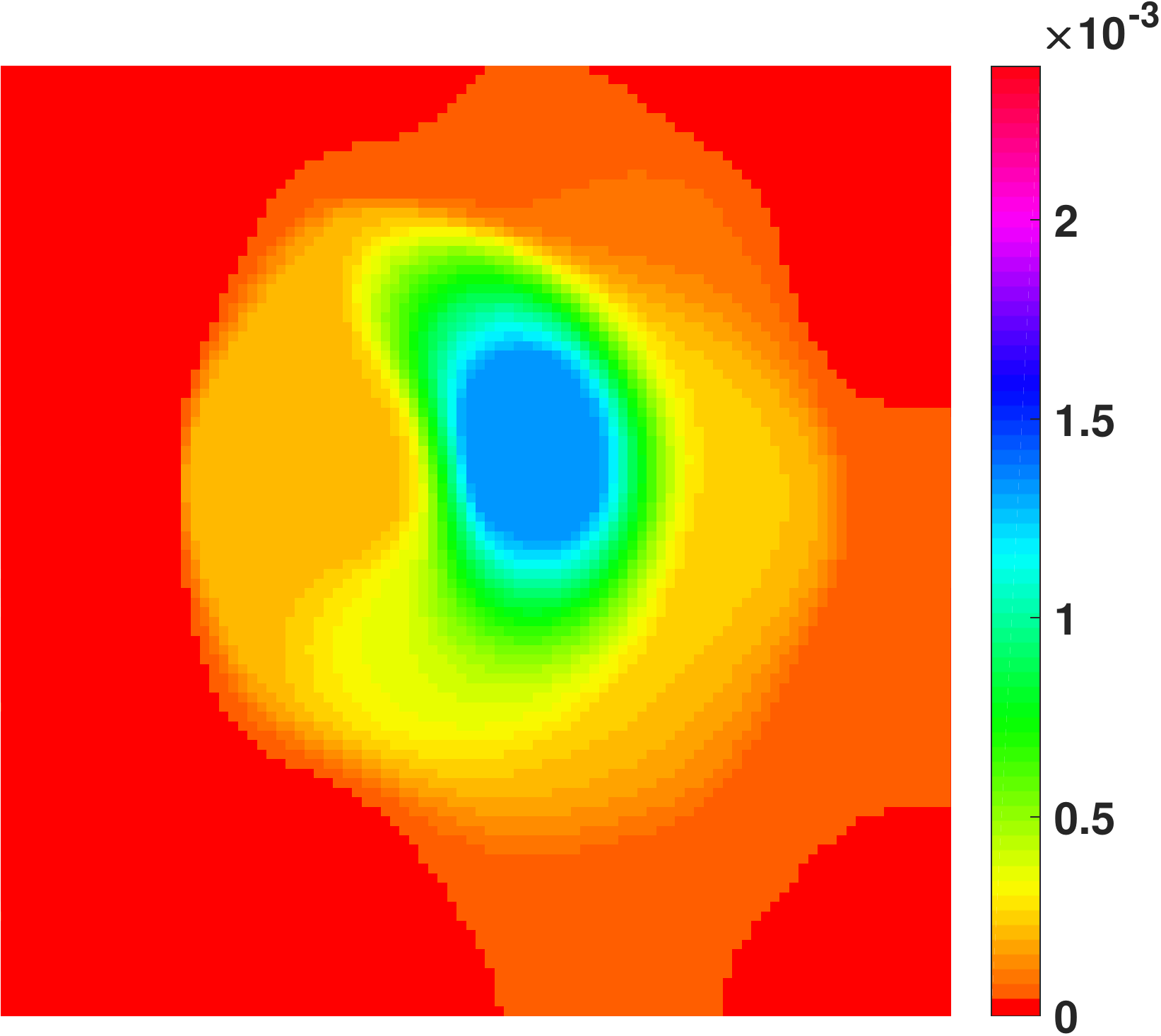} &
\hspace*{-0cm}\includegraphics[width = 4.5cm]{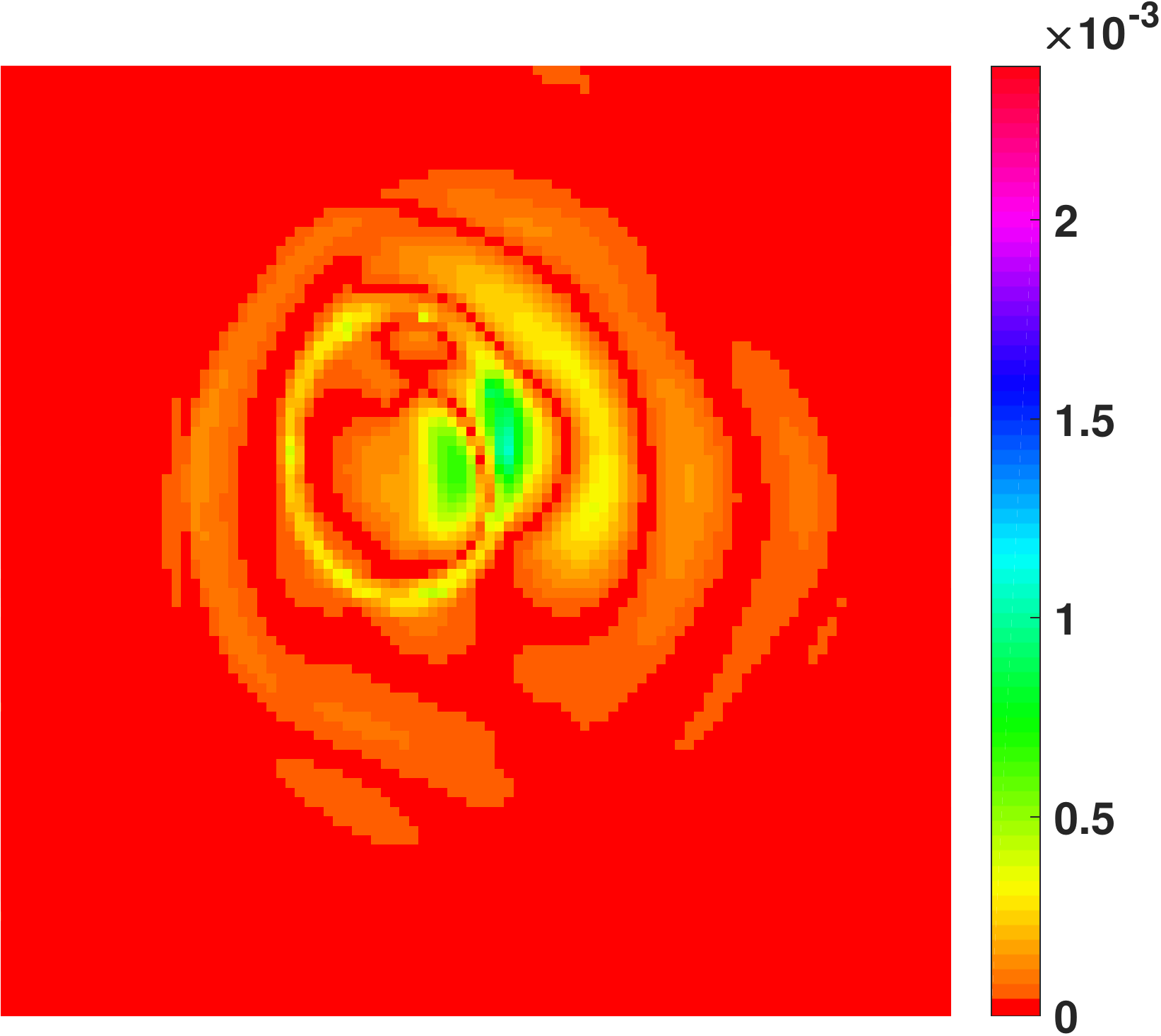} &
\hspace*{-0cm}\includegraphics[width = 4.5cm]{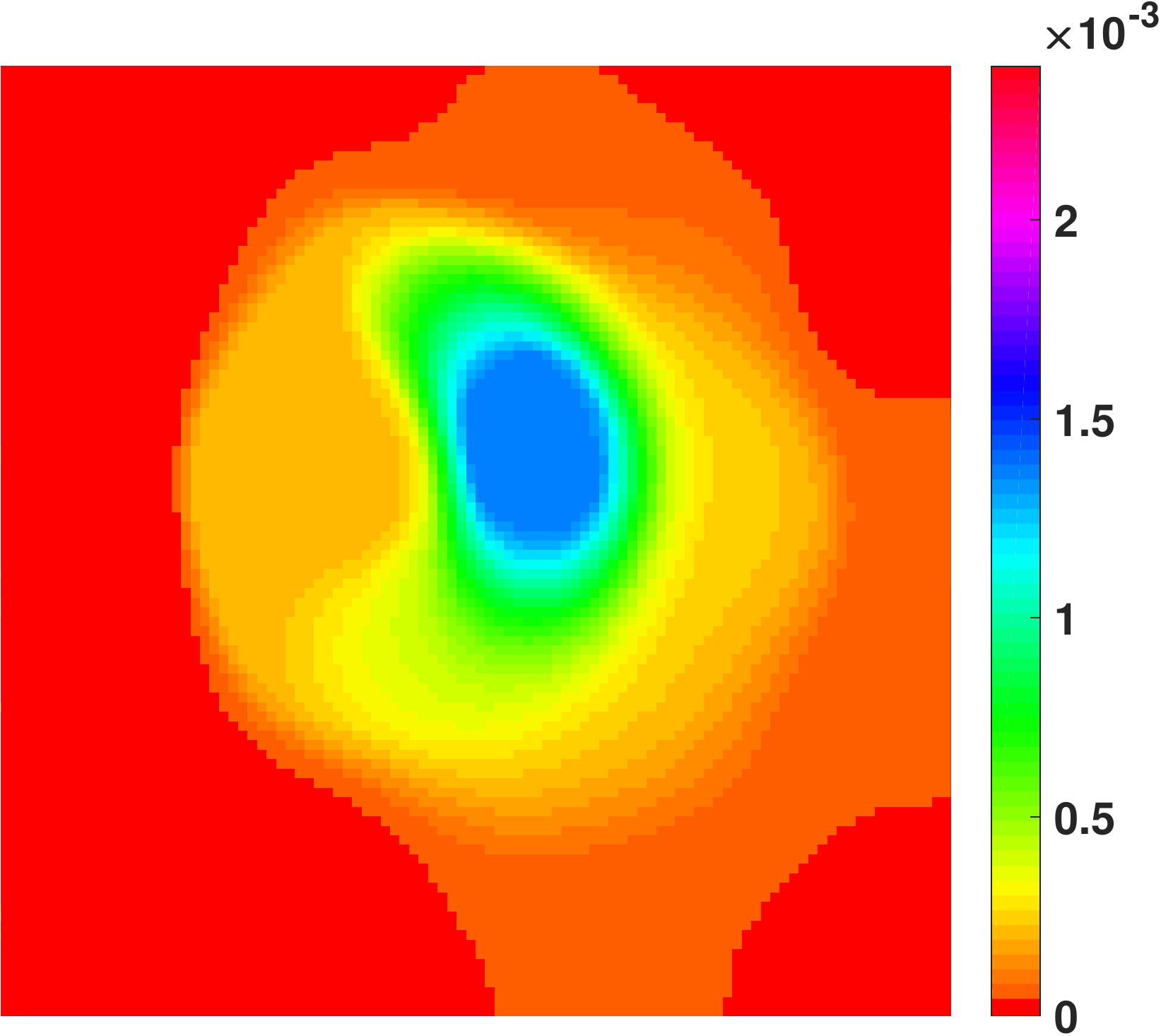} &
\hspace*{-0cm}\includegraphics[width = 4.5cm]{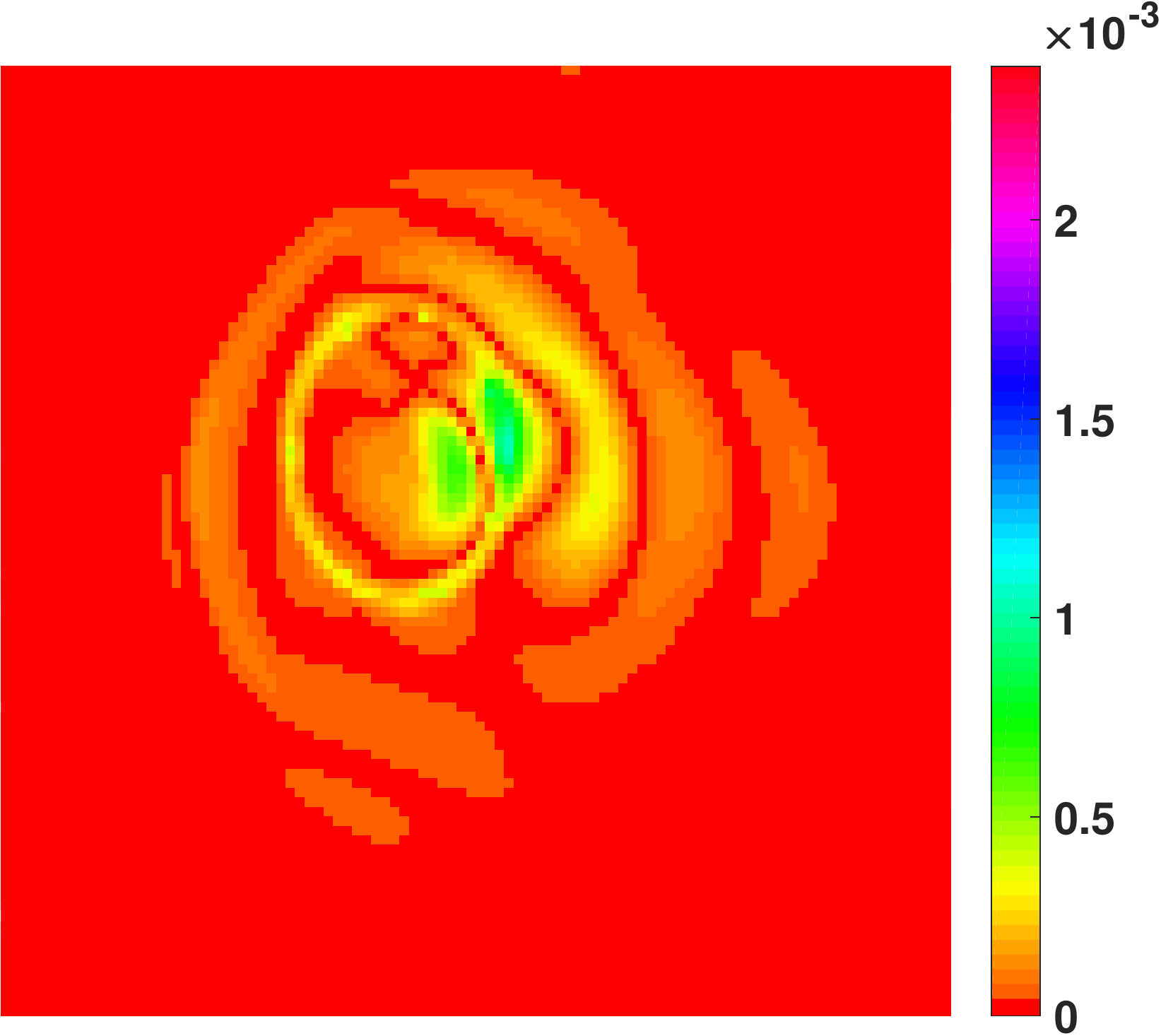} \\
\hspace*{-0.0cm}\includegraphics[width = 4.5cm]{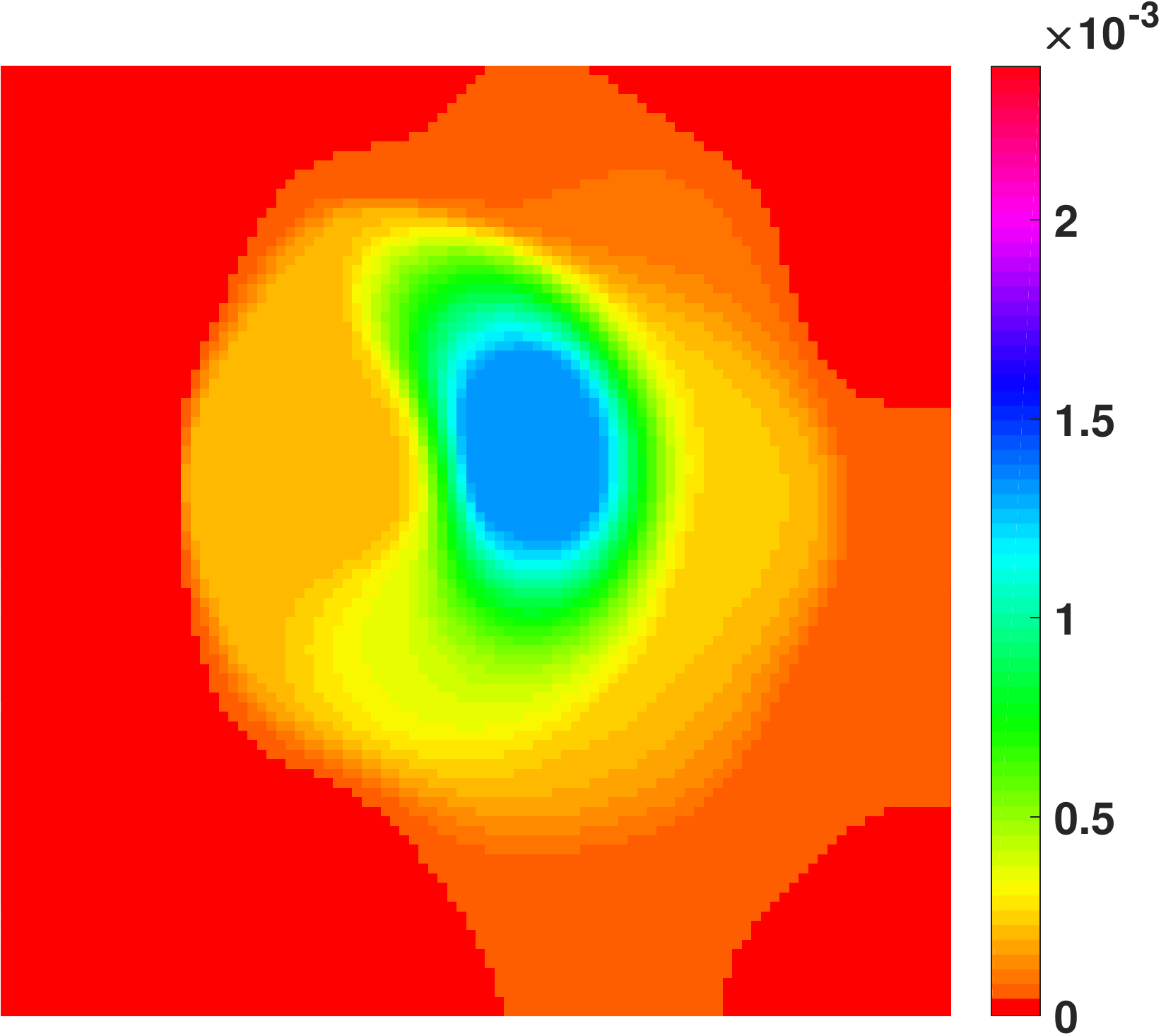} &
\hspace*{-0.0cm}\includegraphics[width = 4.5cm]{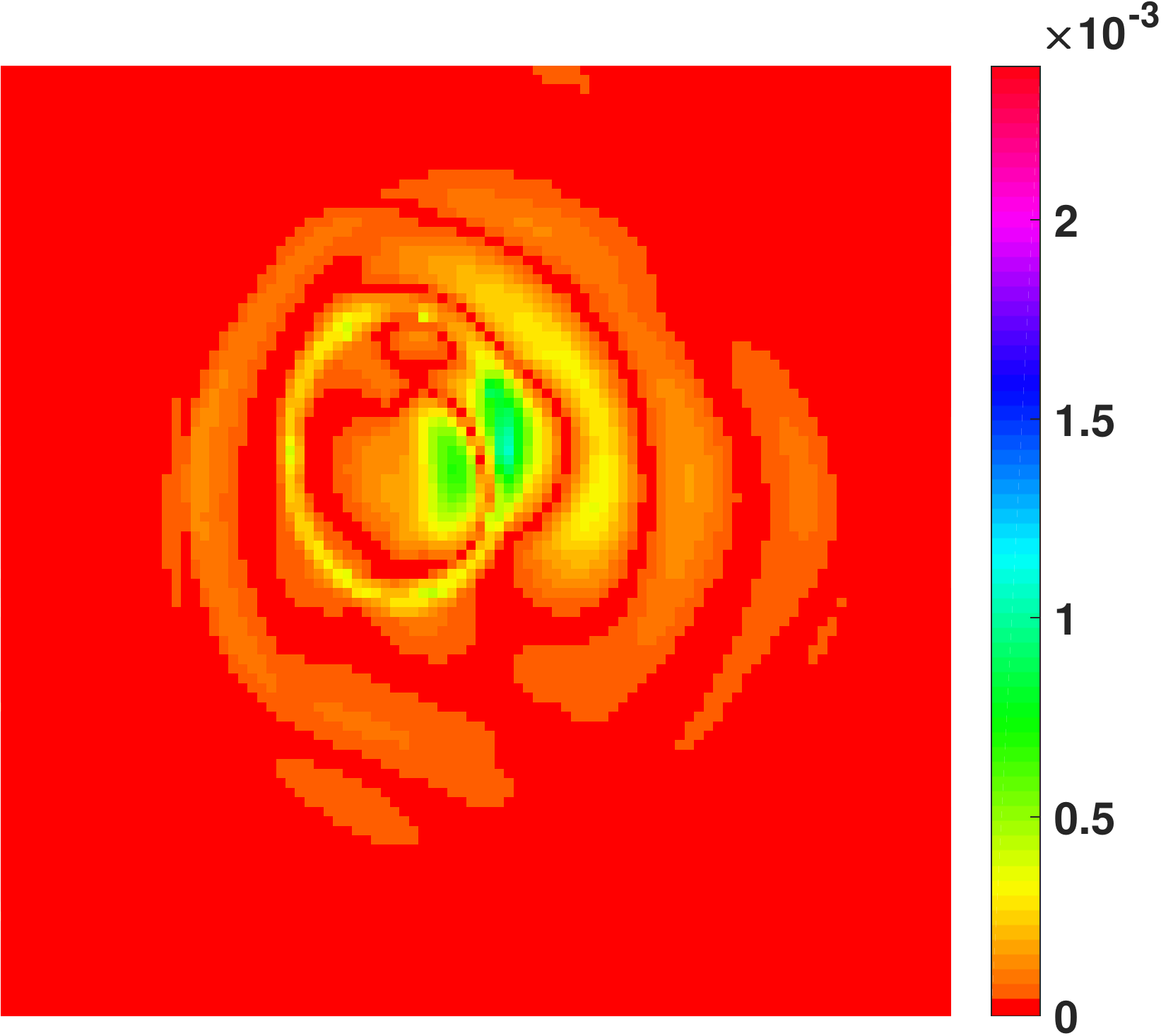} &
\hspace*{-0.0cm}\includegraphics[width = 4.5cm]{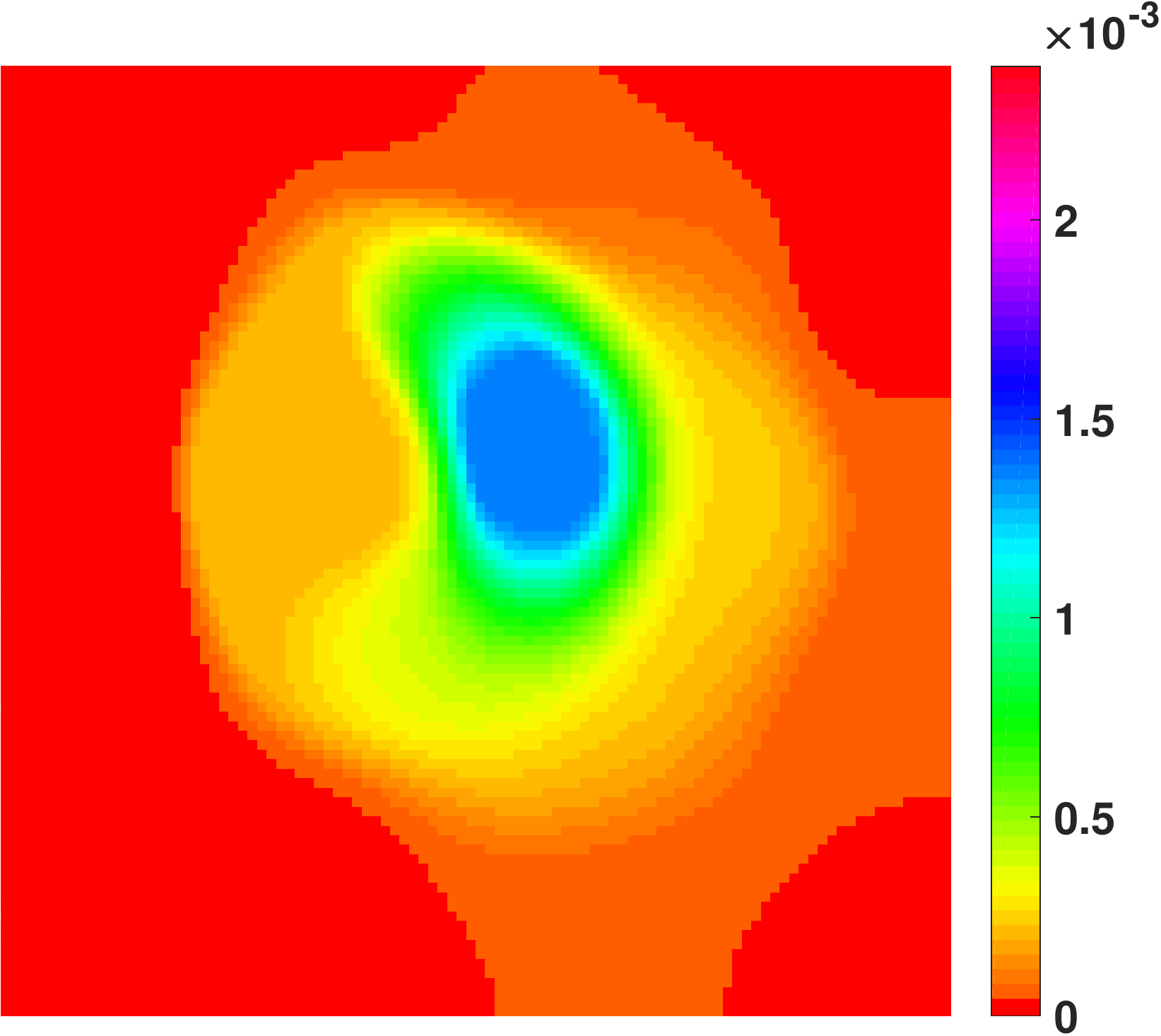} &
\hspace*{-0.0cm}\includegraphics[width = 4.5cm]{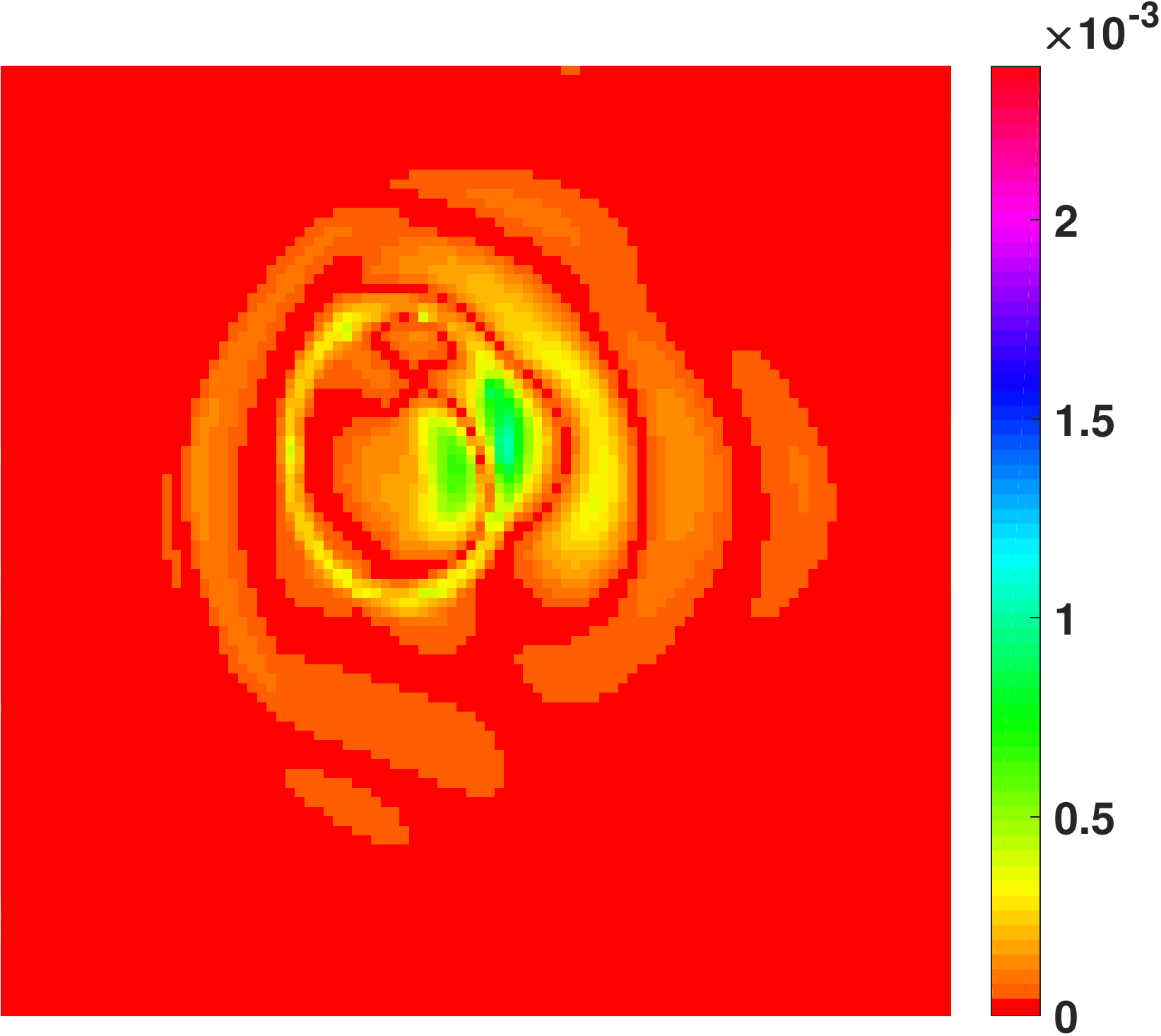} \\
\hspace*{-0.0cm}\includegraphics[width = 4.5cm]{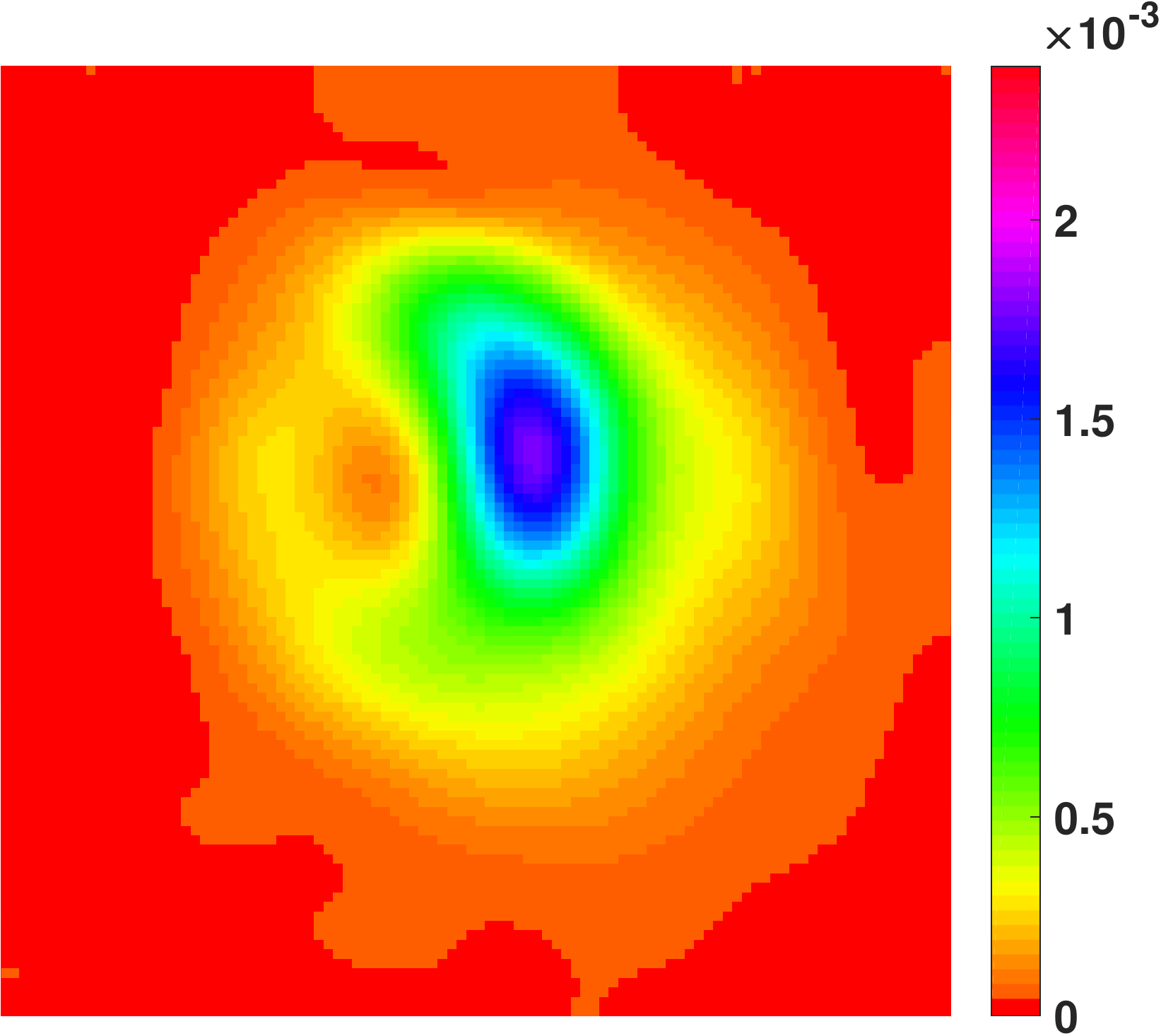} &
\hspace*{-0.0cm}\includegraphics[width = 4.5cm]{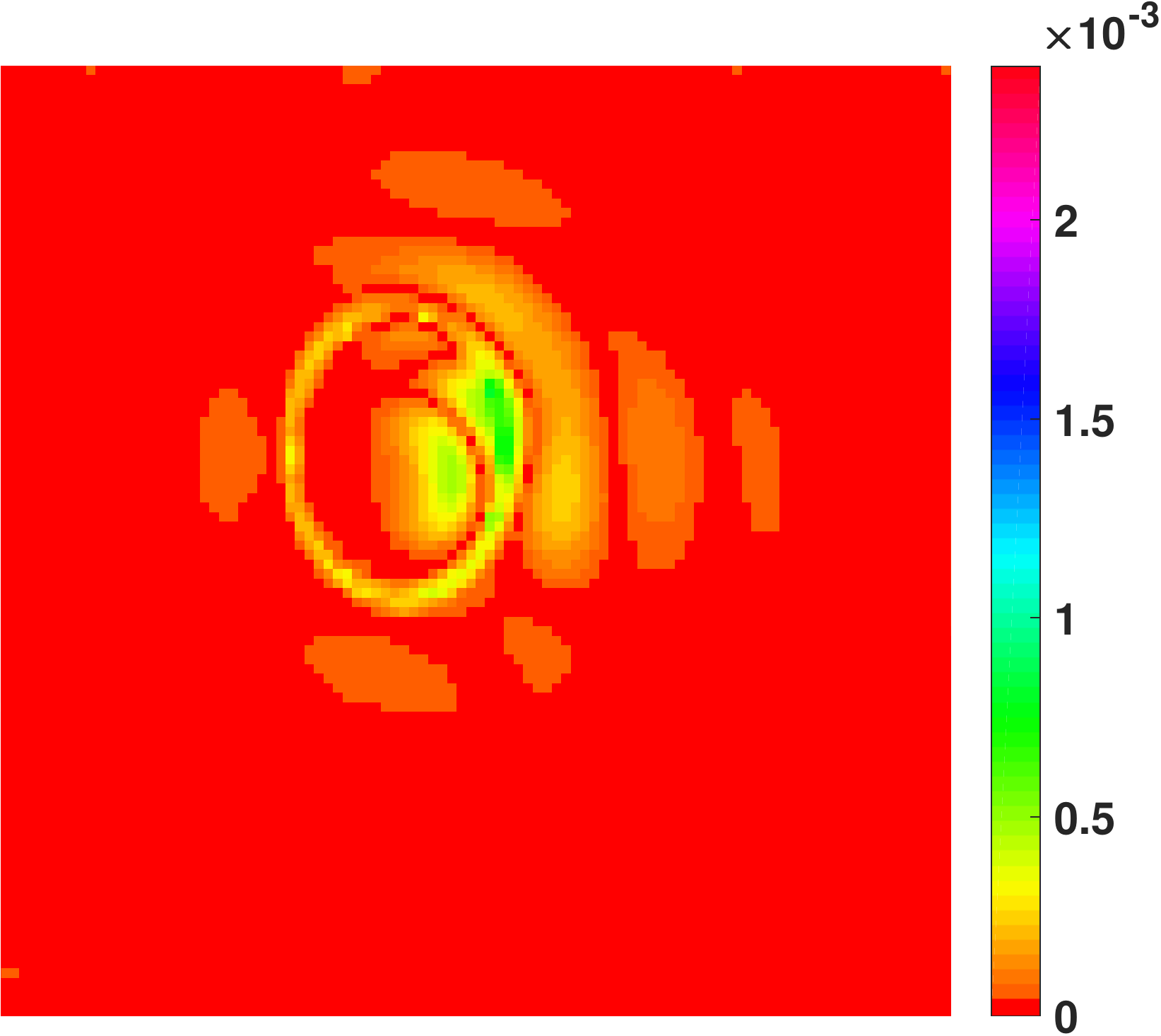} &
\hspace*{-0.0cm}\includegraphics[width = 4.5cm]{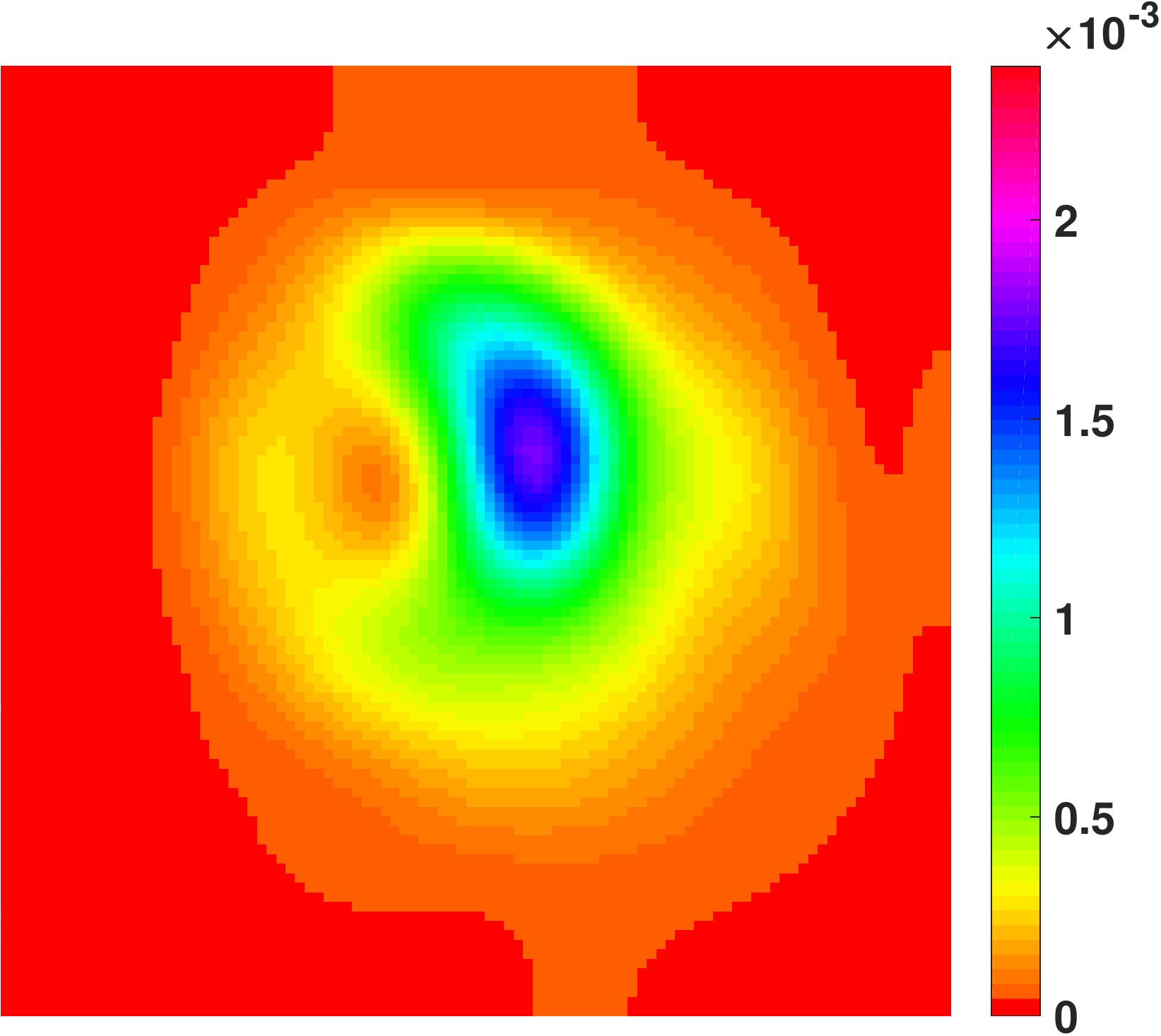} &
\hspace*{-0.0cm}\includegraphics[width = 4.5cm]{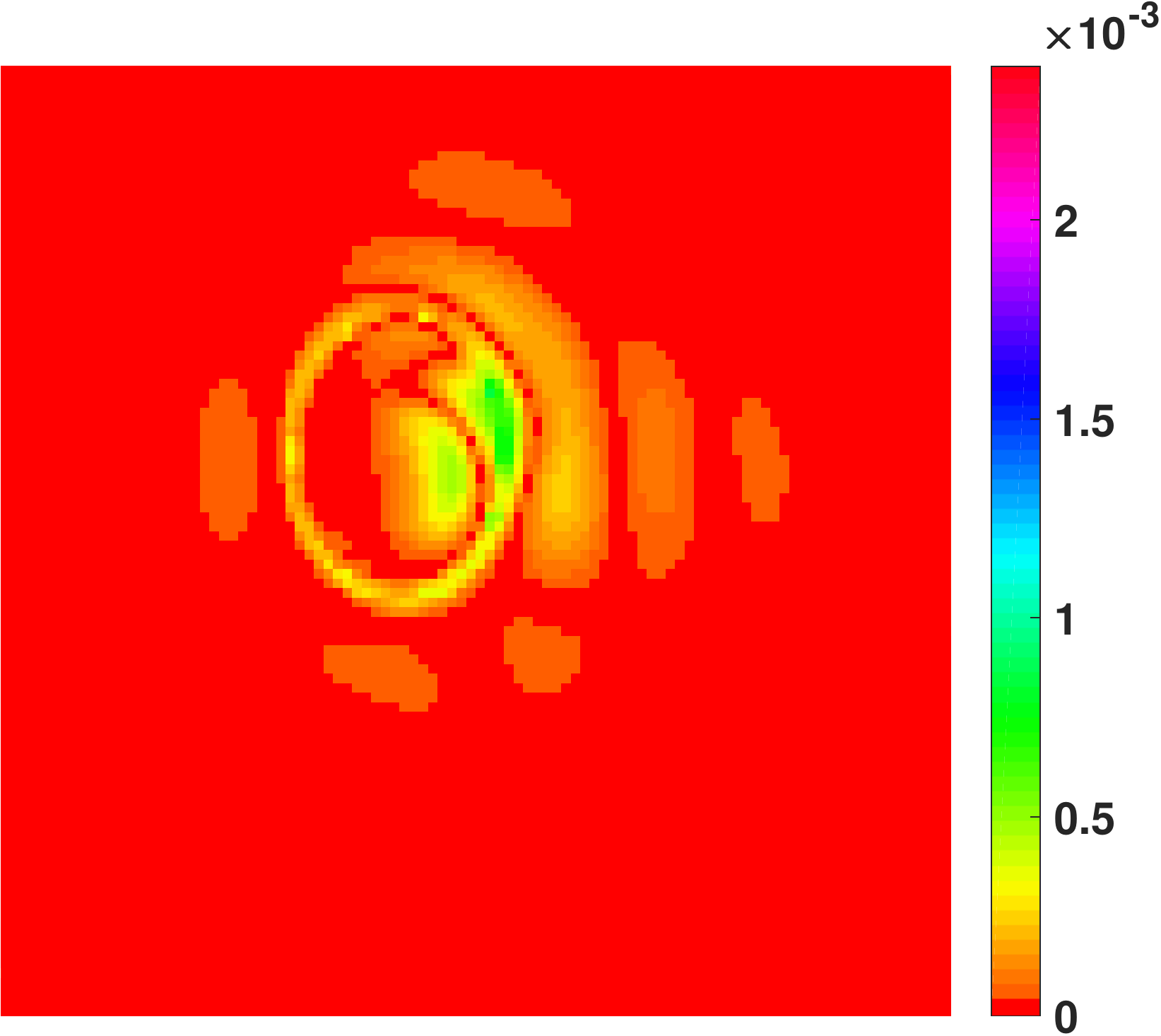} \\
\end{tabular}
\caption{Results for the Stokes $I$ image corresponding to the forward-jet model. First row shows the ground-truth image, whereas the second row shows the \textsc{cs-clean} reconstructed image followed by its error image. Third and fourth row show the results for the TV and $\ell_1$+ TV  problems, respectively. For these rows, the first two columns show the reconstructed and the error images obtained by without imposing the polarization constraint in the reconstruction process, whereas the corresponding images in the case of imposing this constraint are shown in the last two columns.
Similarly, column-wise, last row displays the reconstructed image for Polarized SARA without constraint and its error image; reconstructed image for Polarized SARA and its error image. The shown images correspond to the best results obtained over 5 performed simulations for each case. All the images are shown in linear scale, normalized to the scale of the corresponding ground truth image.}
\label{fig:rec_images_avery_I}
\end{figure*}

\begin{figure*}
\centering
\begin{tabular}{@{}c@{}c@{}c@{}c@{}}
\hspace*{-0.0cm}\includegraphics[trim ={2cm 0 0 0.2cm},clip,width=5.1cm]{true_P_new_avery-eps-converted-to.pdf} \\
\hspace*{-0.0cm}\includegraphics[trim ={2cm 0 0 0.2cm},clip,width=5.1cm]{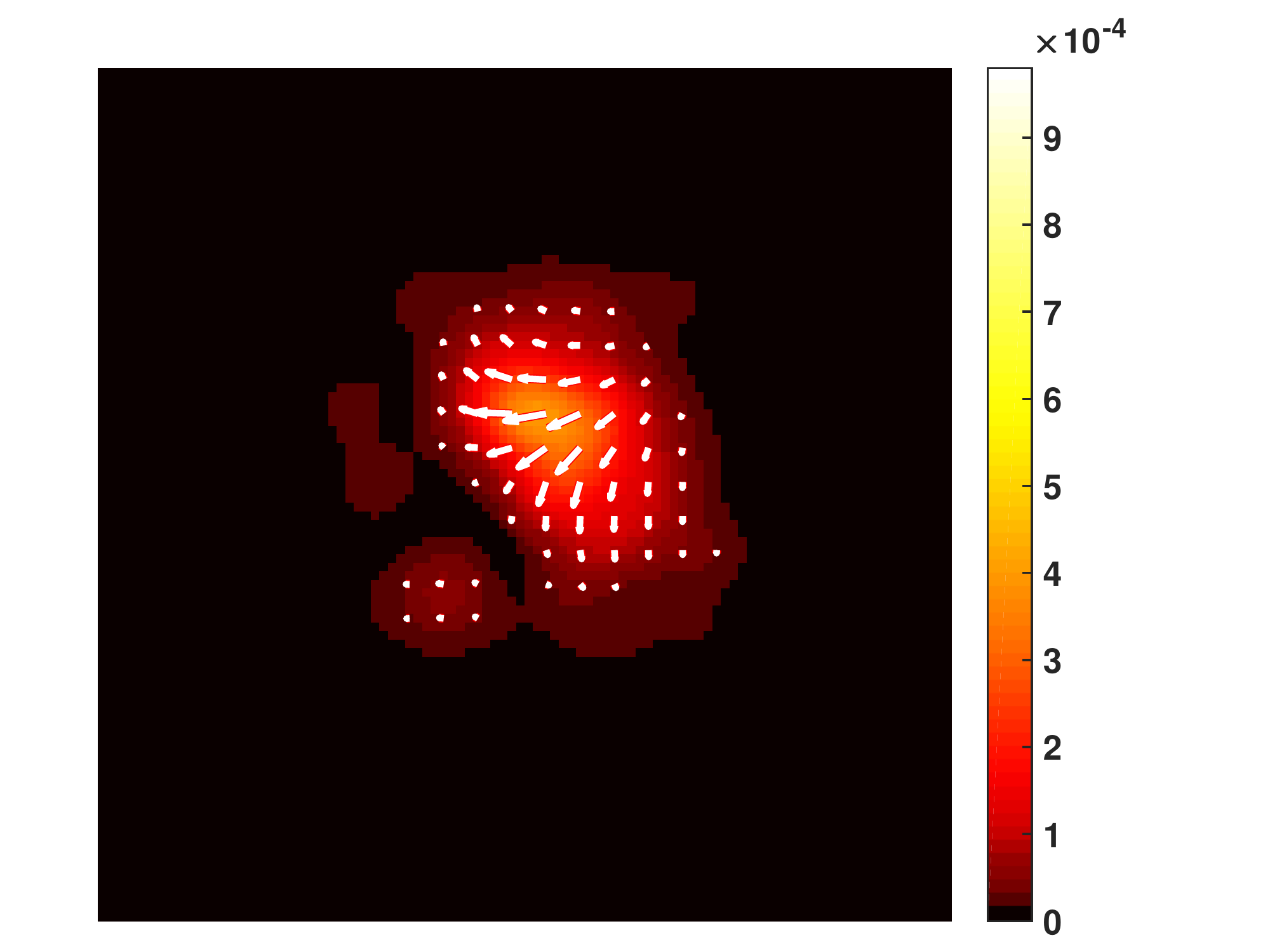} &
\hspace*{-0.55cm}\includegraphics[trim ={2cm 0 0 0.2cm},clip,width=5.1cm]{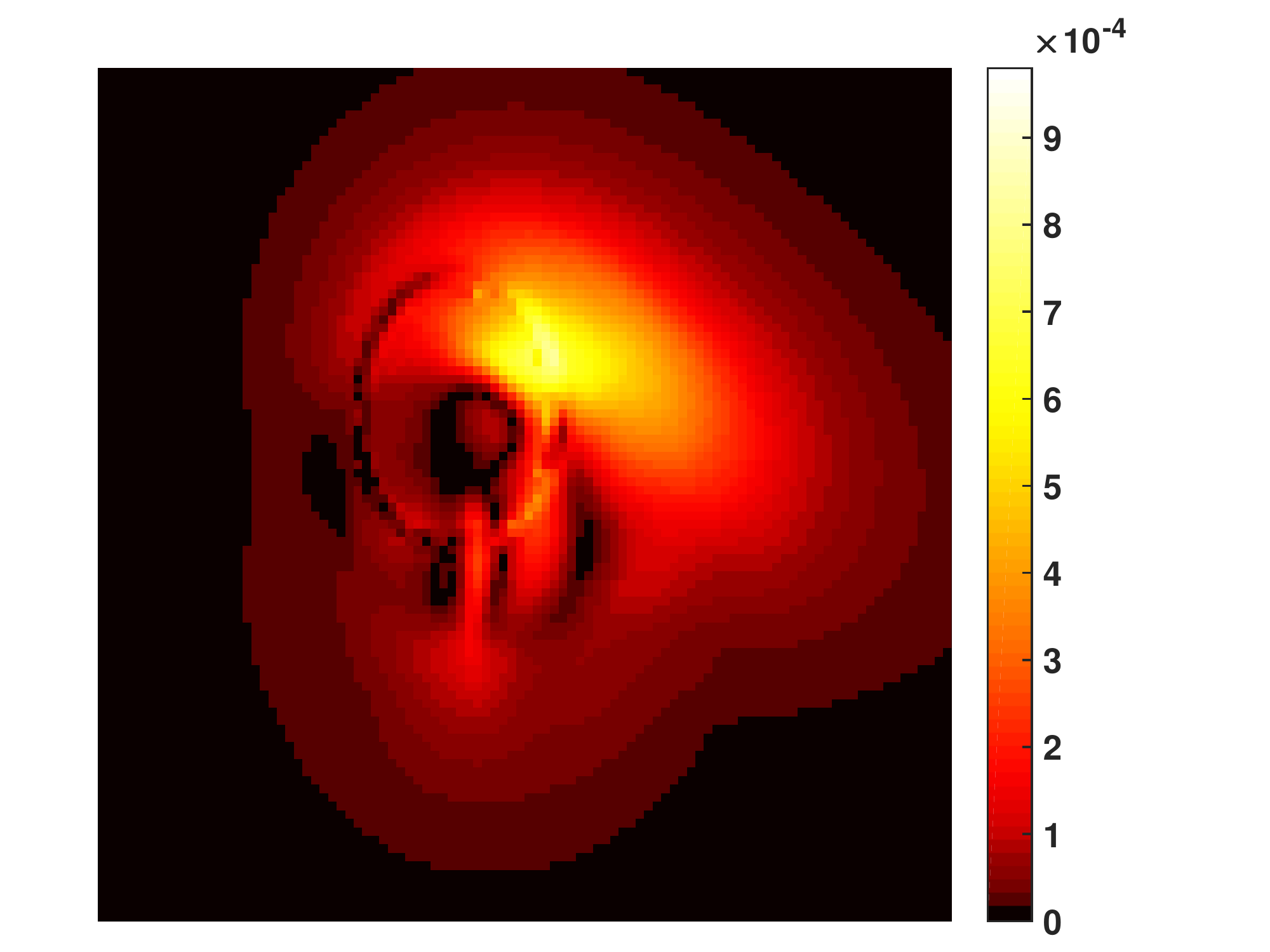} \\
\hspace*{-0.0cm}\includegraphics[trim ={2cm 0 0 0.2cm},clip,width=5.1cm]{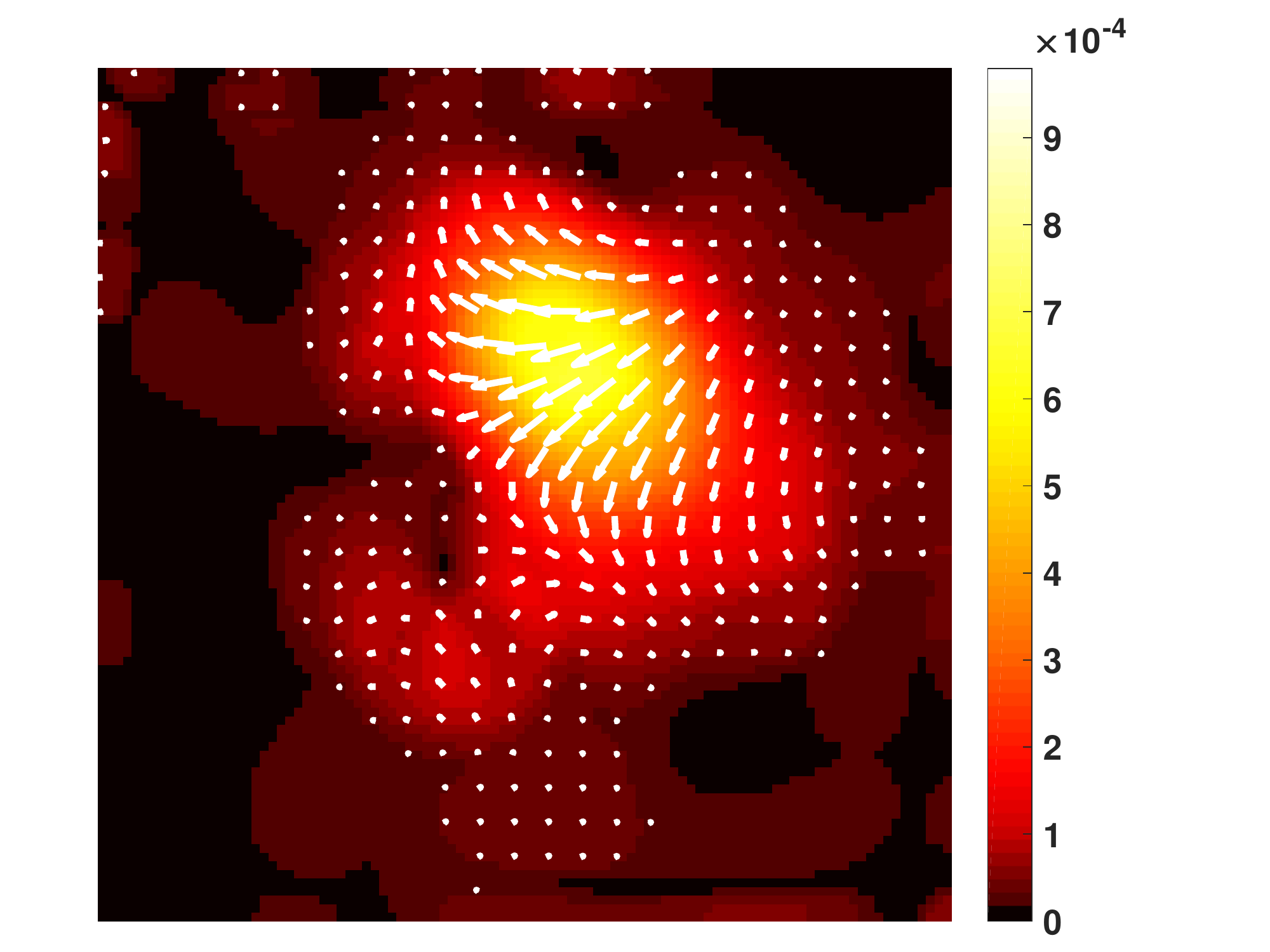} &
\hspace*{-0.55cm}\includegraphics[trim ={2cm 0 0 0.2cm},clip,width=5.1cm]{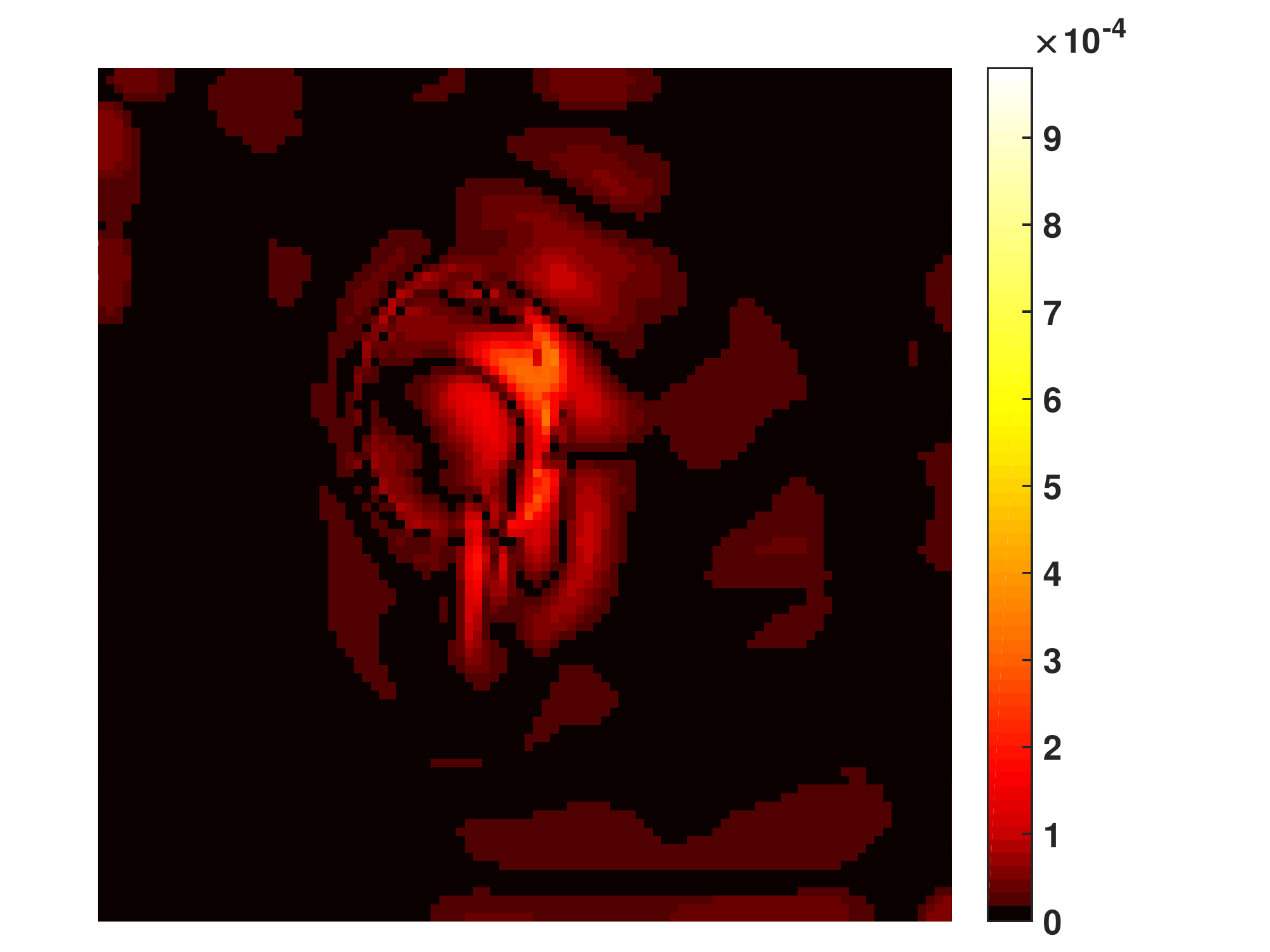} &
\hspace*{-0.55cm}\includegraphics[trim ={2cm 0 0 0.2cm},clip,width=5.1cm]{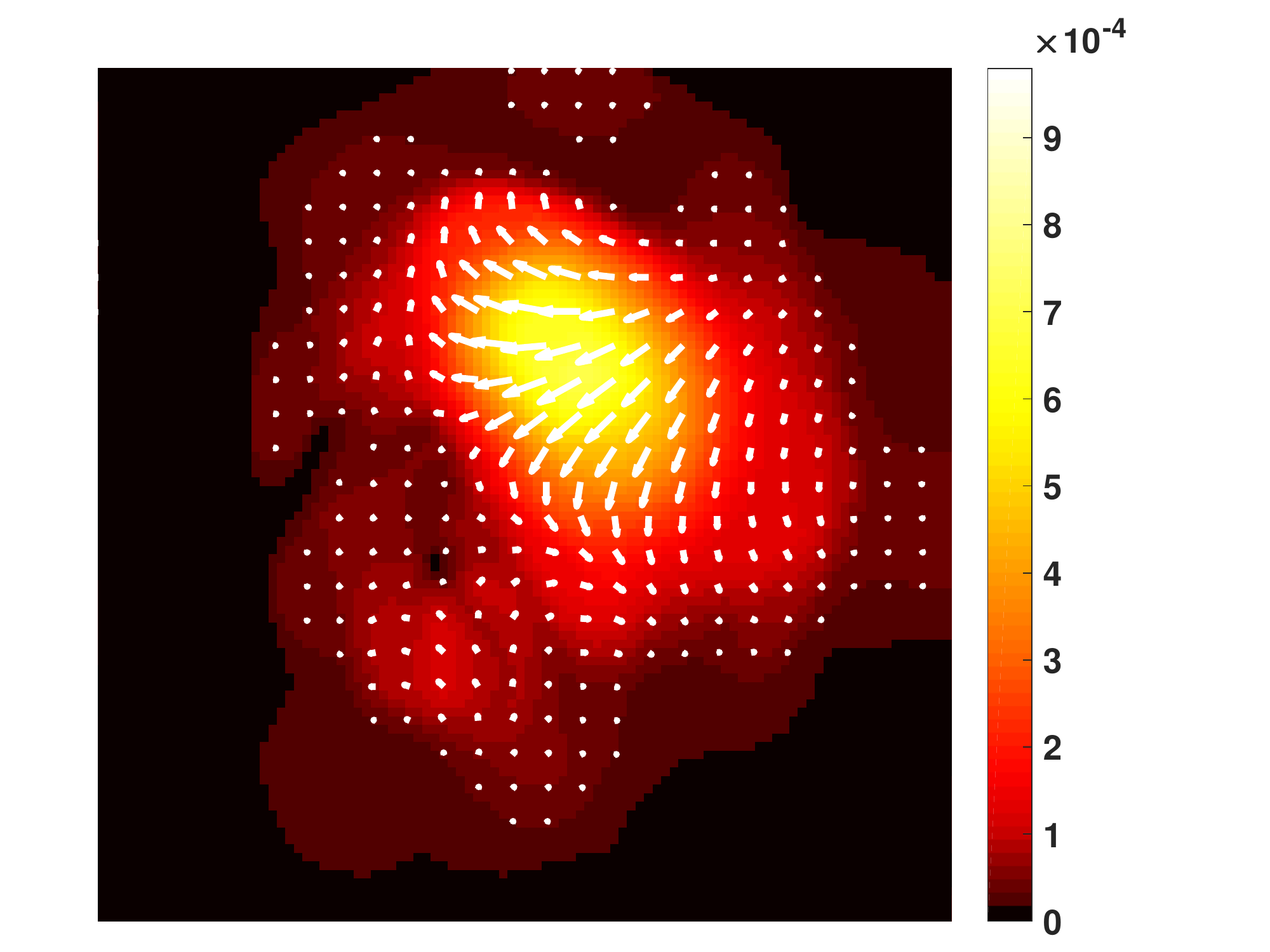} &
\hspace*{-0.55cm}\includegraphics[trim ={2cm 0 0 0.2cm},clip,width=5.1cm]{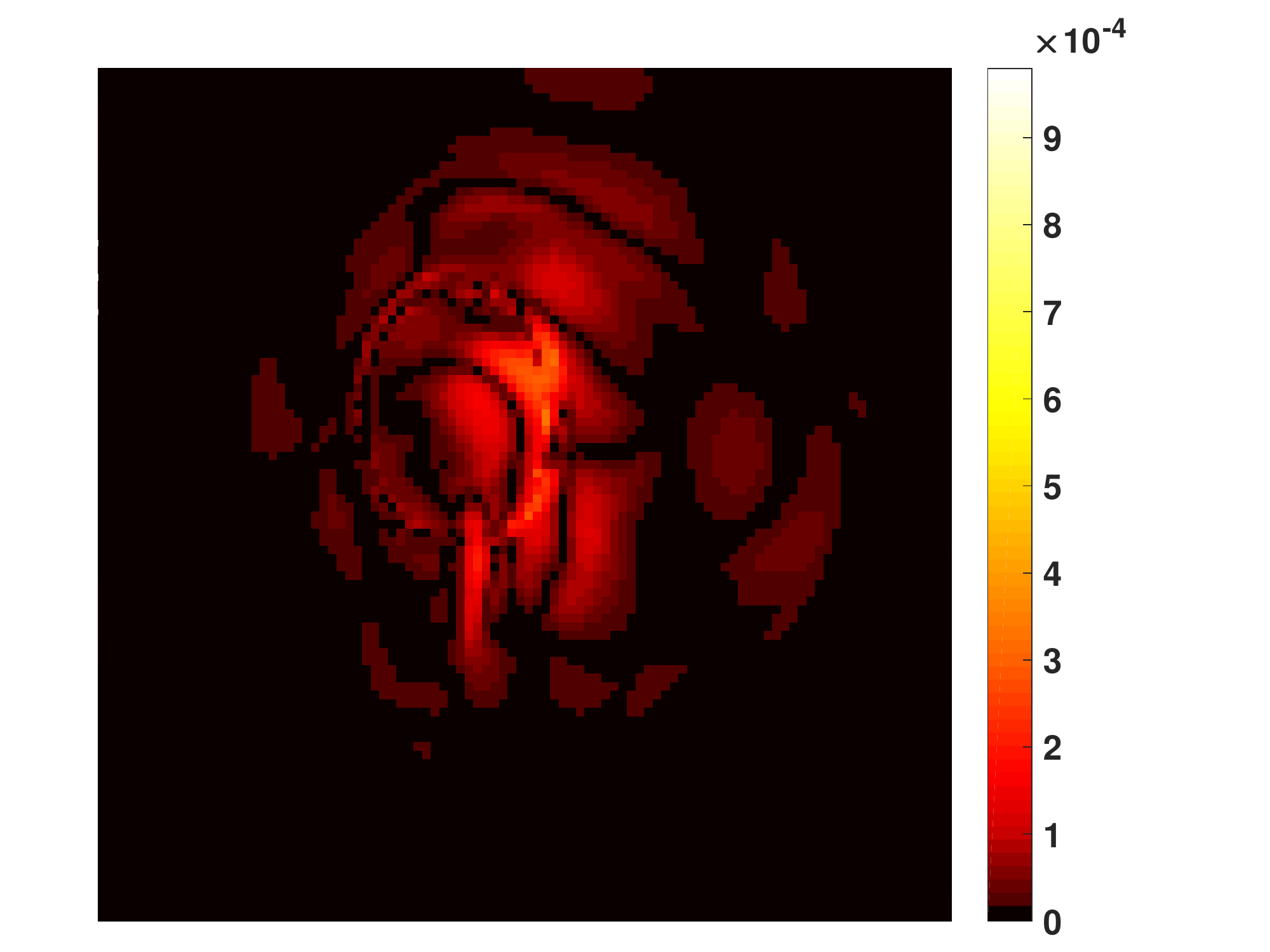}  \\
\hspace*{-0.0cm}\includegraphics[trim ={2cm 0 0 0.2cm},clip,width=5.1cm]{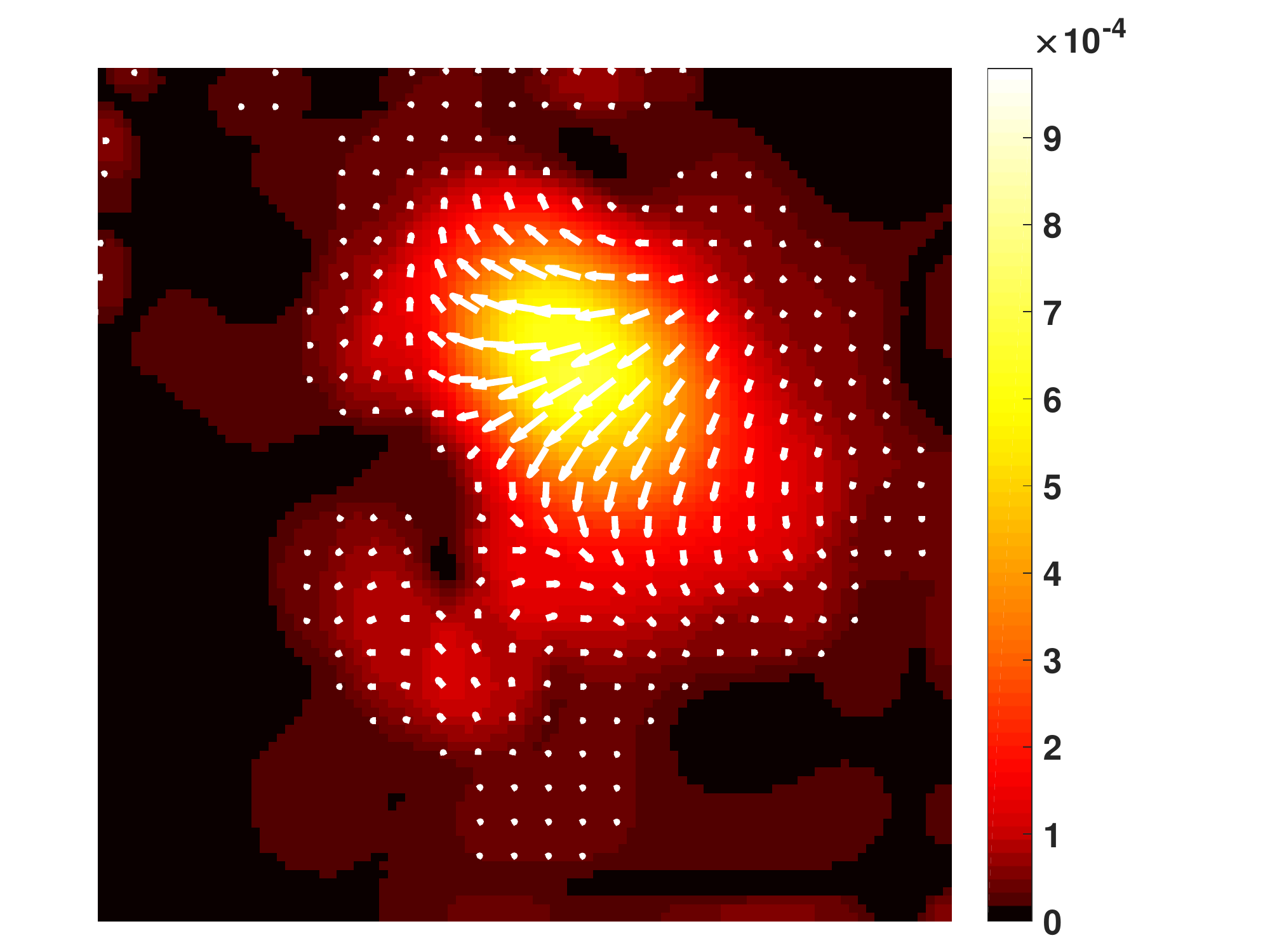} &
\hspace*{-0.55cm}\includegraphics[trim ={2cm 0 0 0.2cm},clip,width=5.1cm]{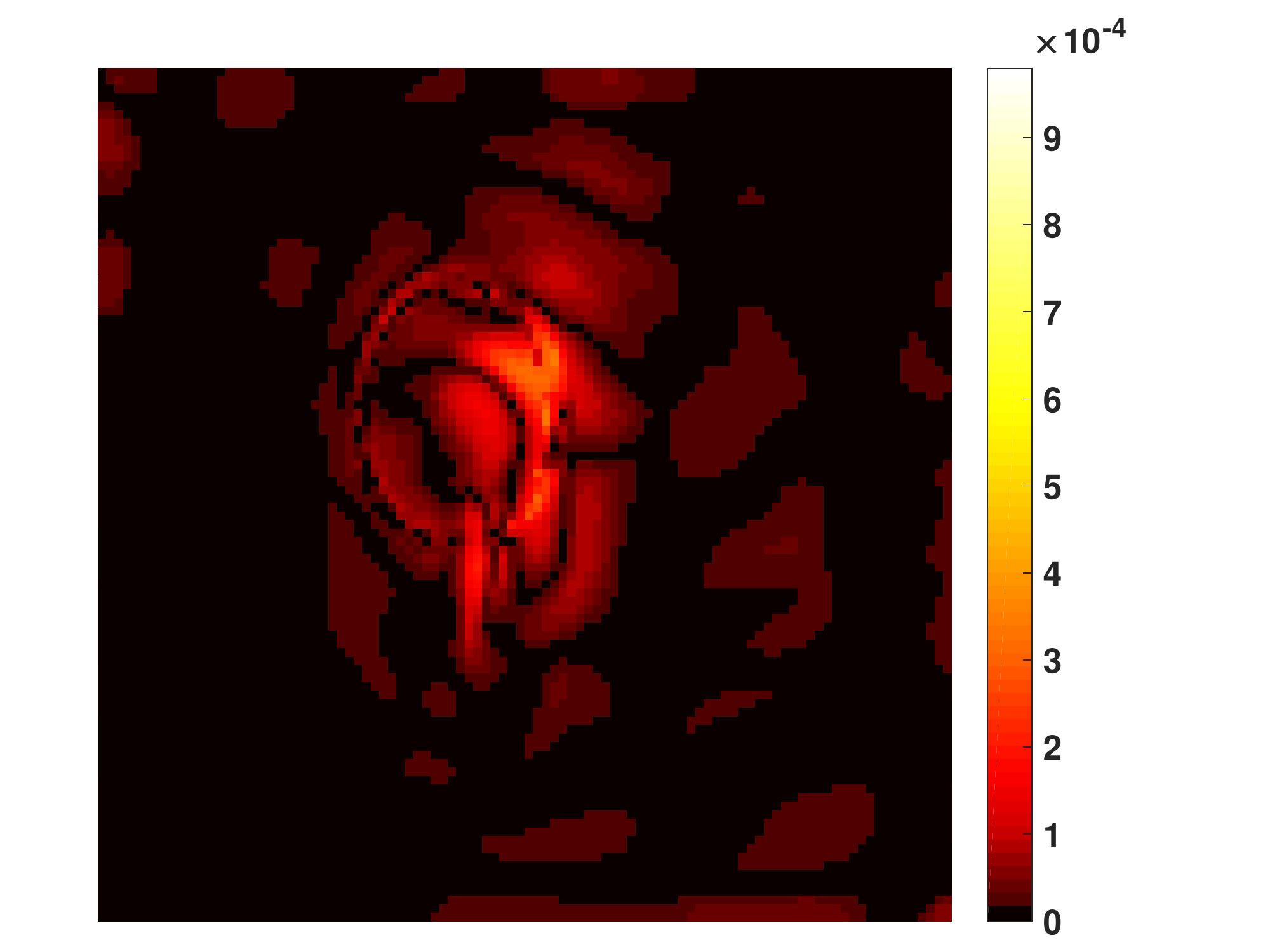} &
\hspace*{-0.55cm}\includegraphics[trim ={2cm 0 0 0.2cm},clip,width=5.1cm]{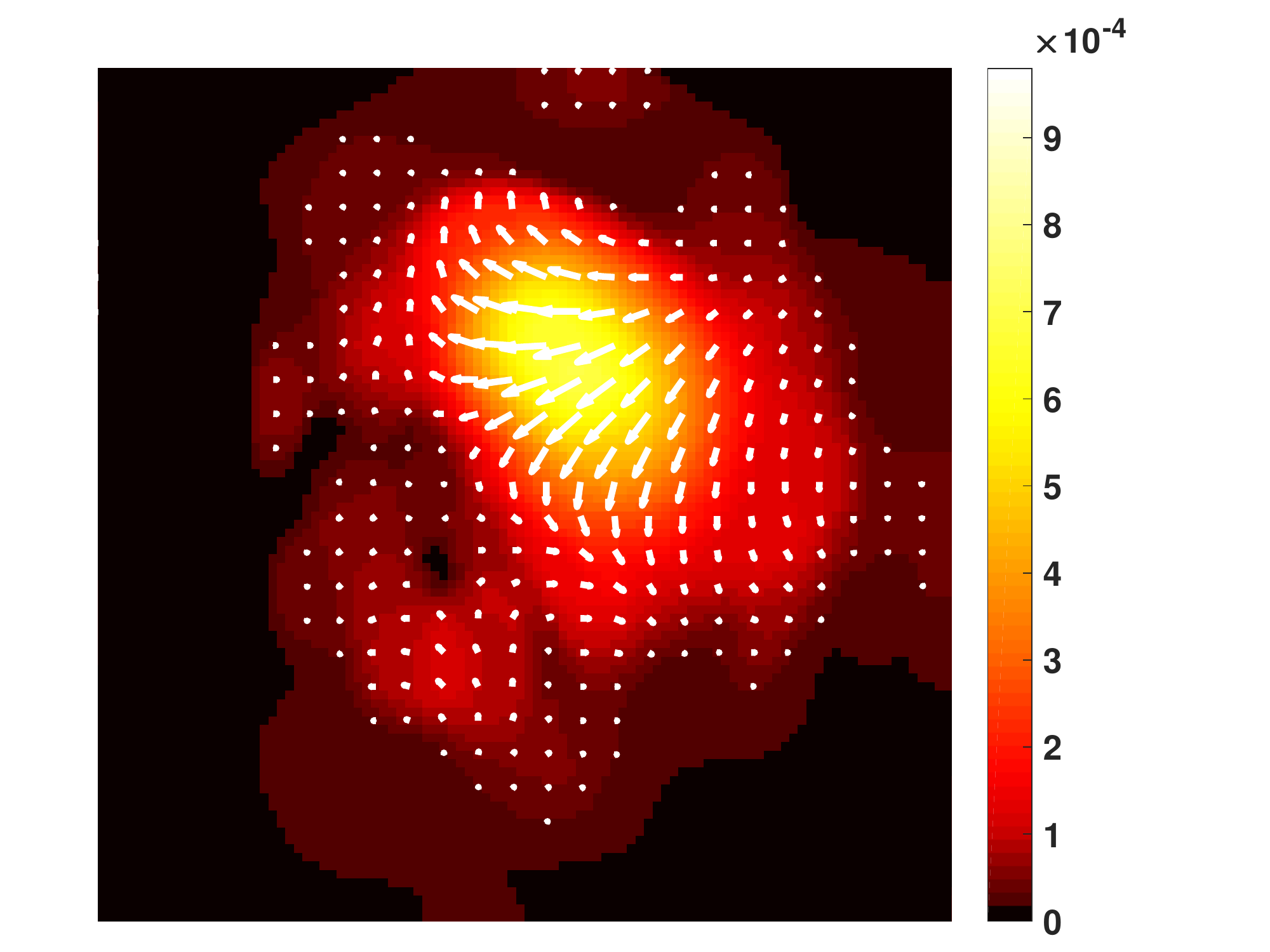} &
\hspace*{-0.55cm}\includegraphics[trim ={2cm 0 0 0.2cm},clip,width=5.1cm]{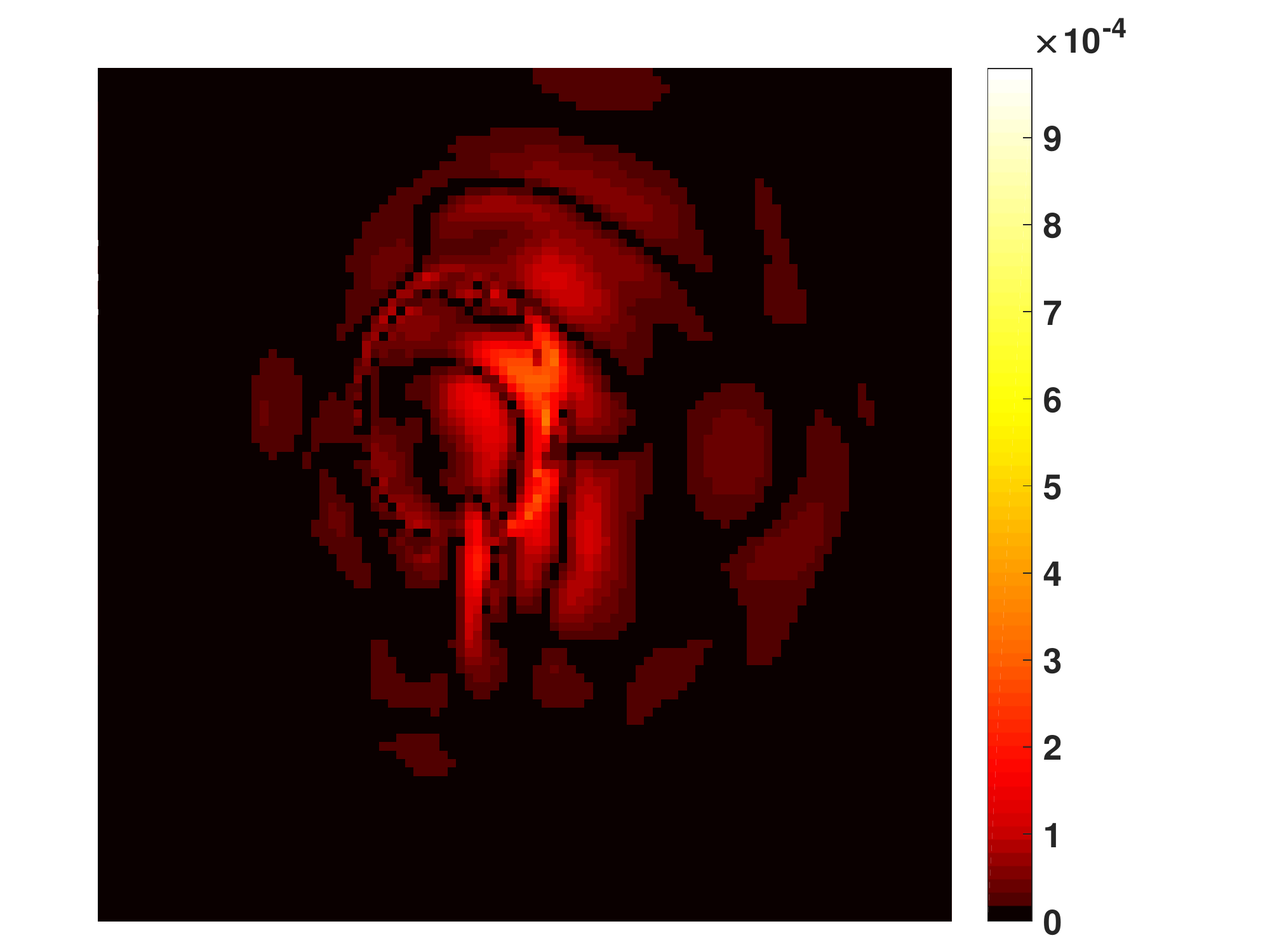} \\
\hspace*{-0.0cm}\includegraphics[trim ={2cm 0 0 0.2cm},clip,width=5.1cm]{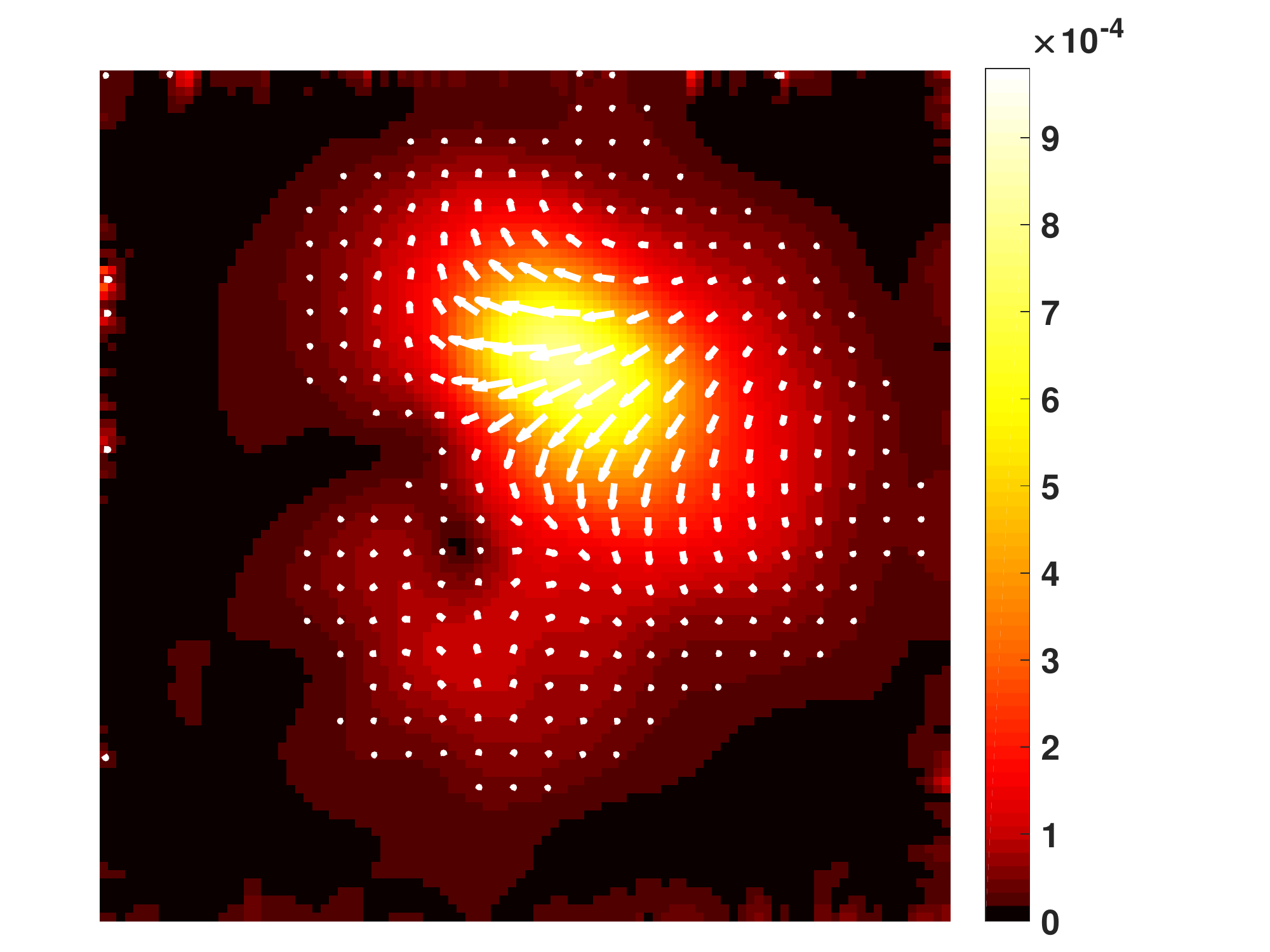} &
\hspace*{-0.55cm}\includegraphics[trim ={2cm 0 0 0.2cm},clip,width=5.1cm]{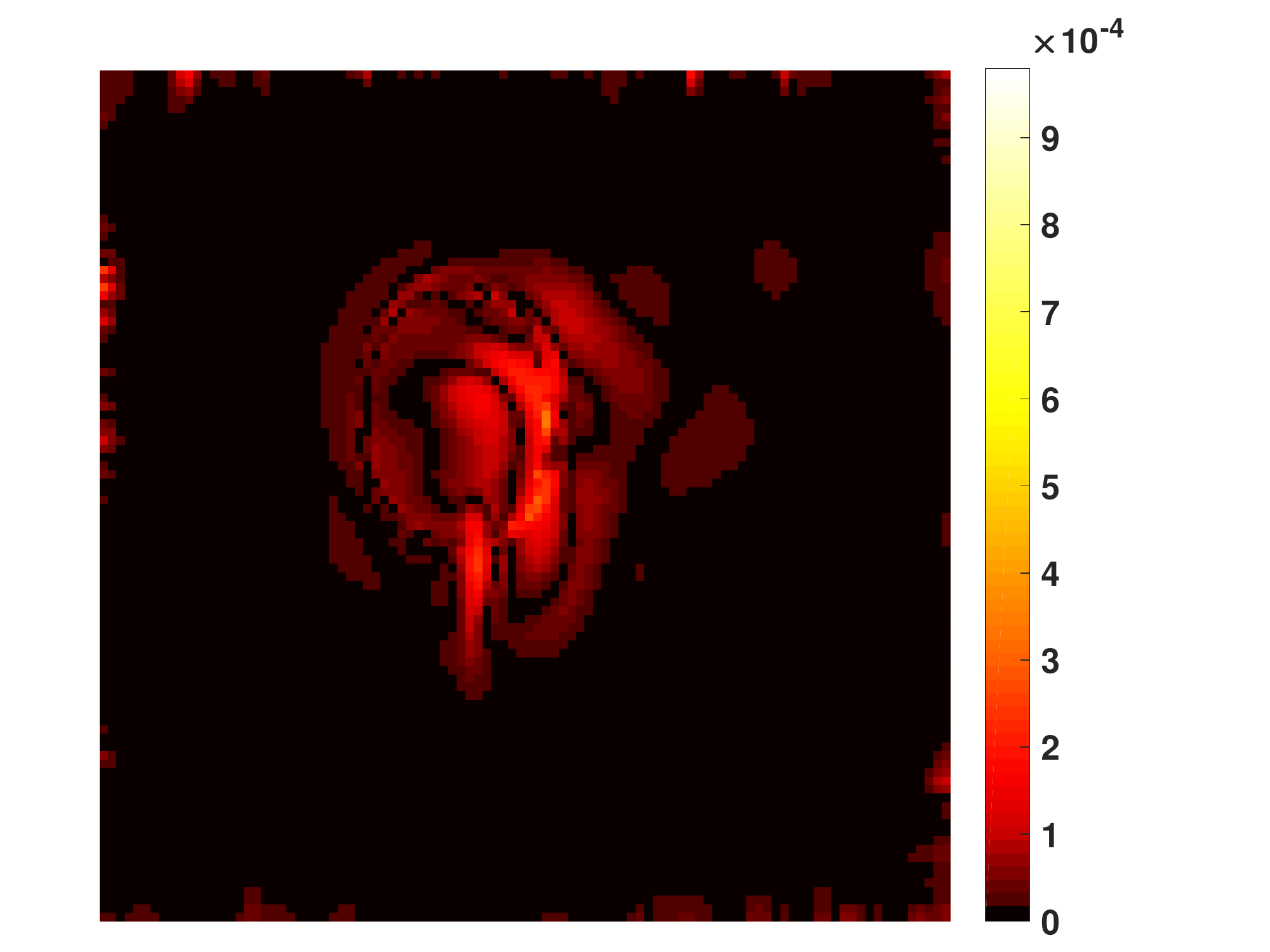} &
\hspace*{-0.55cm}\includegraphics[trim ={2cm 0 0 0.2cm},clip,width=5.1cm]{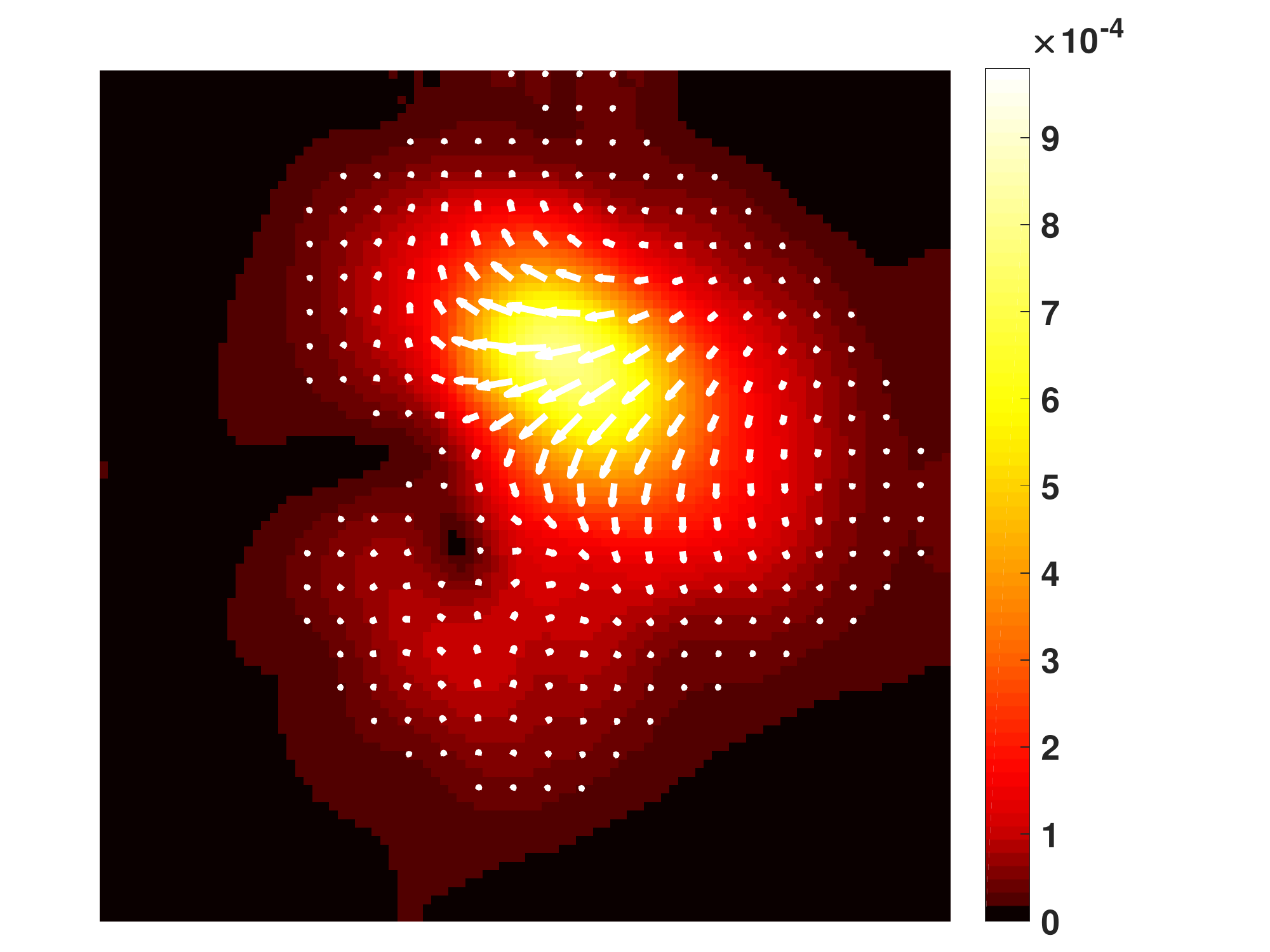} &
\hspace*{-0.55cm}\includegraphics[trim ={2cm 0 0 0.2cm},clip,width=5.1cm]{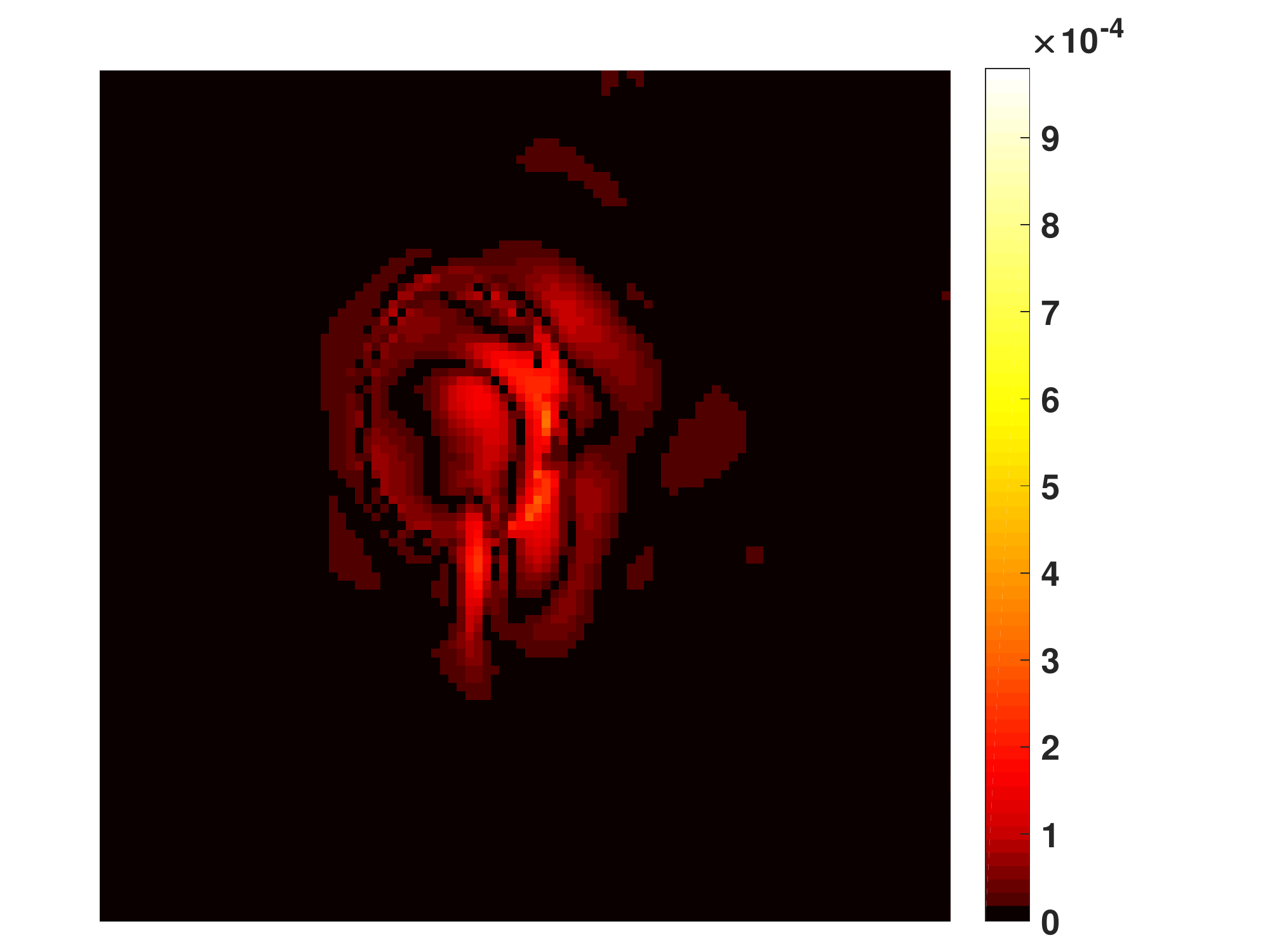} 
\end{tabular}
\vspace*{-0.2cm}
\caption{Results for the linear polarization image corresponding to the forward-jet model. First row shows the ground-truth image, whereas the second row shows the \textsc{cs-clean} reconstructed image followed by its error image. Third and fourth row show the results for the TV and $\ell_1$+ TV problems, respectively. For these rows, the first two columns show the reconstructed and the error images obtained by without imposing the polarization constraint in the reconstruction process, whereas the corresponding images in the case of imposing this constraint are shown in the last two columns. 
Similarly, column-wise, last row displays the reconstructed image for Polarized SARA without constraint and its error image; reconstructed image for Polarized SARA and its error image. The reconstructed images correspond to the linear polarization intensity images, overlaid by the white bars representing the EVPA. The shown images correspond to the best results  obtained over 5 performed simulations for each case. All the images are shown in linear scale, normalized to the scale of the corresponding ground-truth image.}
\label{fig:rec_images_avery_P}
\end{figure*}

\begin{figure*}
\centering
\begin{tabular}{@{}c@{}c@{}c@{}c@{}}
\hspace*{-0.0cm}\includegraphics[width = 4.5cm]{true_I_new_jason-eps-converted-to.pdf} \\
\hspace*{-0cm}\includegraphics[width = 4.5cm]{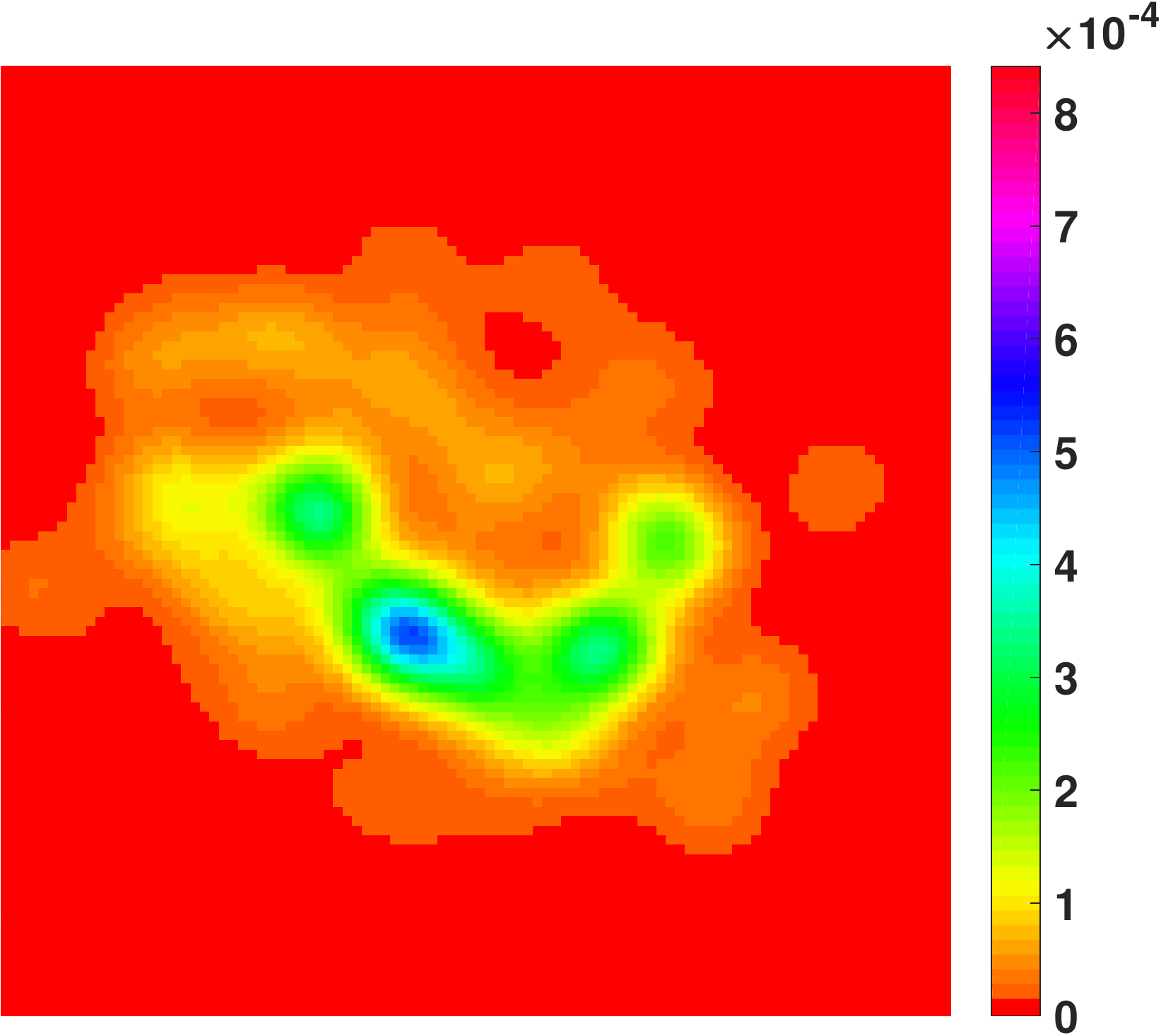} &
\hspace*{-0cm}\includegraphics[width = 4.5cm]{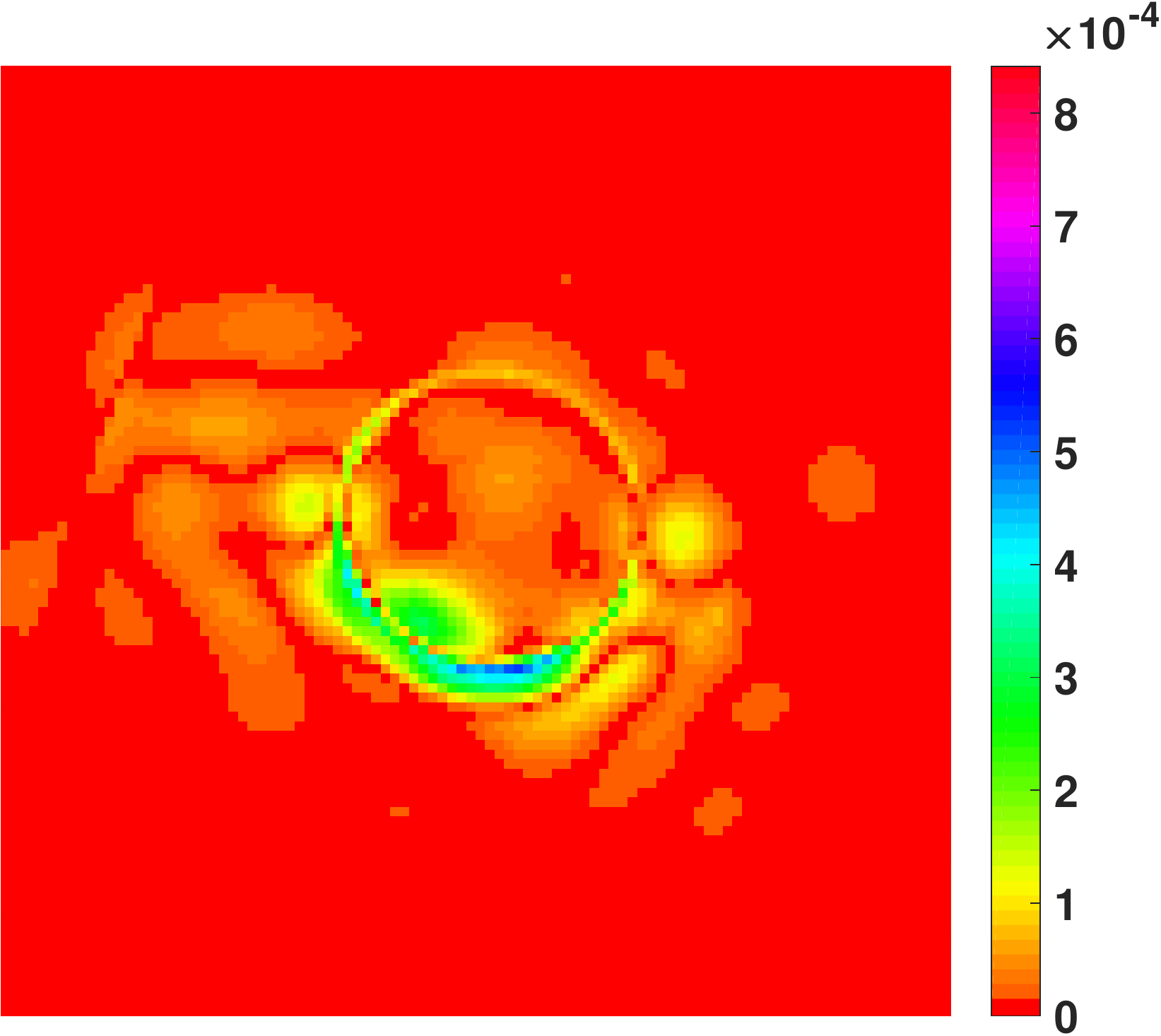} \\
\hspace*{-0.0cm}\includegraphics[width = 4.5cm]{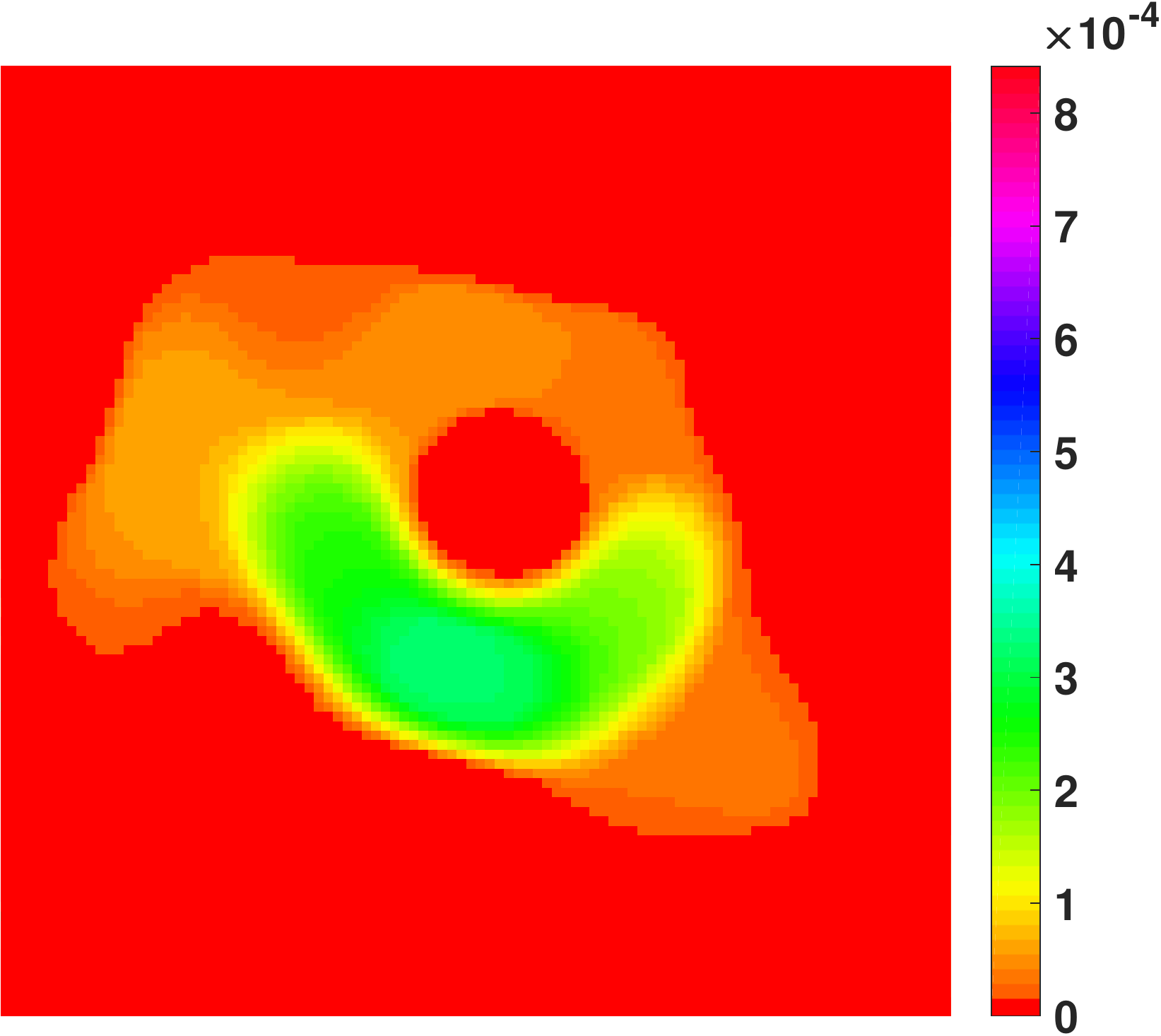} &
\hspace*{-0cm}\includegraphics[width = 4.5cm]{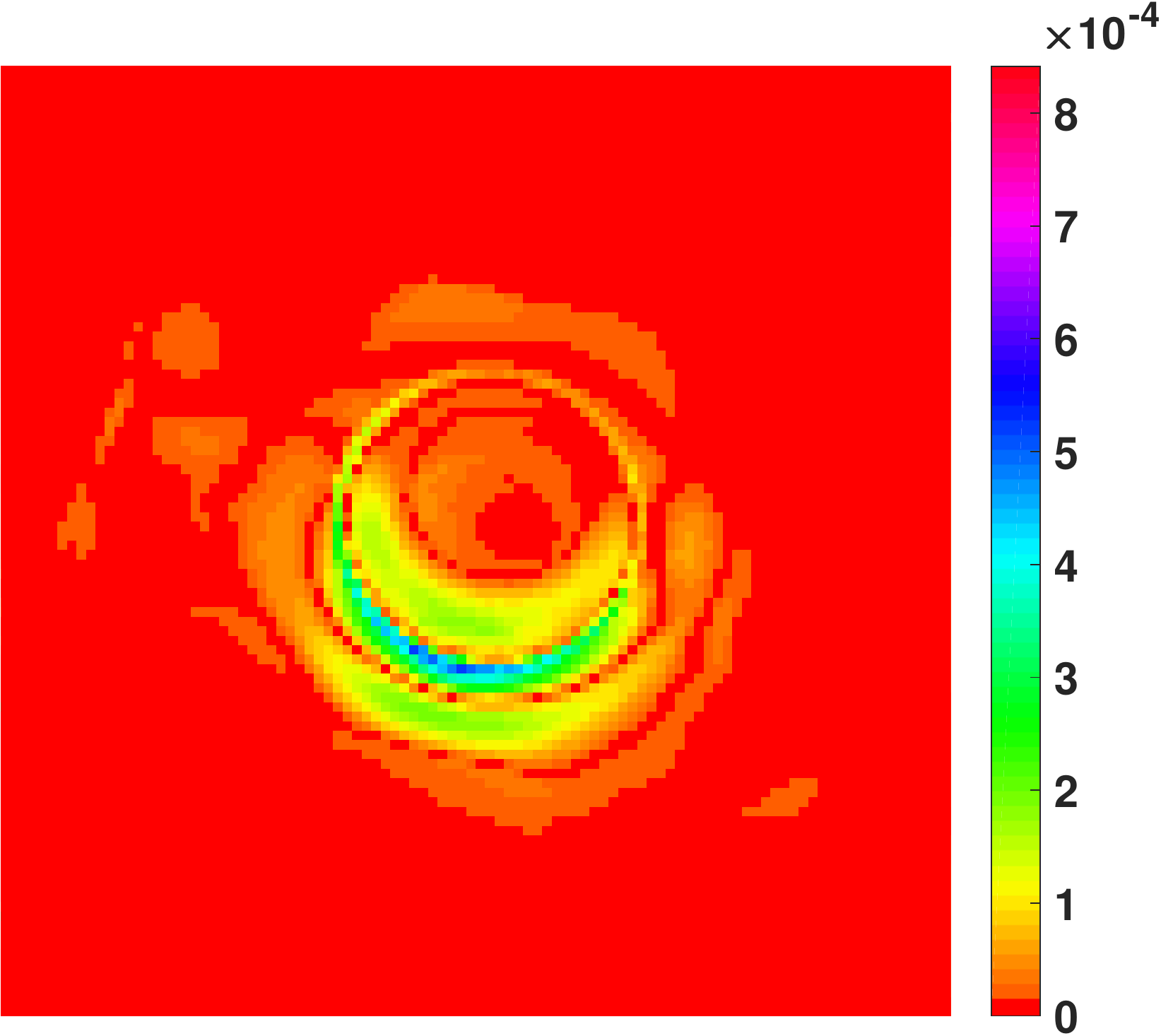} &
\hspace*{-0cm}\includegraphics[width = 4.5cm]{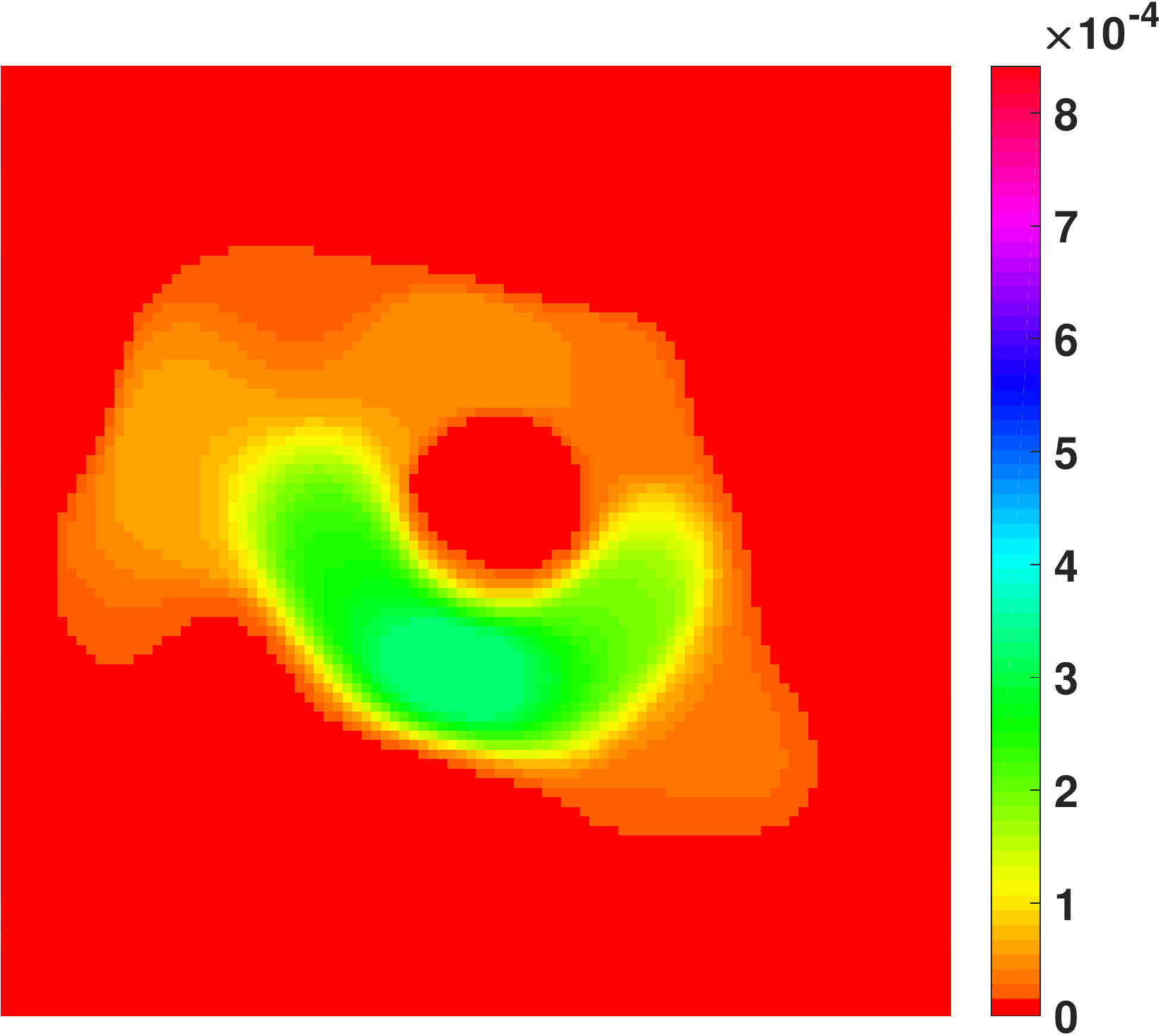} &
\hspace*{-0cm}\includegraphics[width = 4.5cm]{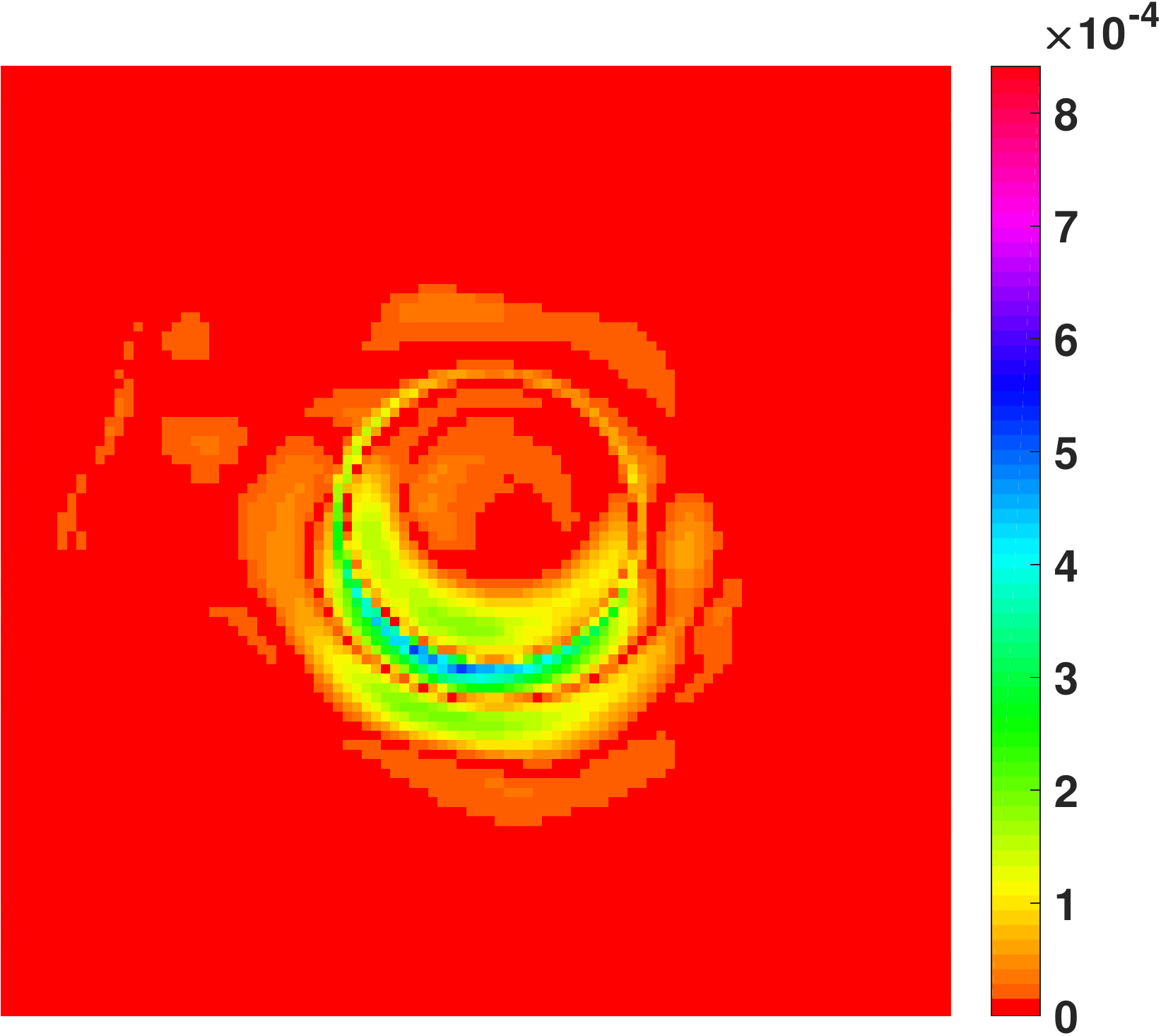} \\
\hspace*{-0.0cm}\includegraphics[width = 4.5cm]{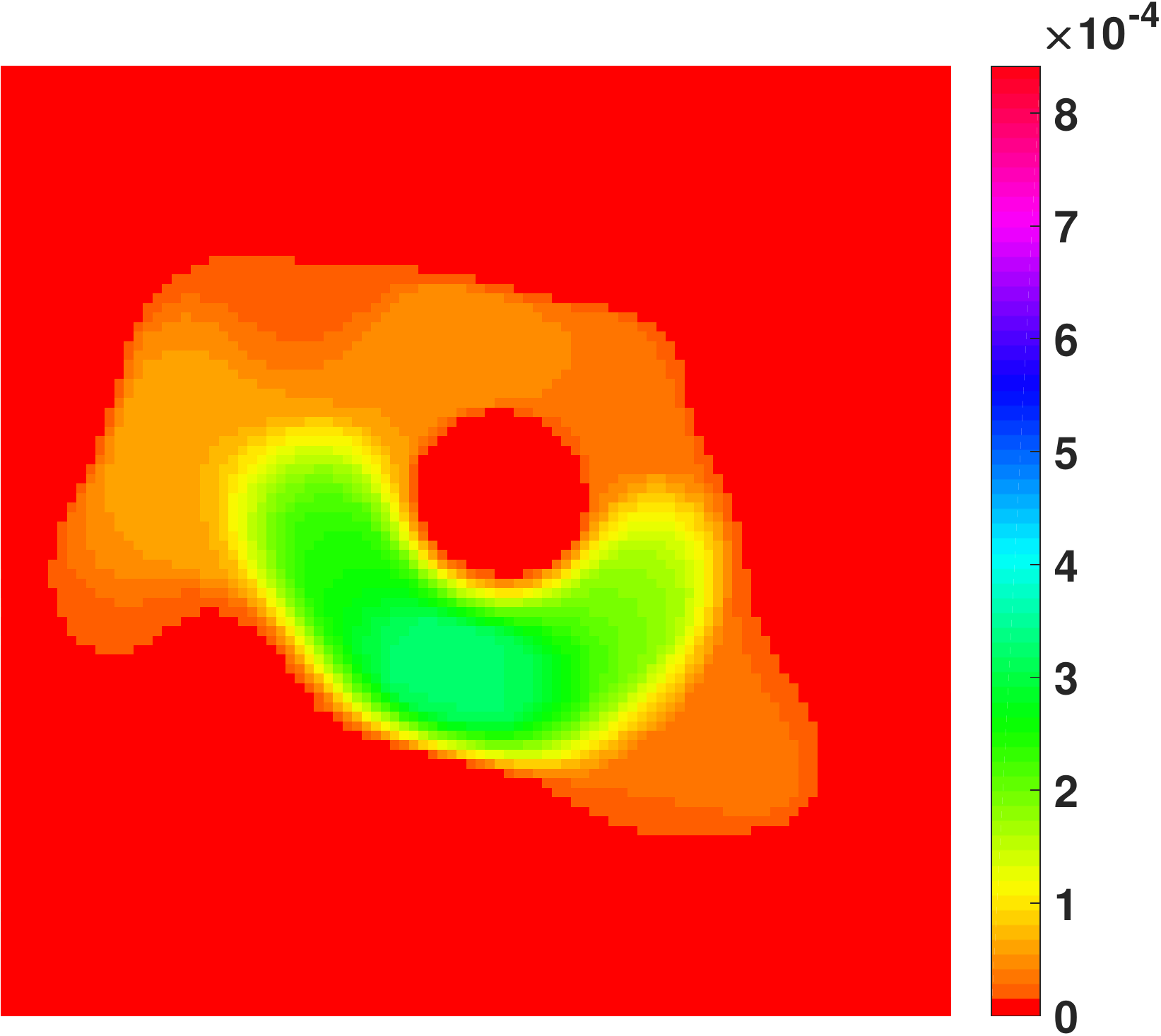} &
\hspace*{-0.0cm}\includegraphics[width = 4.5cm]{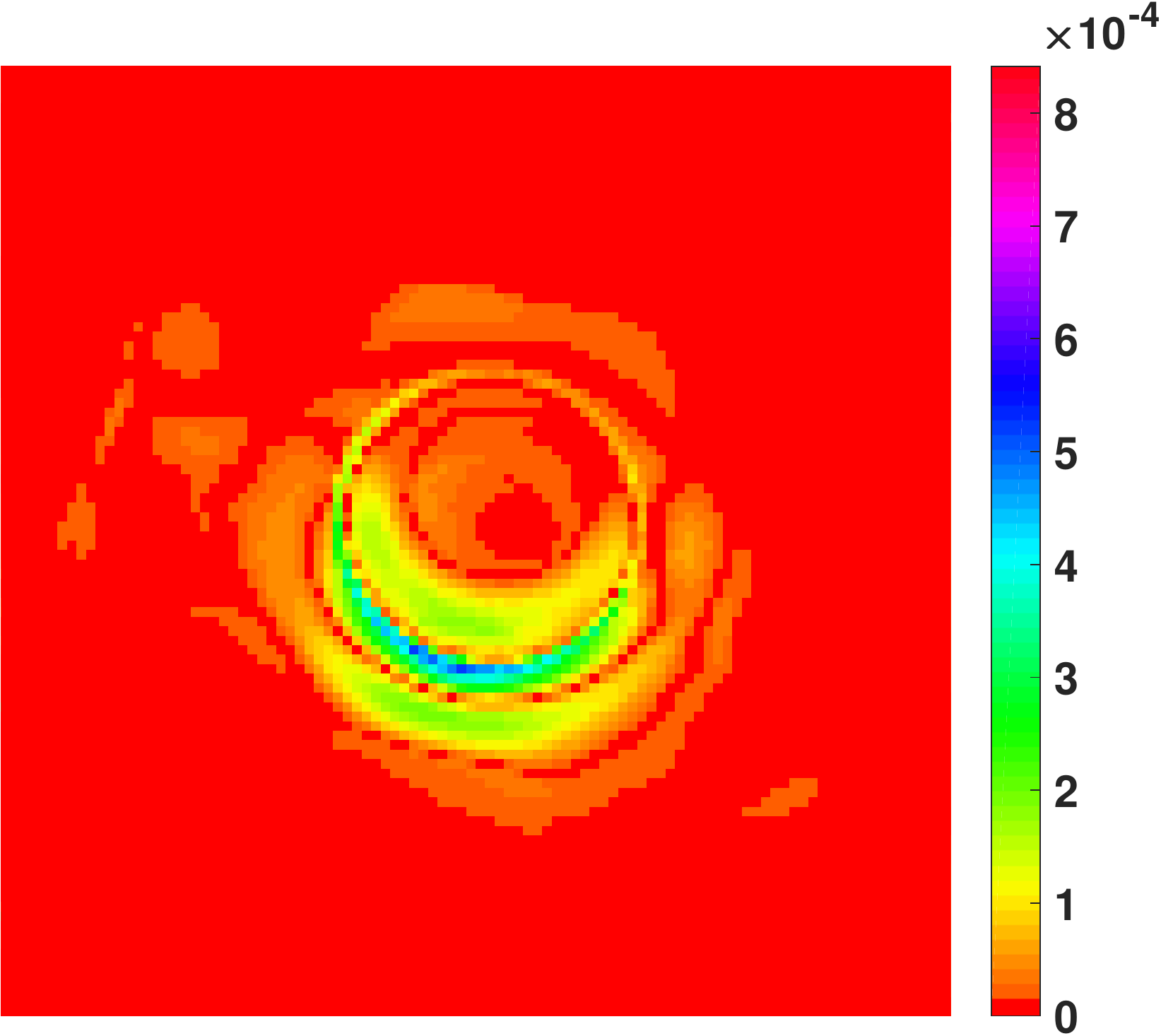} &
\hspace*{-0.0cm}\includegraphics[width = 4.5cm]{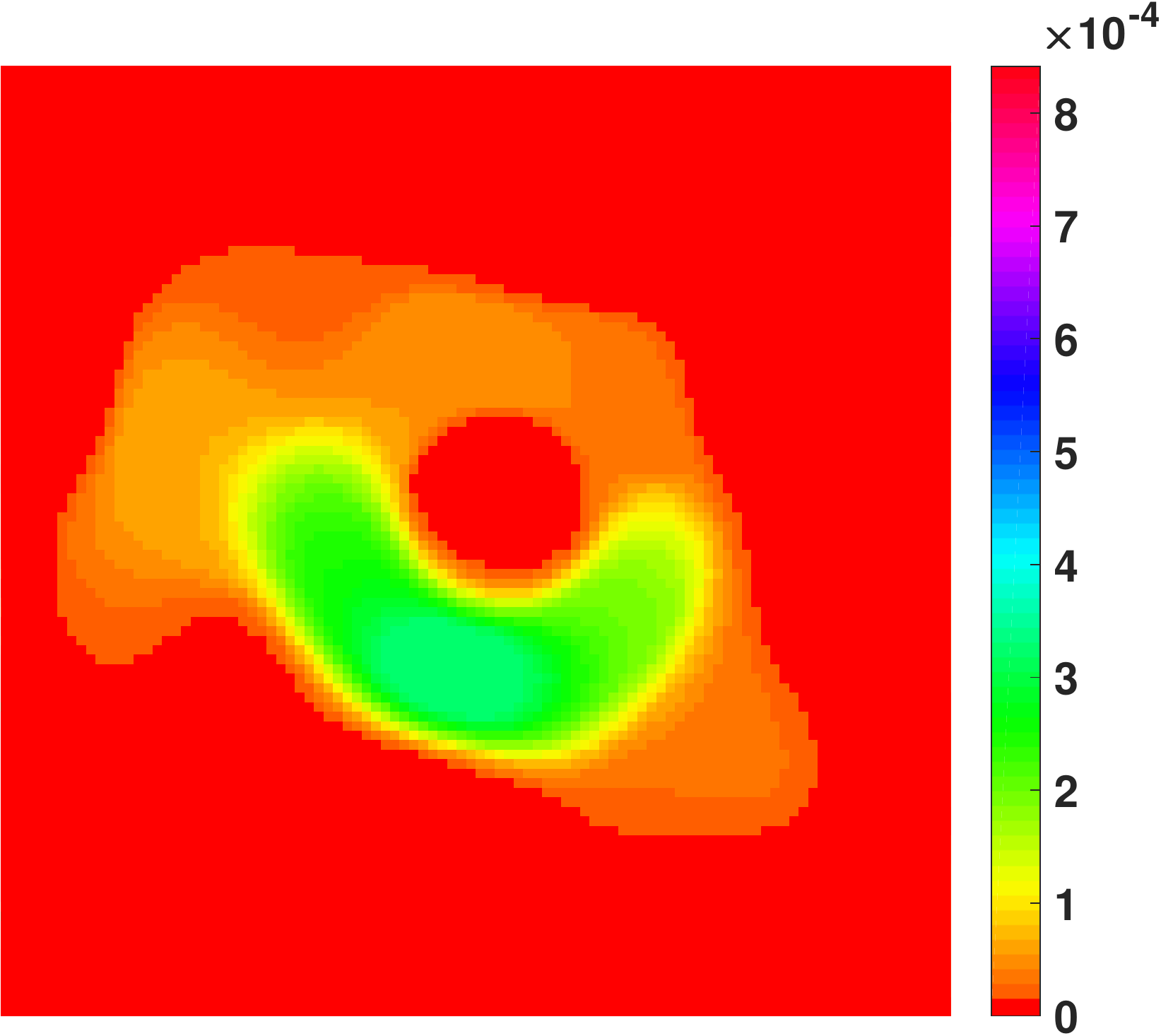} &
\hspace*{-0.0cm}\includegraphics[width = 4.5cm]{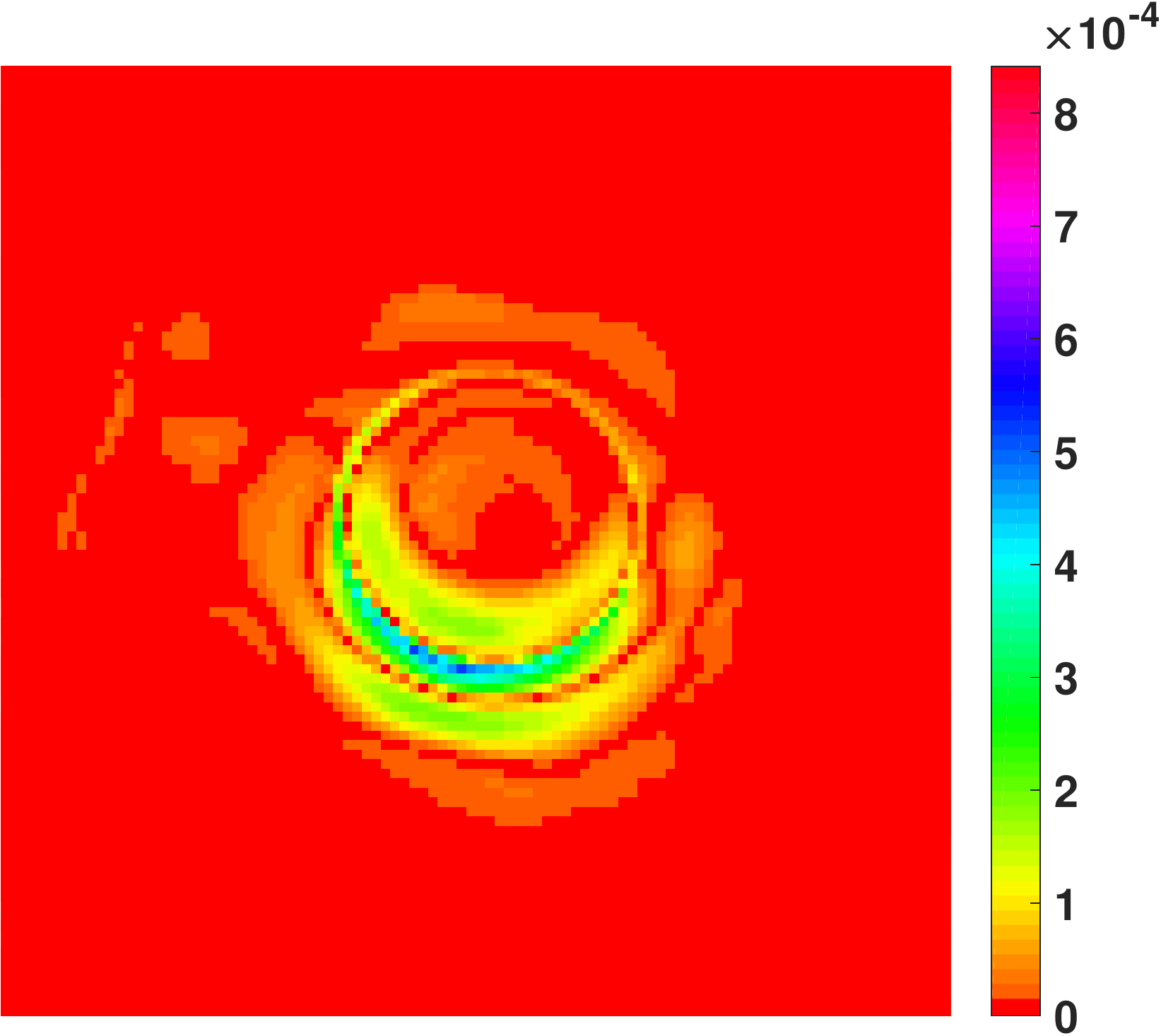} \\
\hspace*{-0.0cm}\includegraphics[width = 4.5cm]{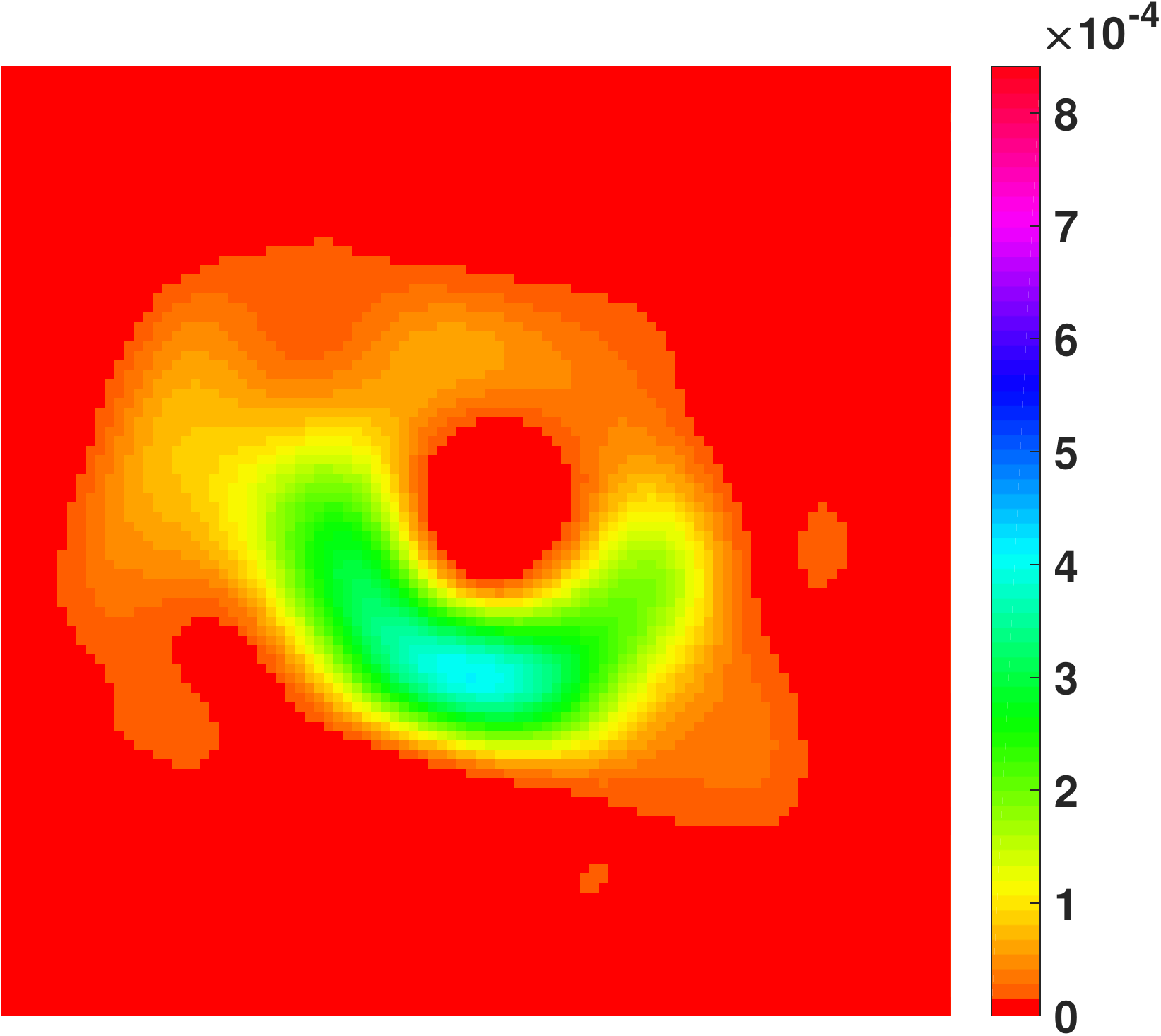} &
\hspace*{-0.0cm}\includegraphics[width = 4.5cm]{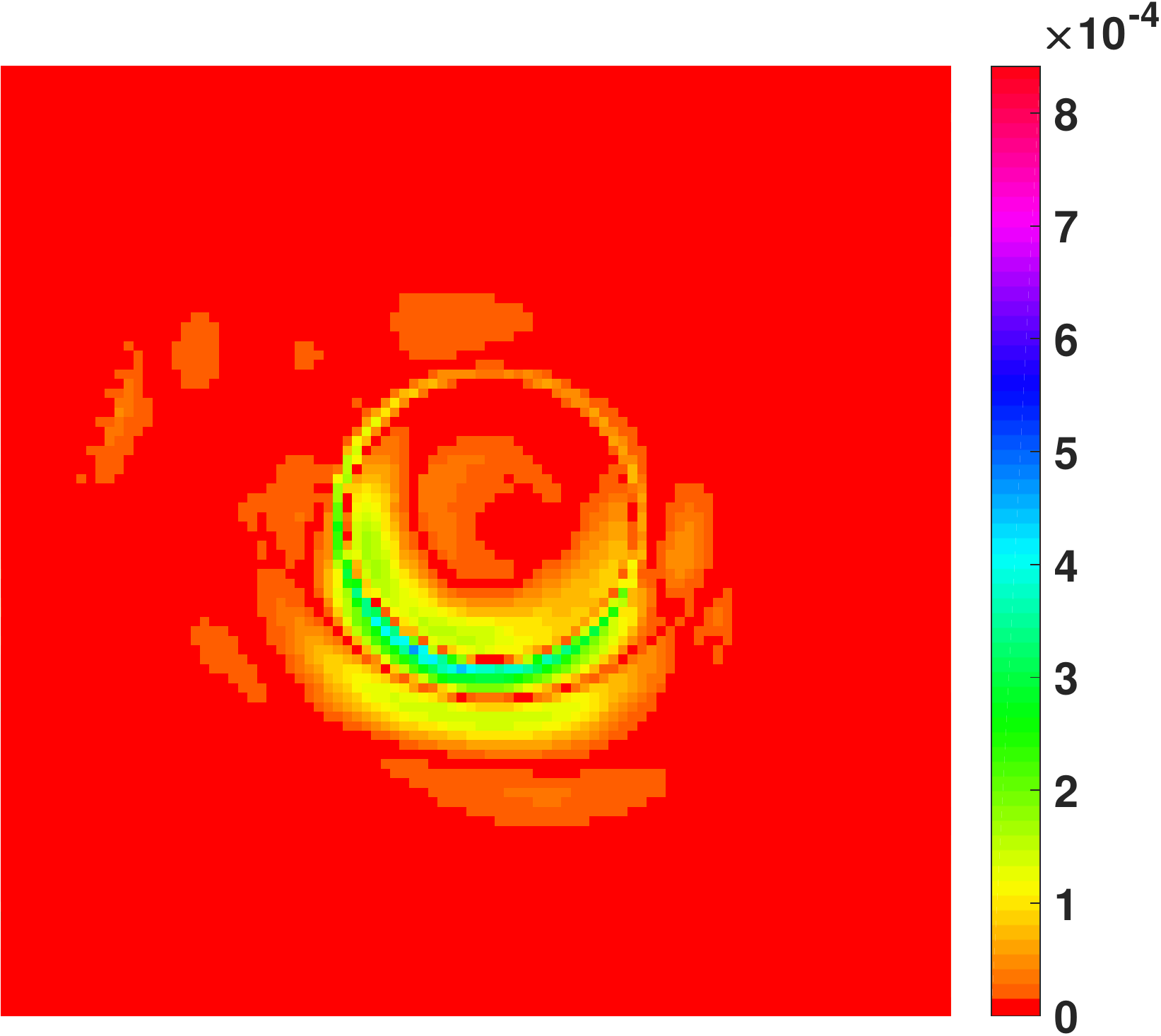} &
\hspace*{-0.0cm}\includegraphics[width = 4.5cm]{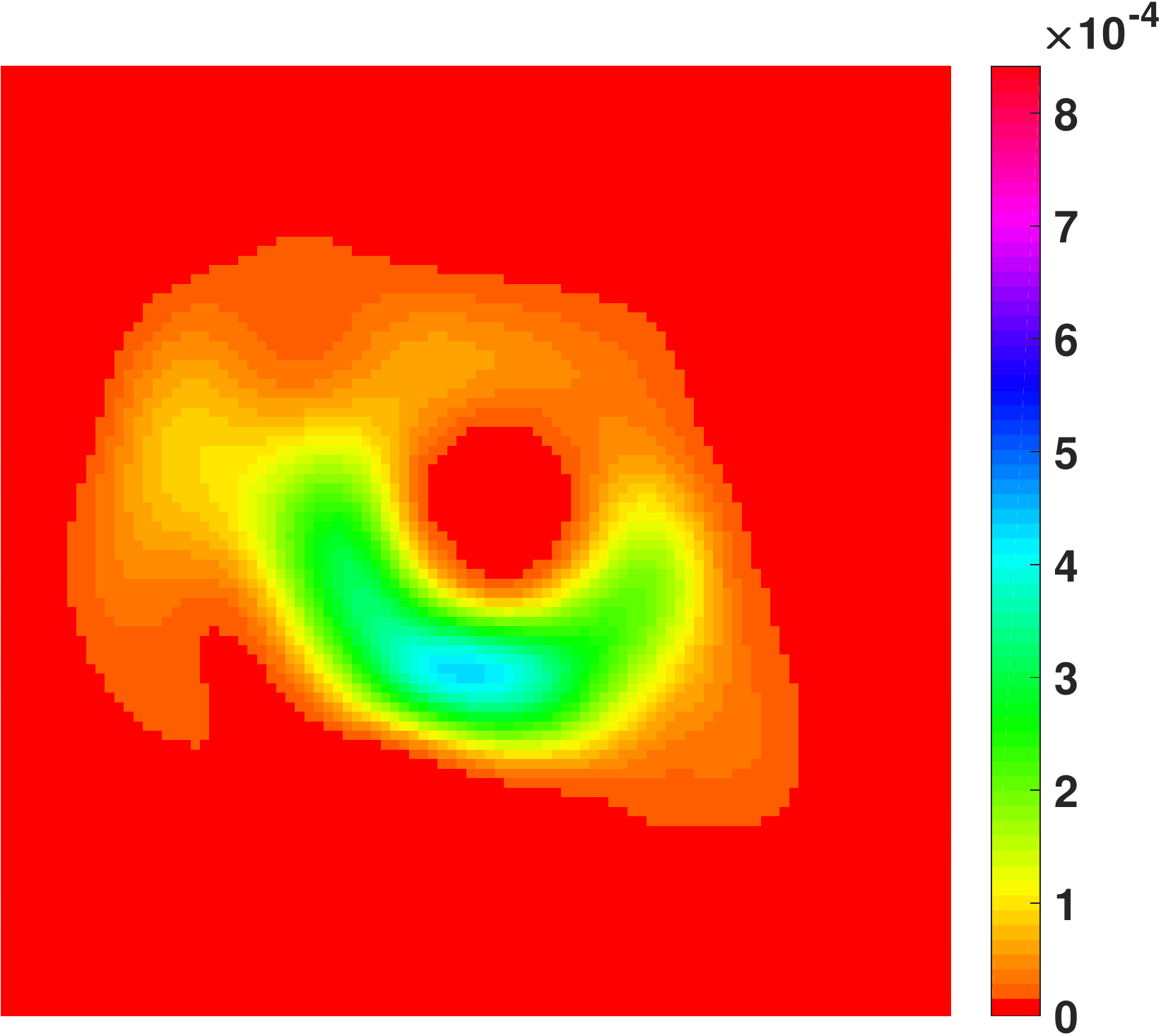} &
\hspace*{-0.0cm}\includegraphics[width = 4.5cm]{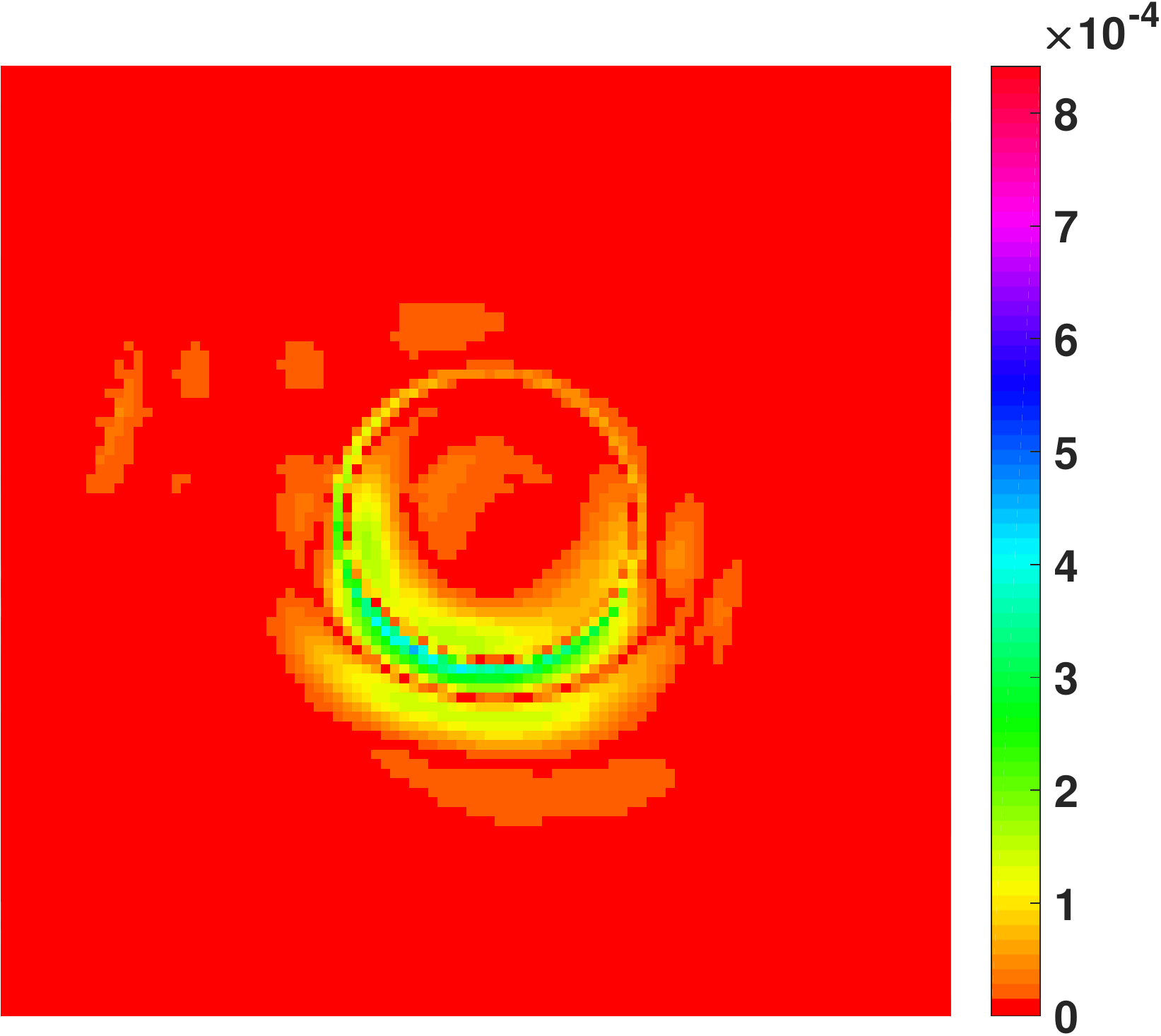} 
\end{tabular}
\caption{Results for the Stokes $I$ image corresponding to the counter-jet model. First row shows the ground-truth image, whereas the second row shows the \textsc{cs-clean} reconstructed image followed by its error image. 
Third and fourth row show the results for the TV and $\ell_1$+ TV  problems, respectively. For these rows, the first two columns show the reconstructed and the error images obtained by without imposing the polarization constraint in the reconstruction process, whereas the corresponding images in the case of imposing this constraint are shown in the last two columns. 
Similarly, column-wise, last row displays the reconstructed image for Polarized SARA without constraint and its error image; reconstructed image for Polarized SARA and its error image. The shown images correspond to the best results  obtained over 5 performed simulations for each case. All the images are shown in linear scale, normalized to the scale of the corresponding ground truth image.}
\label{fig:rec_images_jason_I}
\end{figure*}

\begin{figure*}
\centering
\begin{tabular}{@{}c@{}c@{}c@{}c@{}}
\hspace*{-0.0cm}\includegraphics[trim ={2cm 0 0 0.2cm},clip,width=5.1cm]{true_P_new_jason-eps-converted-to.pdf} \\
\hspace*{-0.0cm}\includegraphics[trim ={2cm 0 0 0.2cm},clip,width=5.1cm]{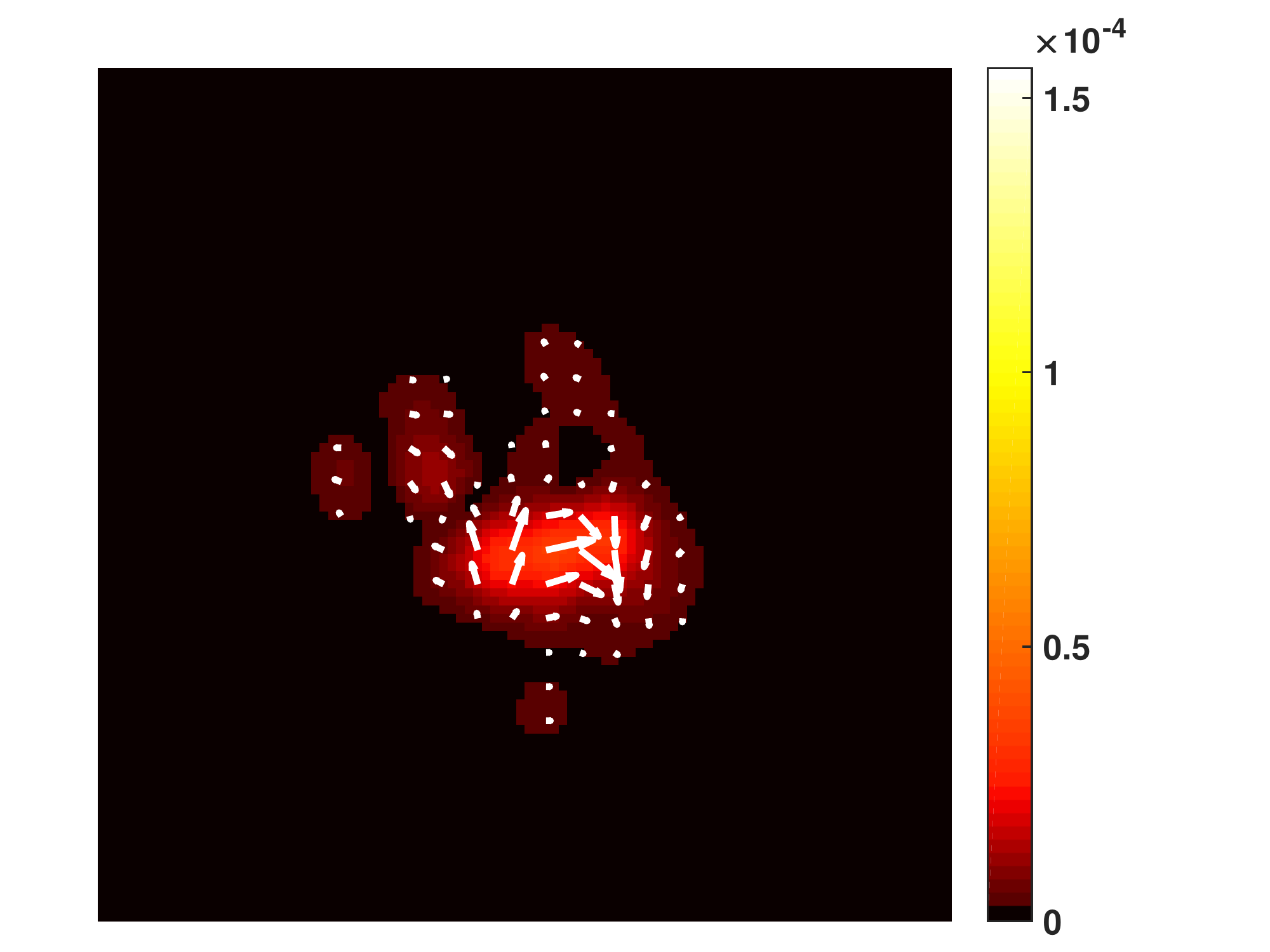} &
\hspace*{-0.55cm}\includegraphics[trim ={2cm 0 0 0.2cm},clip,width=5.1cm]{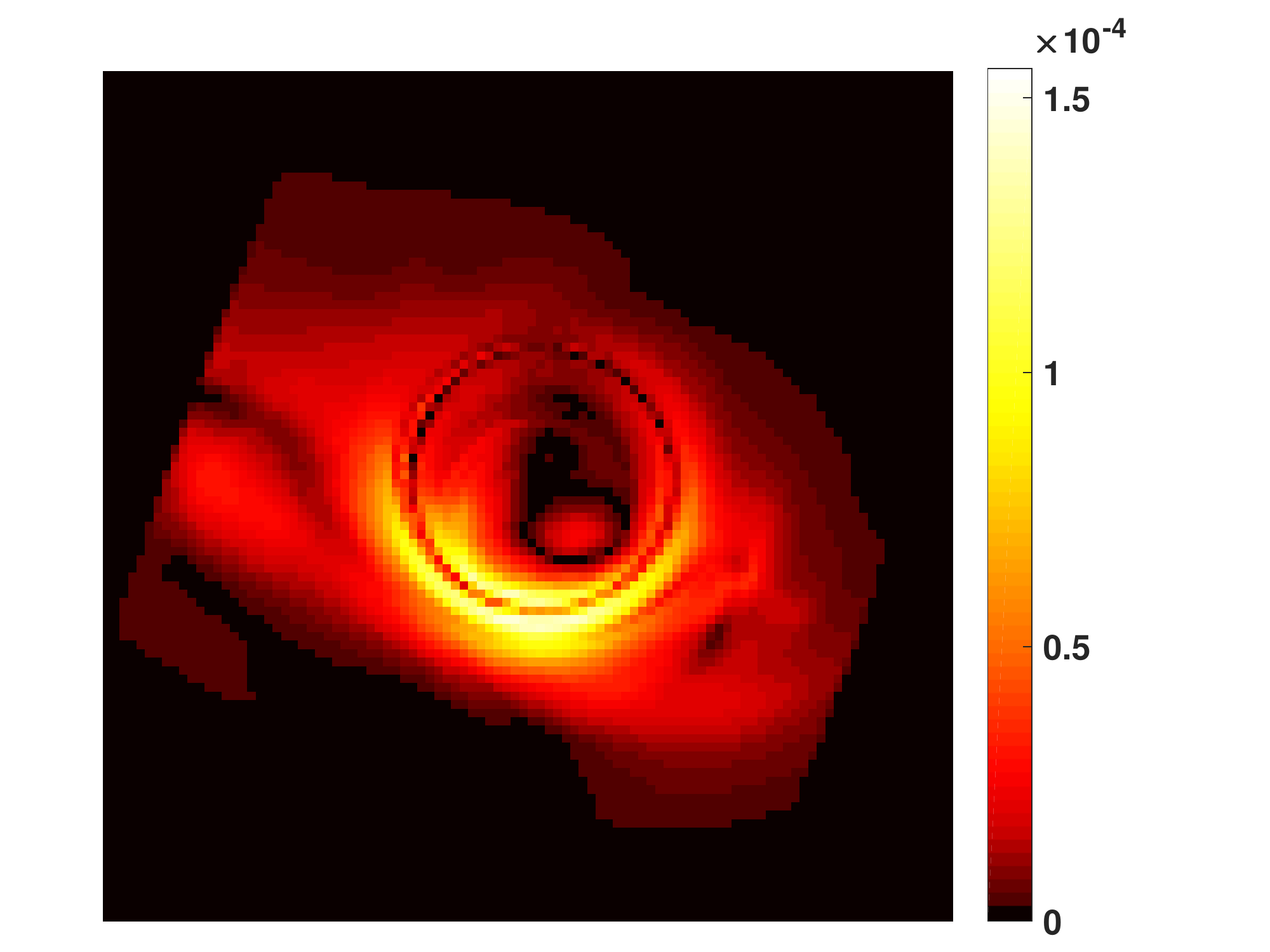} \\
\hspace*{-0.0cm}\includegraphics[trim ={2cm 0 0 0.2cm},clip,width=5.1cm]{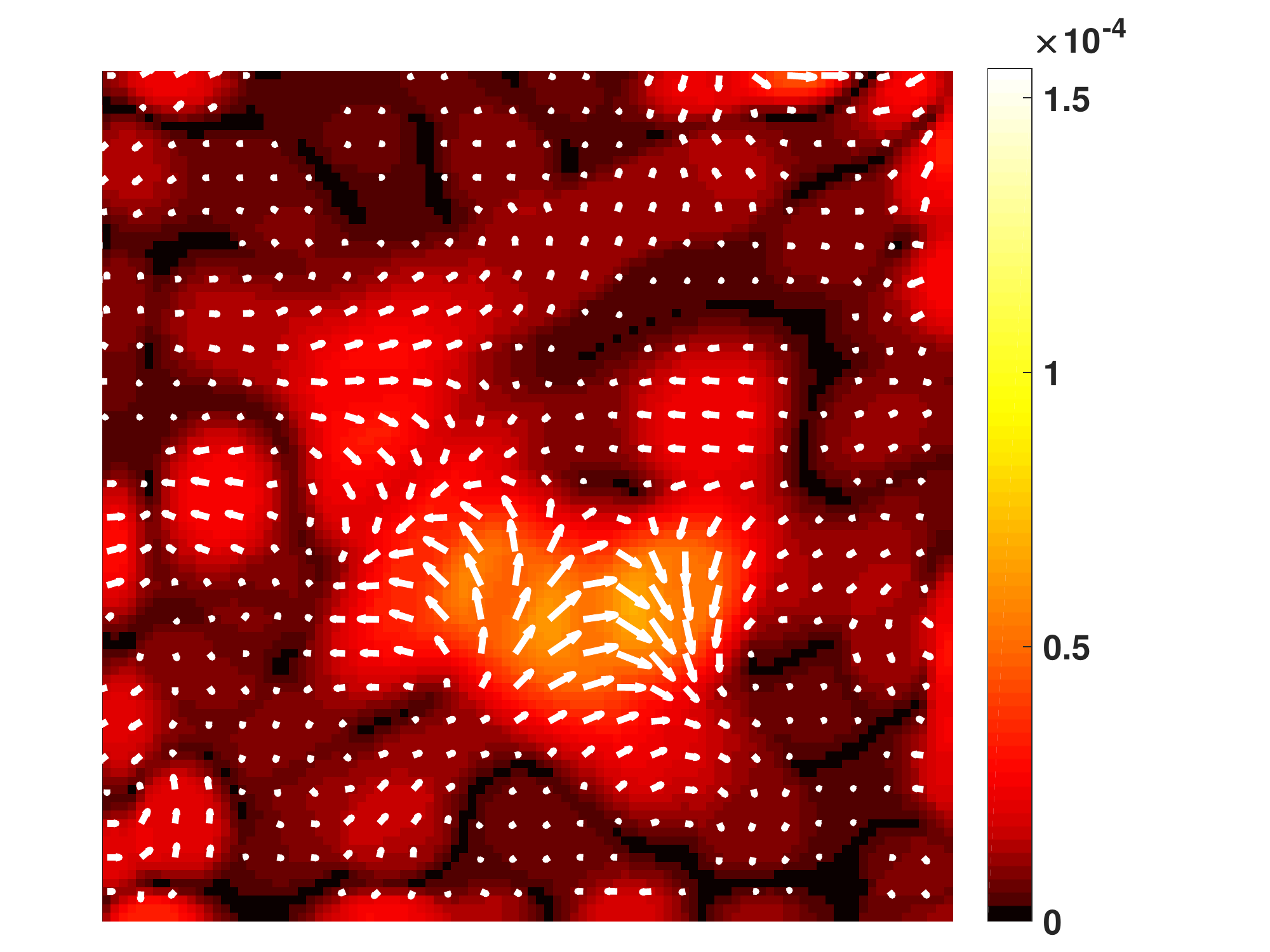} &
\hspace*{-0.55cm}\includegraphics[trim ={2cm 0 0 0.2cm},clip,width=5.1cm]{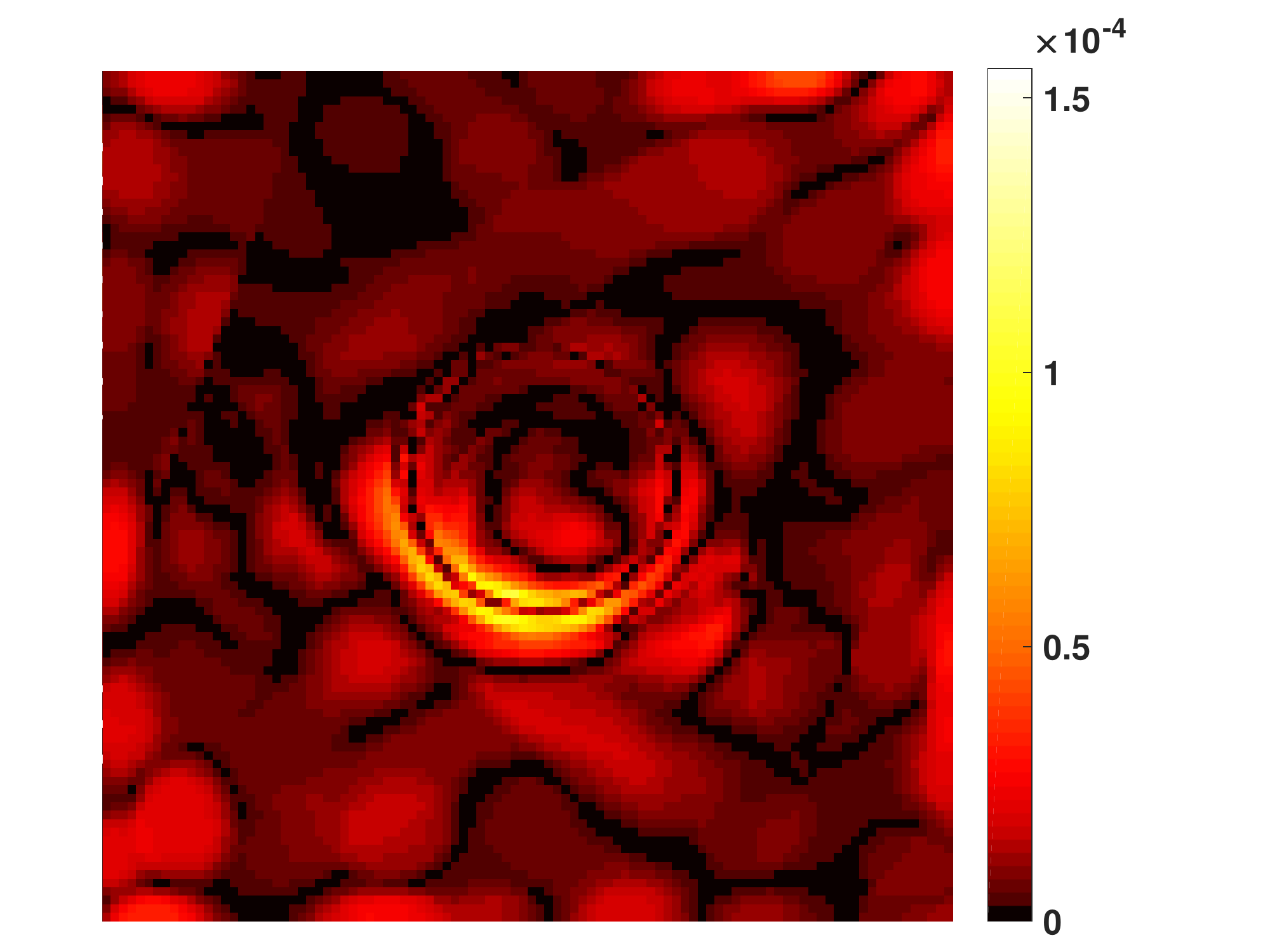} &
\hspace*{-0.55cm}\includegraphics[trim ={2cm 0 0 0.2cm},clip,width=5.1cm]{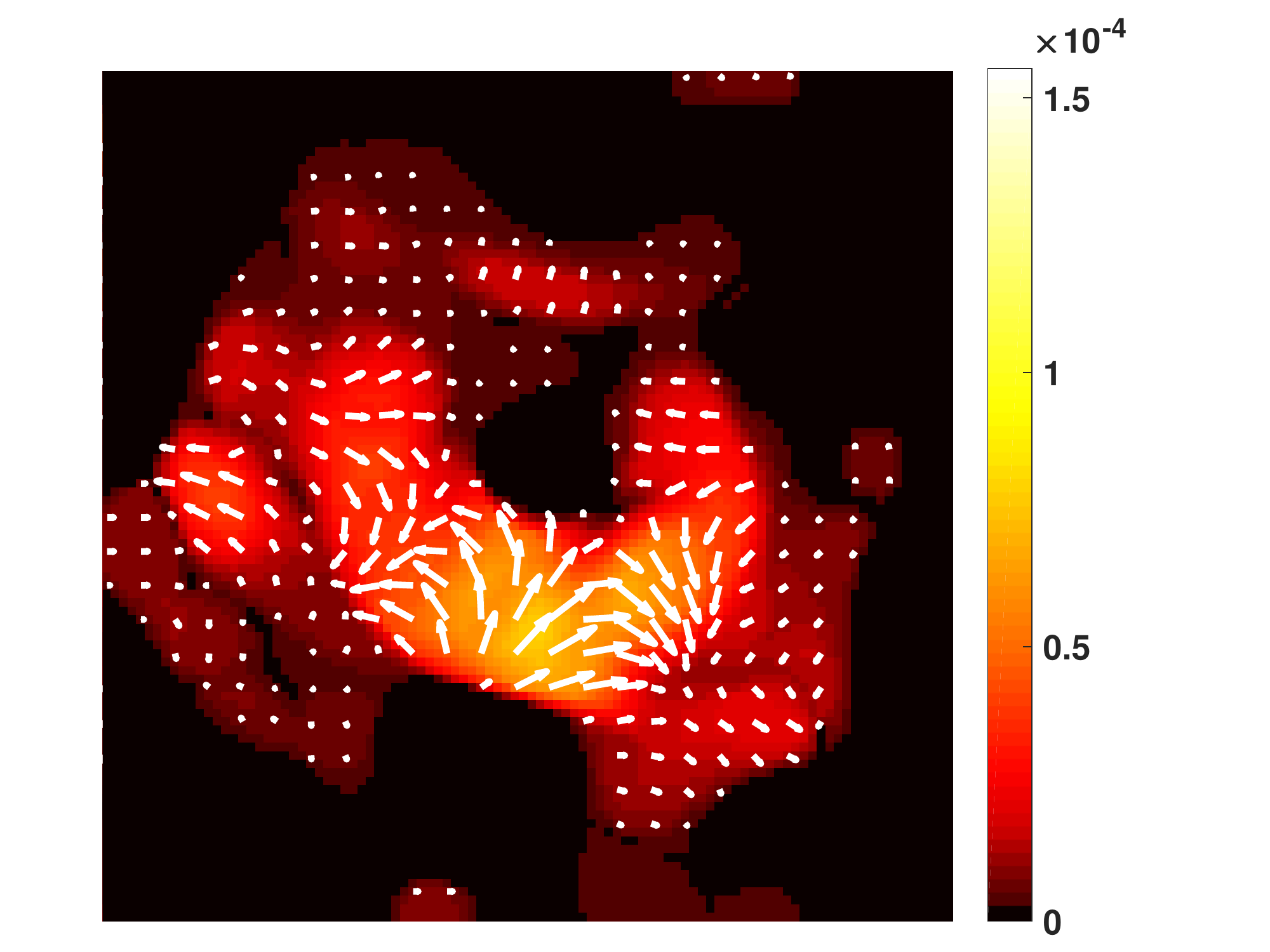} &
\hspace*{-0.55cm}\includegraphics[trim ={2cm 0 0 0.2cm},clip,width=5.1cm]{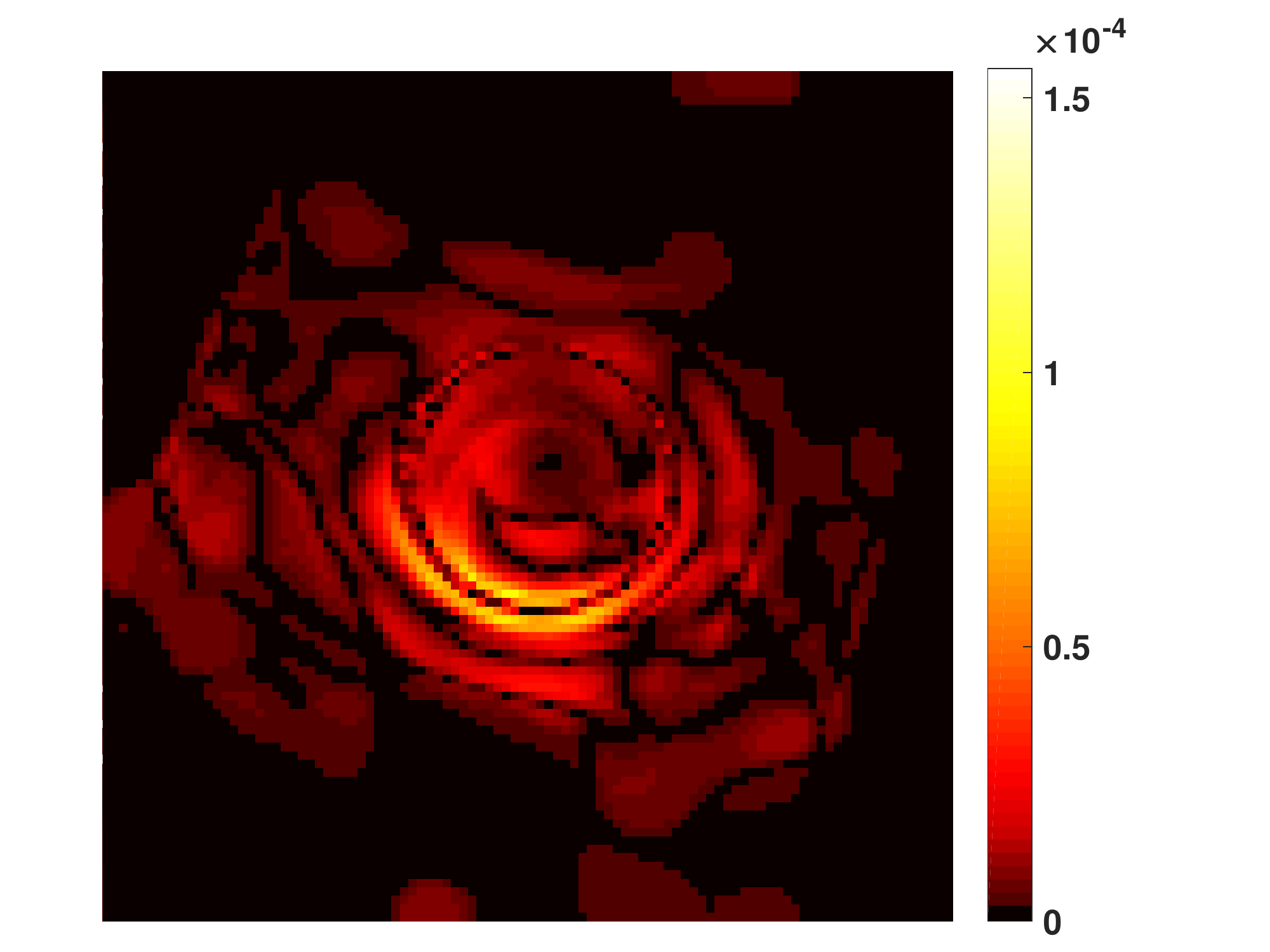}  \\
\hspace*{-0.0cm}\includegraphics[trim ={2cm 0 0 0.2cm},clip,width=5.1cm]{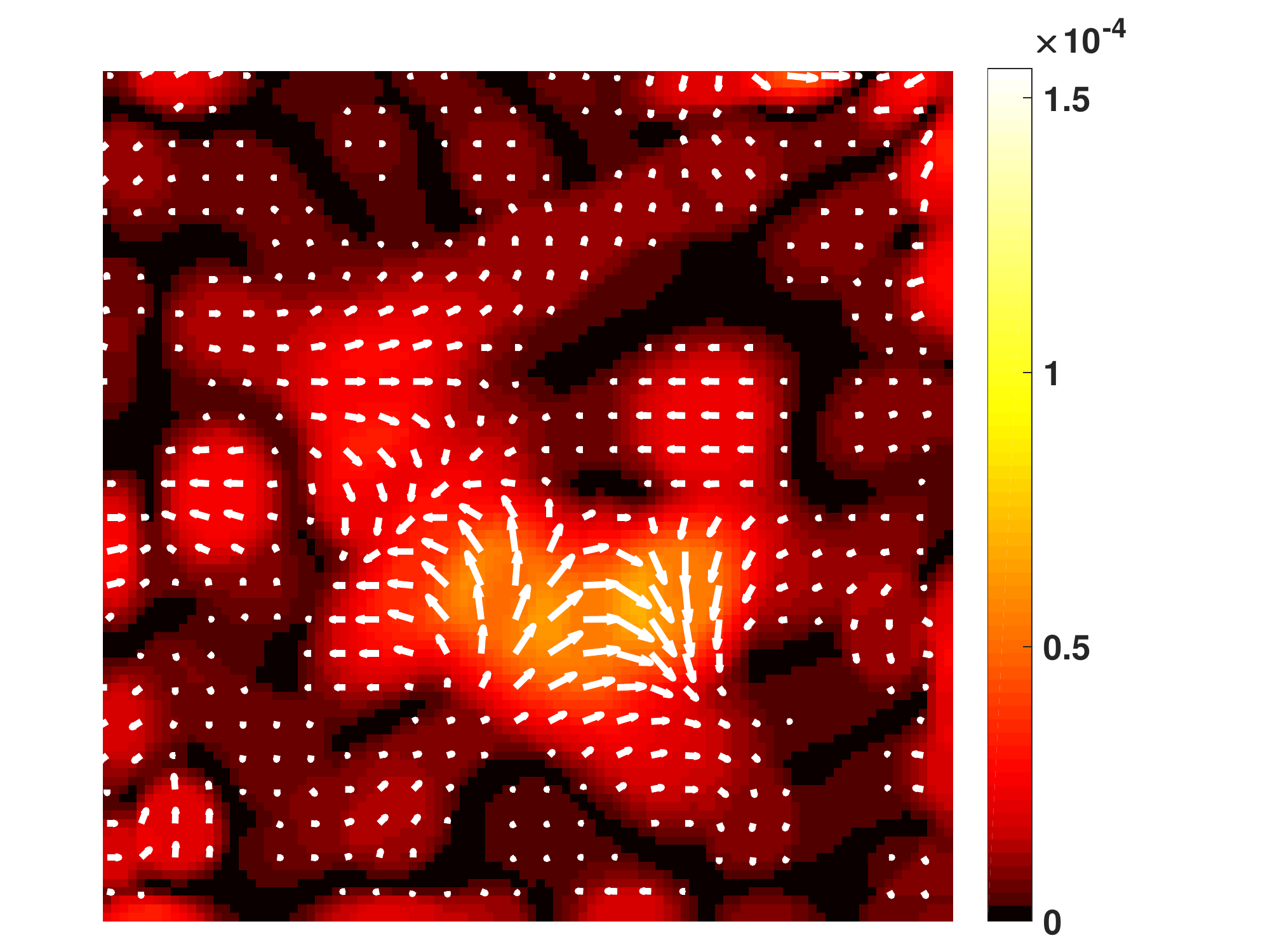} &
\hspace*{-0.55cm}\includegraphics[trim ={2cm 0 0 0.2cm},clip,width=5.1cm]{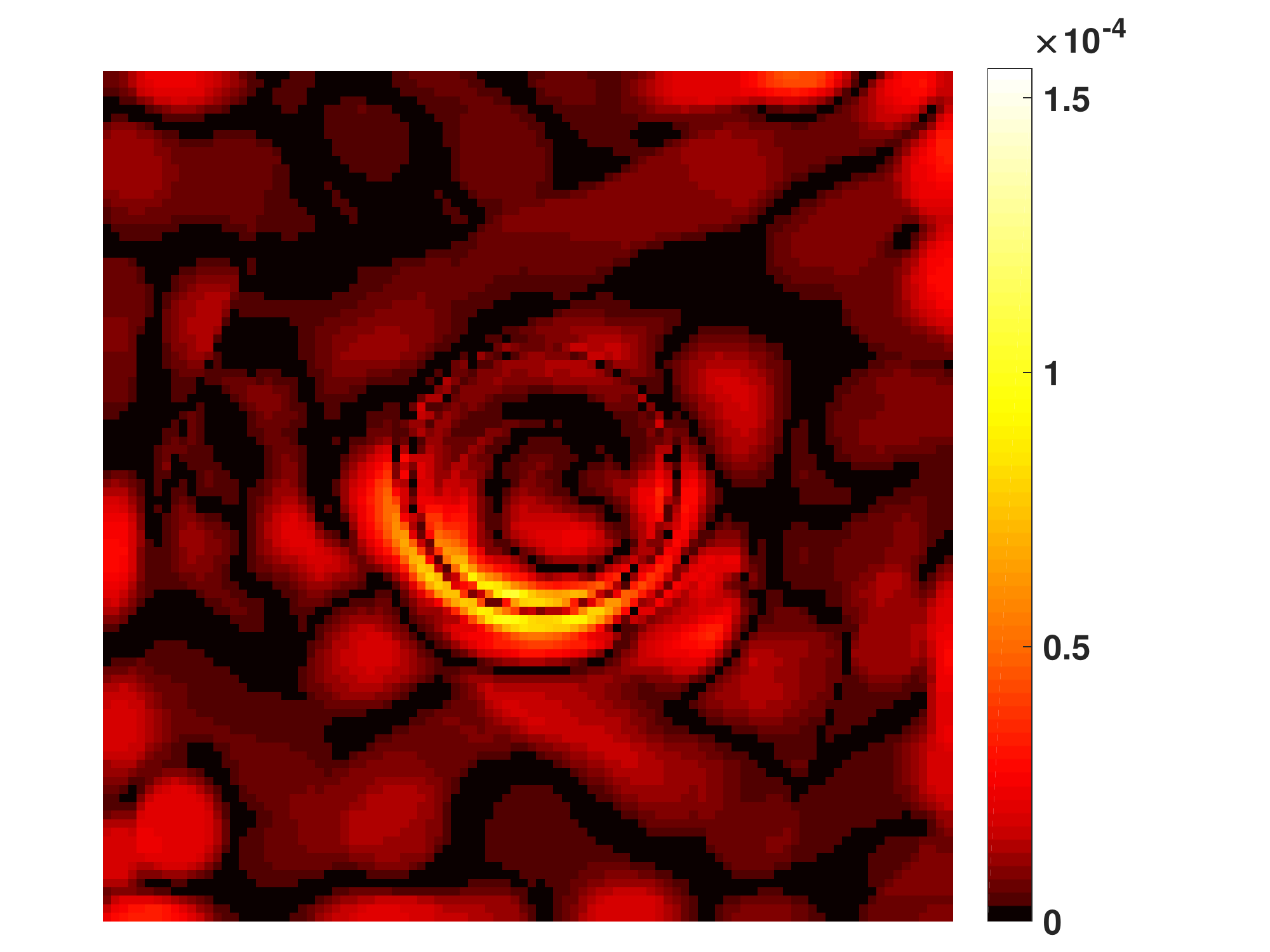} &
\hspace*{-0.55cm}\includegraphics[trim ={2cm 0 0 0.2cm},clip,width=5.1cm]{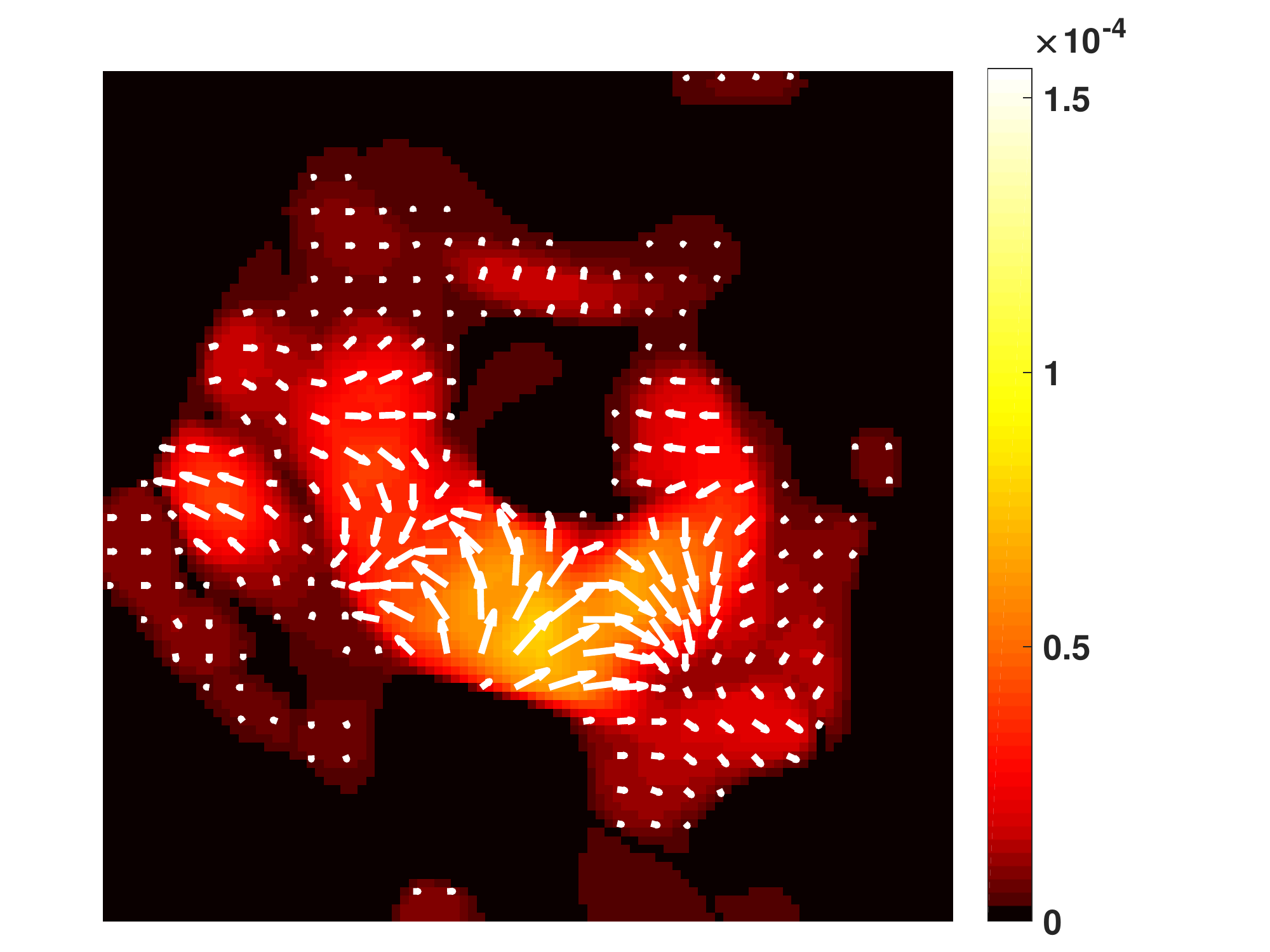} &
\hspace*{-0.55cm}\includegraphics[trim ={2cm 0 0 0.2cm},clip,width=5.1cm]{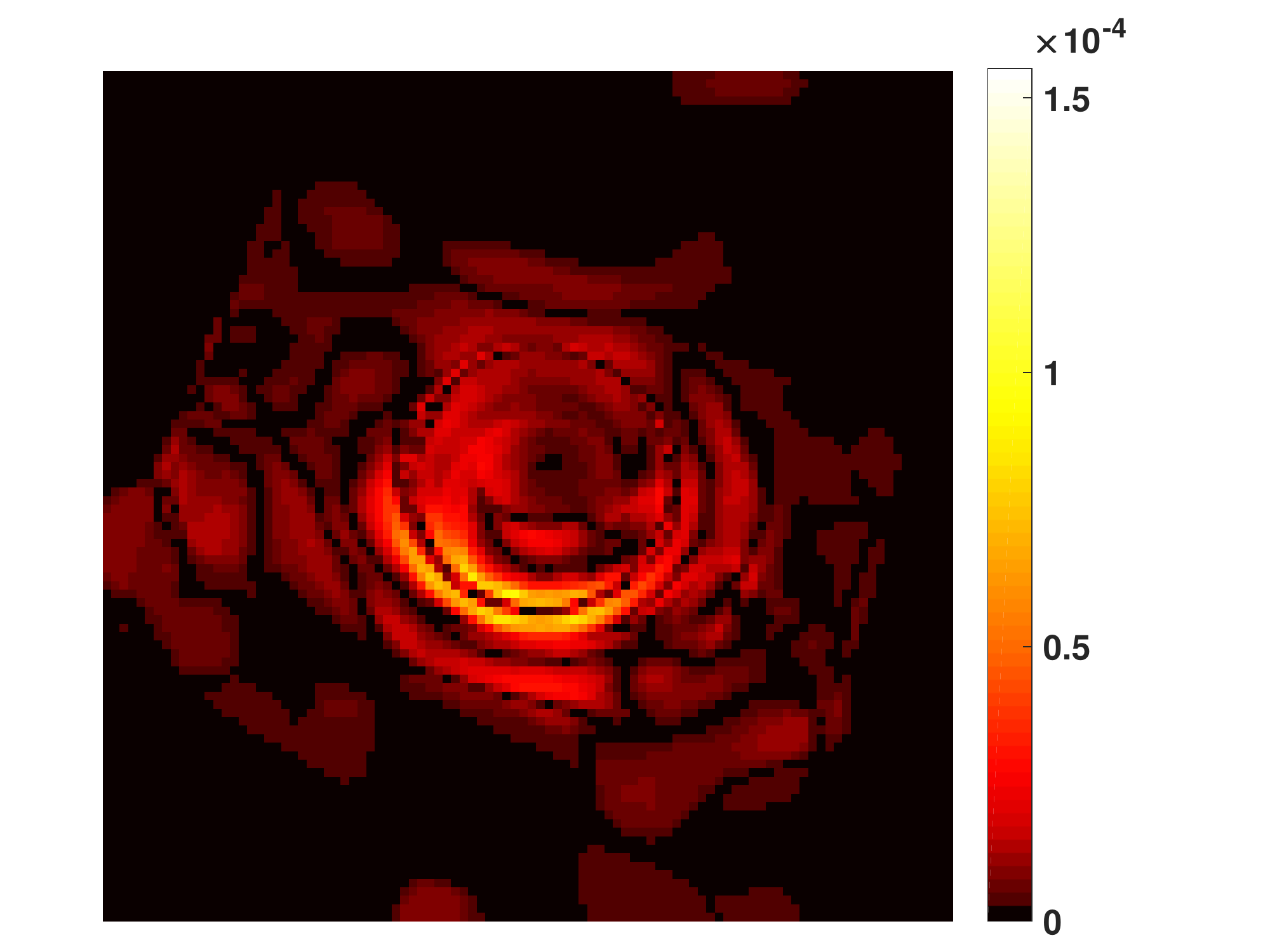} \\
\hspace*{-0.0cm}\includegraphics[trim ={2cm 0 0 0.2cm},clip,width=5.1cm]{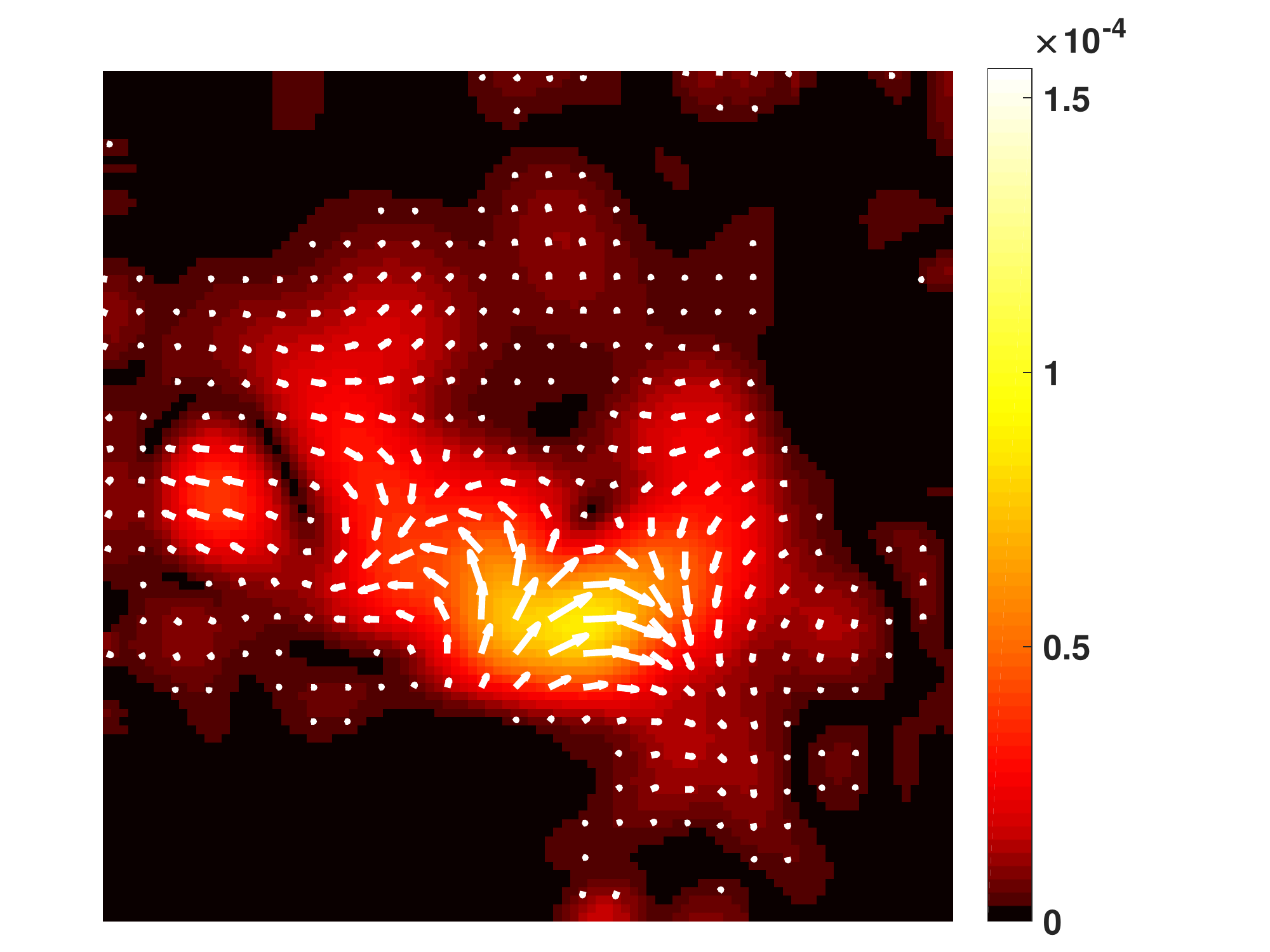} &
\hspace*{-0.55cm}\includegraphics[trim ={2cm 0 0 0.2cm},clip,width=5.1cm]{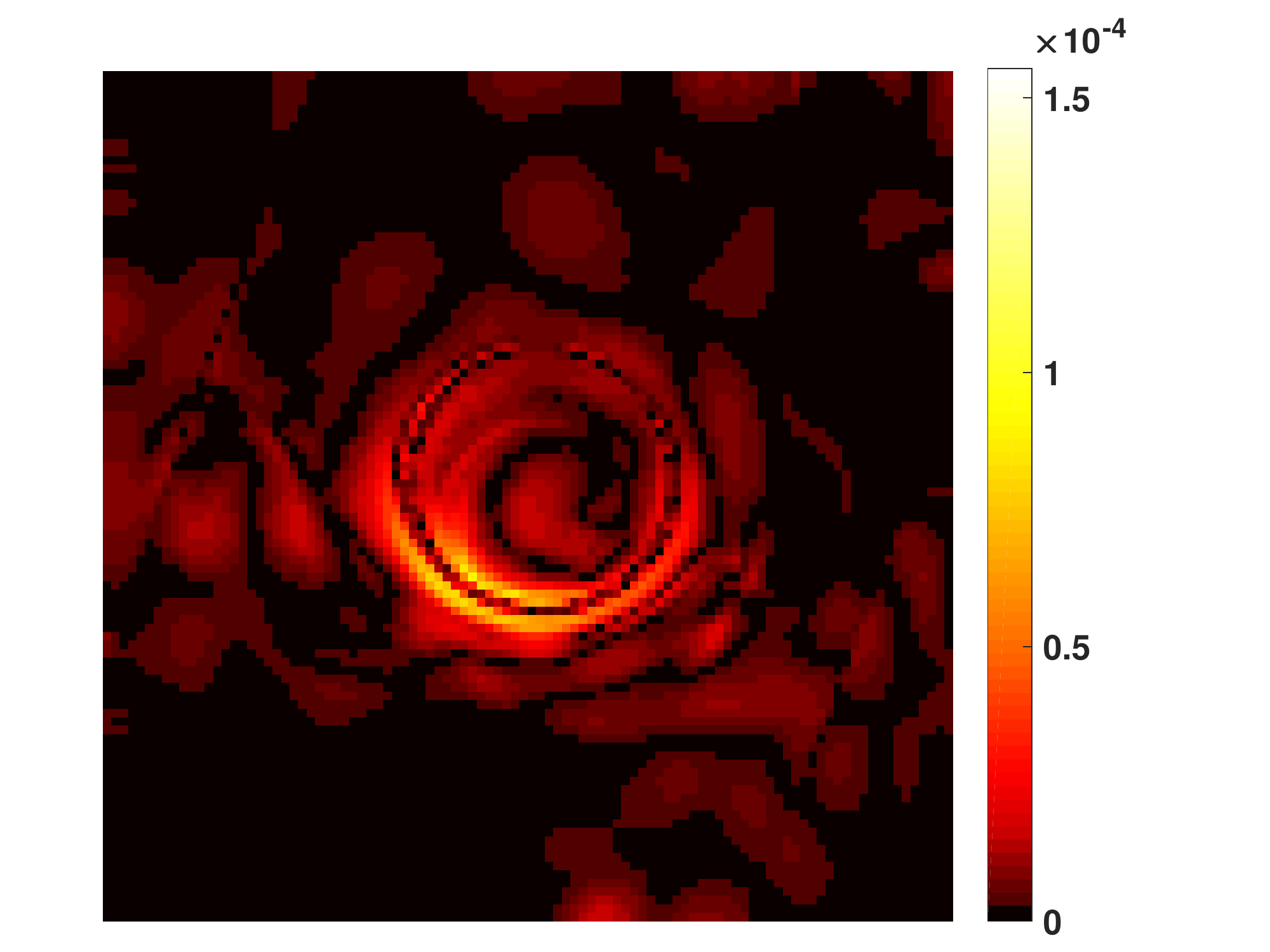} &
\hspace*{-0.55cm}\includegraphics[trim ={2cm 0 0 0.2cm},clip,width=5.1cm]{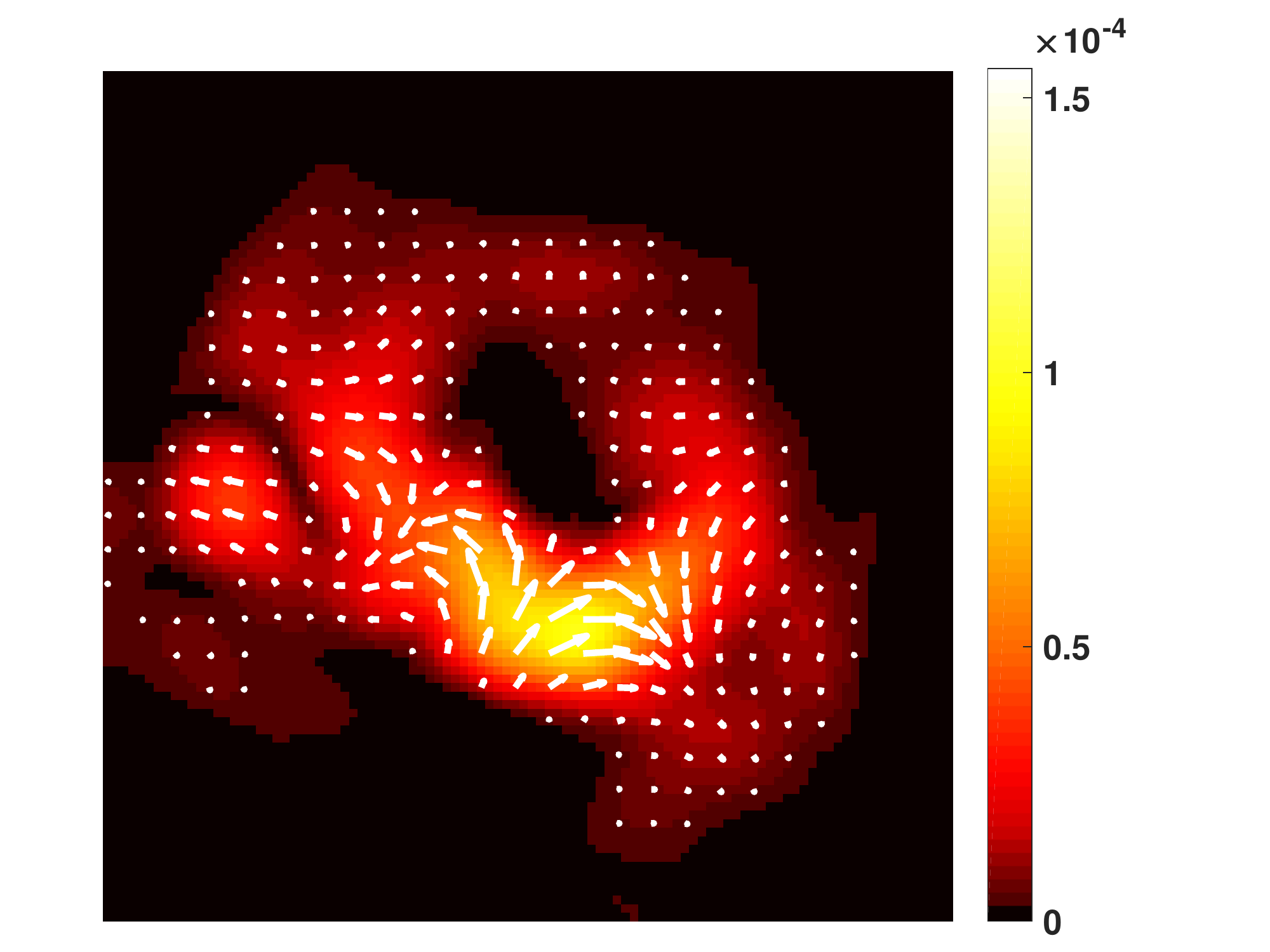} &
\hspace*{-0.55cm}\includegraphics[trim ={2cm 0 0 0.2cm},clip,width=5.1cm]{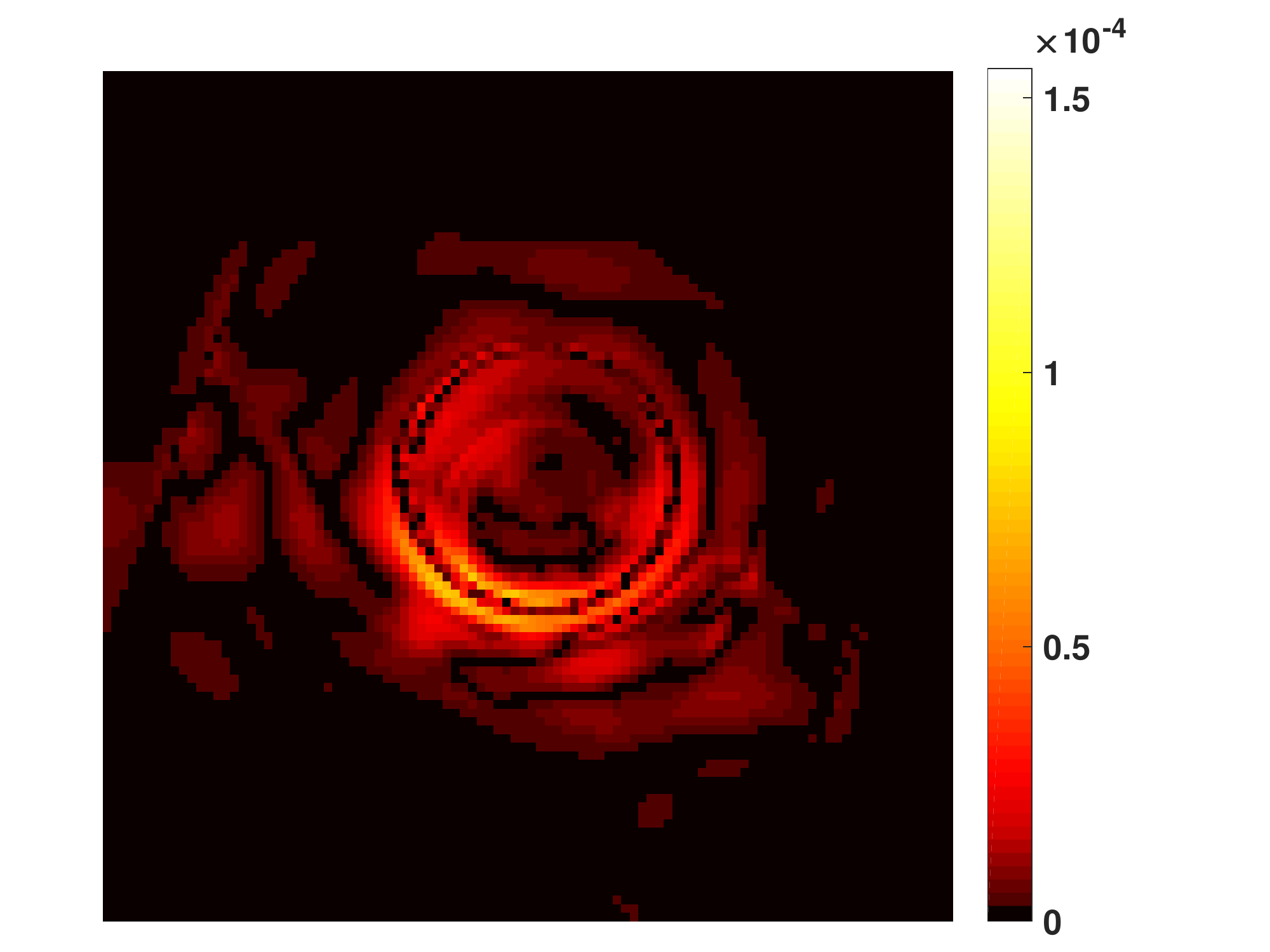} 
\end{tabular}
\vspace*{-0.2cm}
\caption{Results for the linear polarization image corresponding to the counter-jet model. First row shows the ground-truth image, whereas the second row shows the \textsc{cs-clean} reconstructed image followed by its error image. 
Third and fourth row show the results for the TV and $\ell_1$+ TV  problems, respectively. For these rows, the first two columns show the reconstructed and the error images obtained by without imposing the polarization constraint in the reconstruction process, whereas the corresponding images in the case of imposing this constraint are shown in the last two columns. 
Similarly, column-wise, last row displays the reconstructed image for Polarized SARA without constraint and its error image; reconstructed image for Polarized SARA and its error image. The reconstructed images correspond to the linear polarization intensity images, overlaid by the white bars representing the EVPA. The shown images correspond to the best results  obtained over 5 performed simulations for each case. All the images are shown in linear scale, normalized to the scale of the corresponding ground truth image.}
\label{fig:rec_images_jason_P}
\end{figure*}


\begin{figure}
\centering
\begin{tabular}{@{}c@{}c@{}}
\hspace*{-0.0cm}\includegraphics[width = 4.2cm]{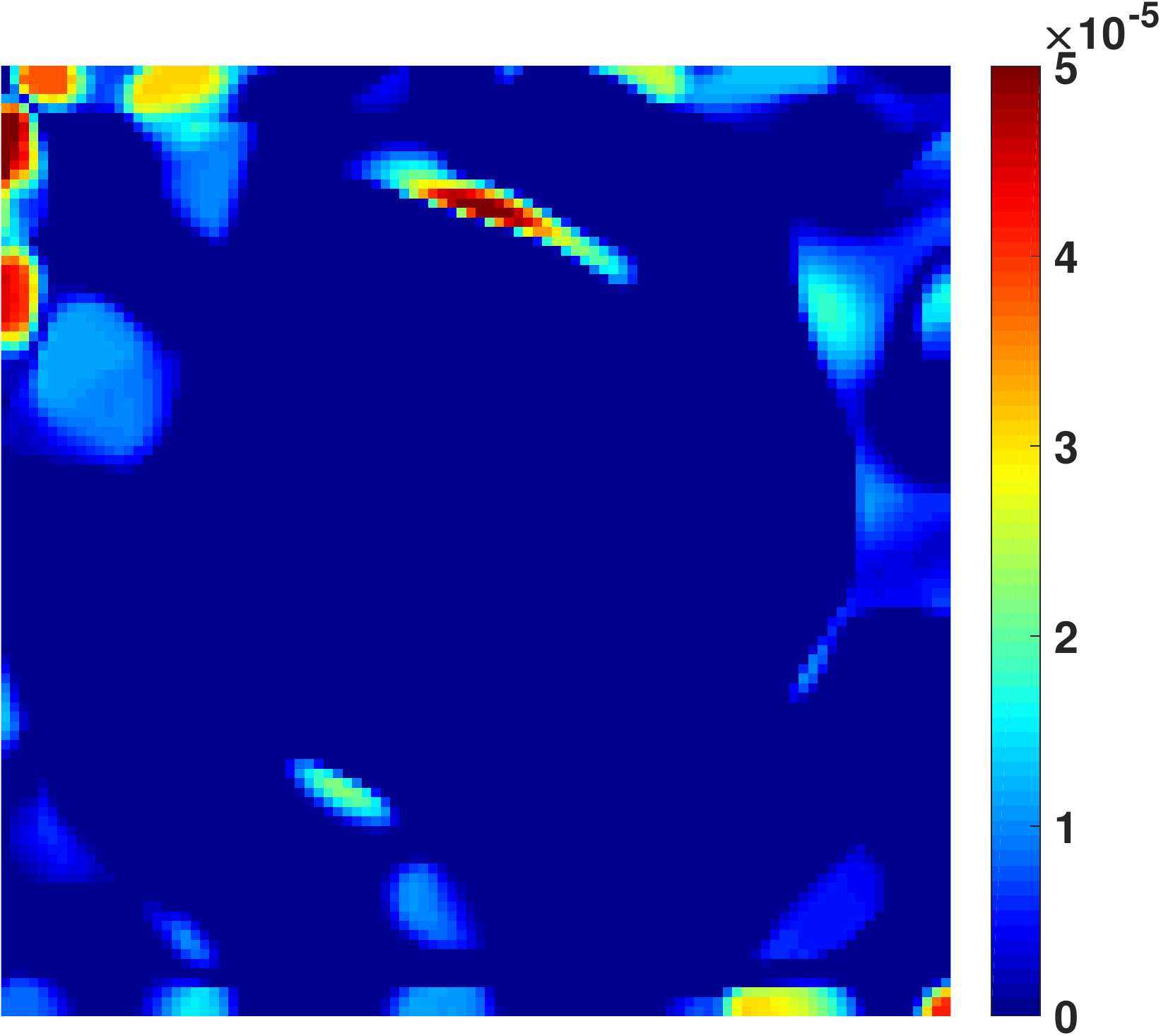} &
\hspace*{-0.0cm}\includegraphics[width = 4.2cm]{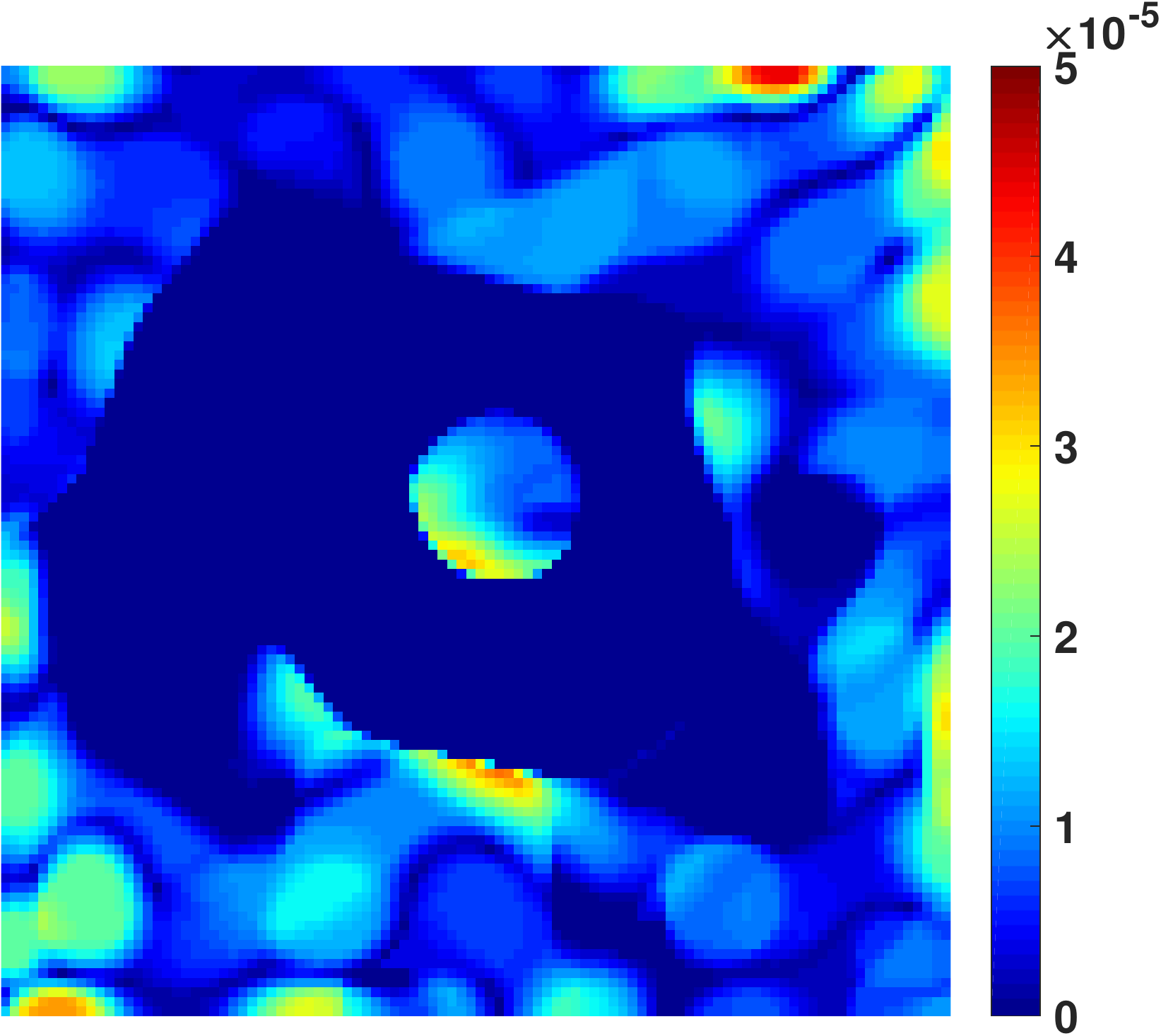} \\
\hspace*{-0.0cm}\includegraphics[width = 4.2cm]{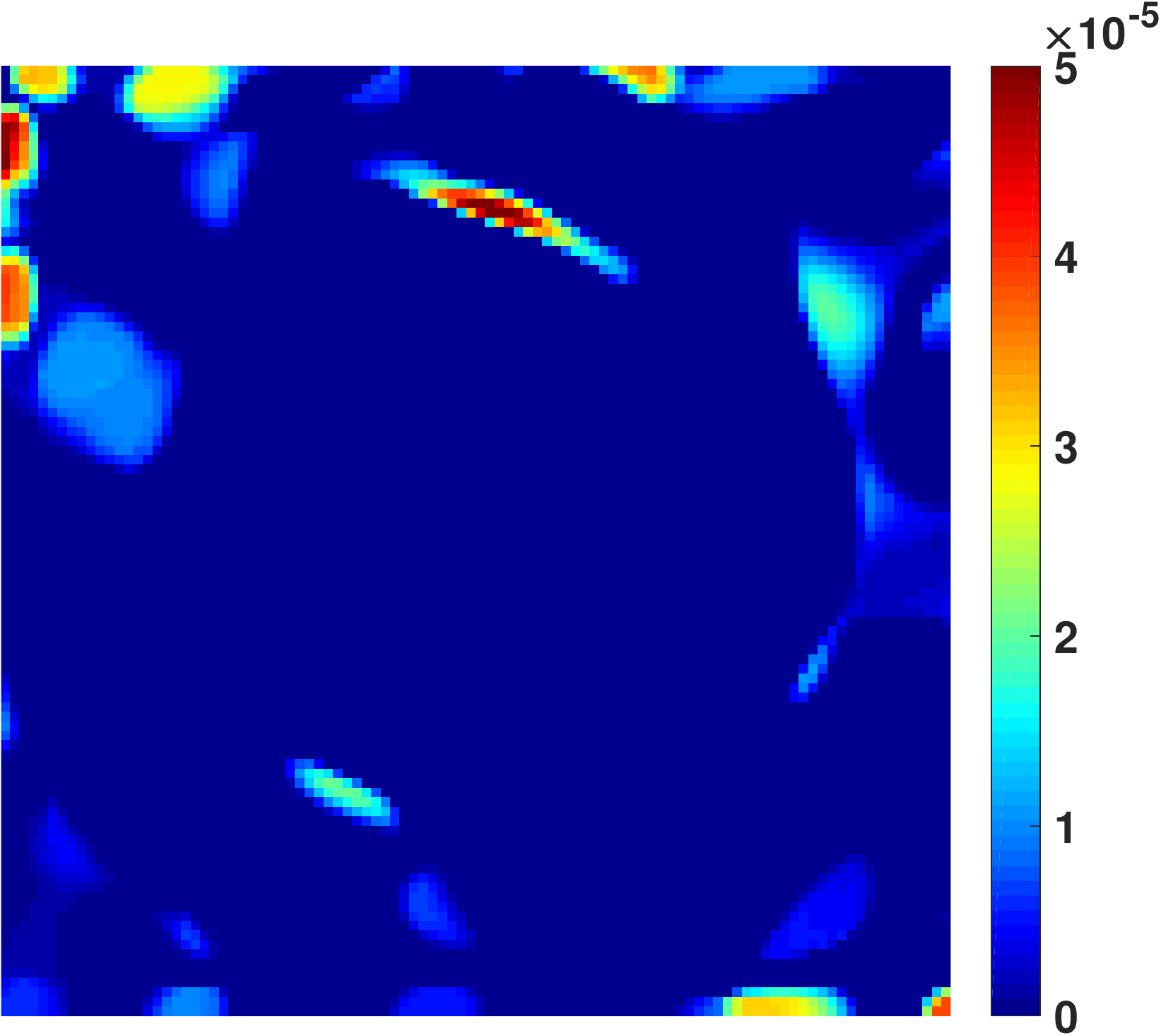} &
\hspace*{-0.0cm}\includegraphics[width = 4.2cm]{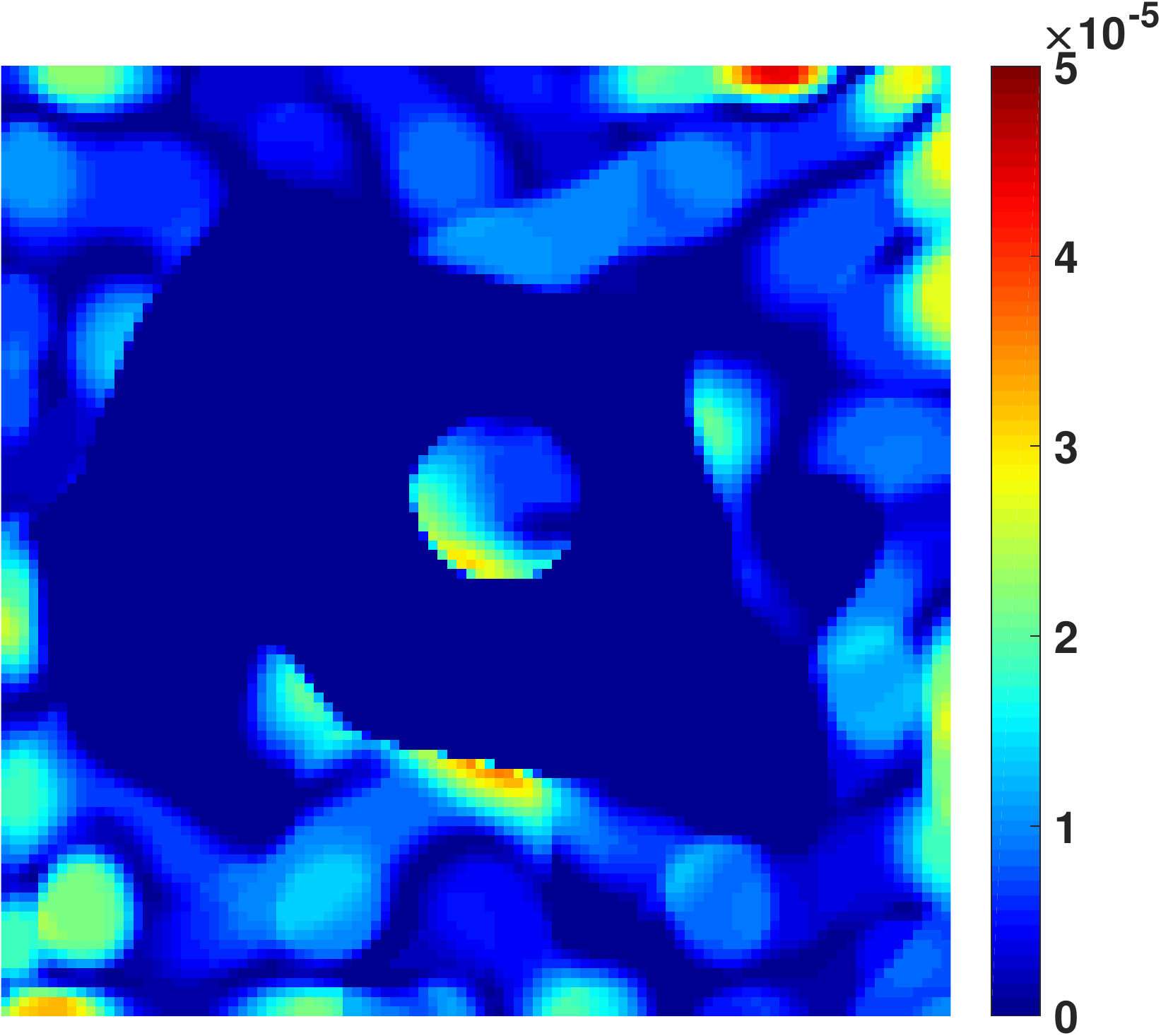} \\
\hspace*{-0.042cm}\includegraphics[width = 4.18cm]{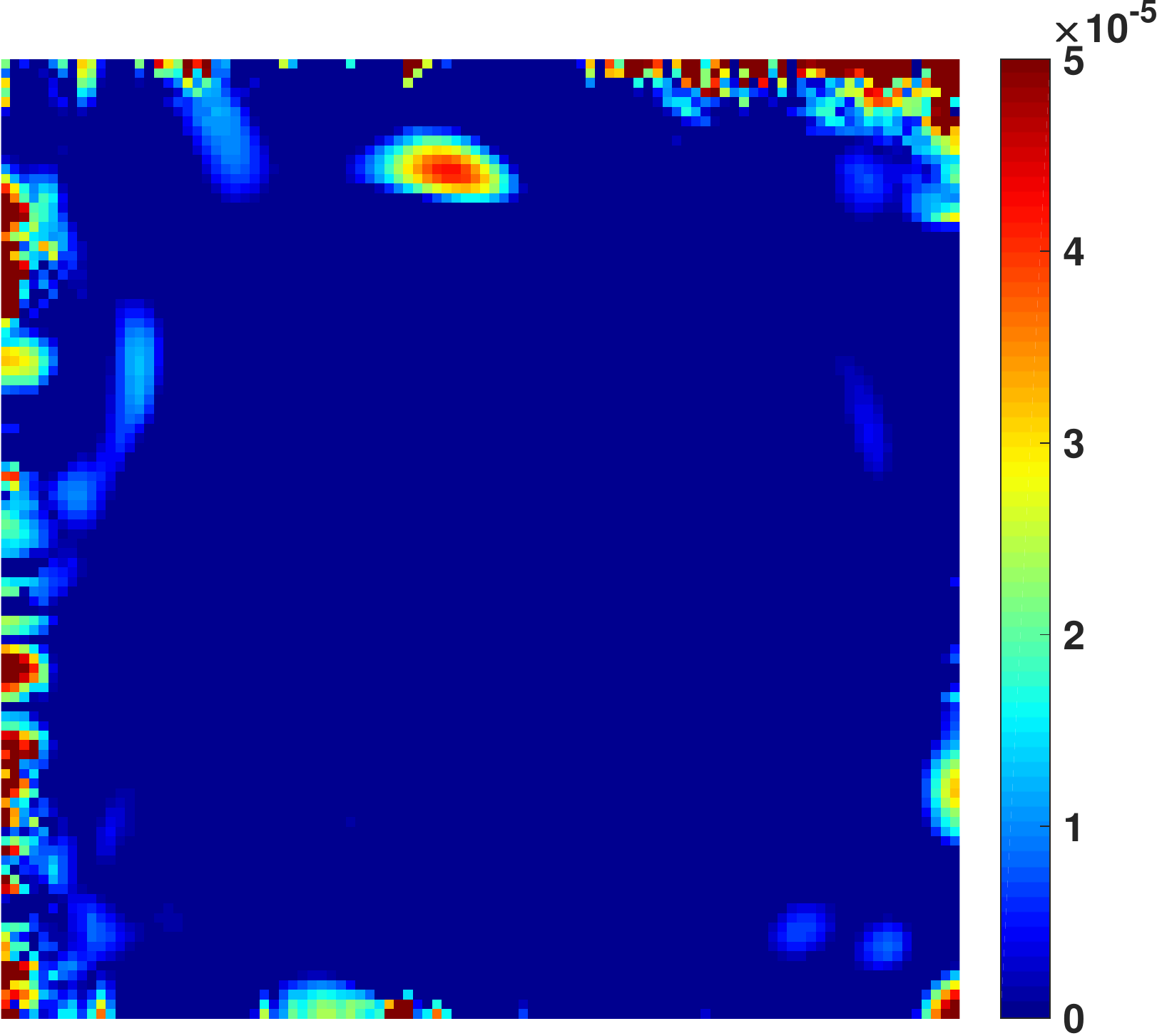} &
\hspace*{-0.048cm}\includegraphics[width = 4.18cm]{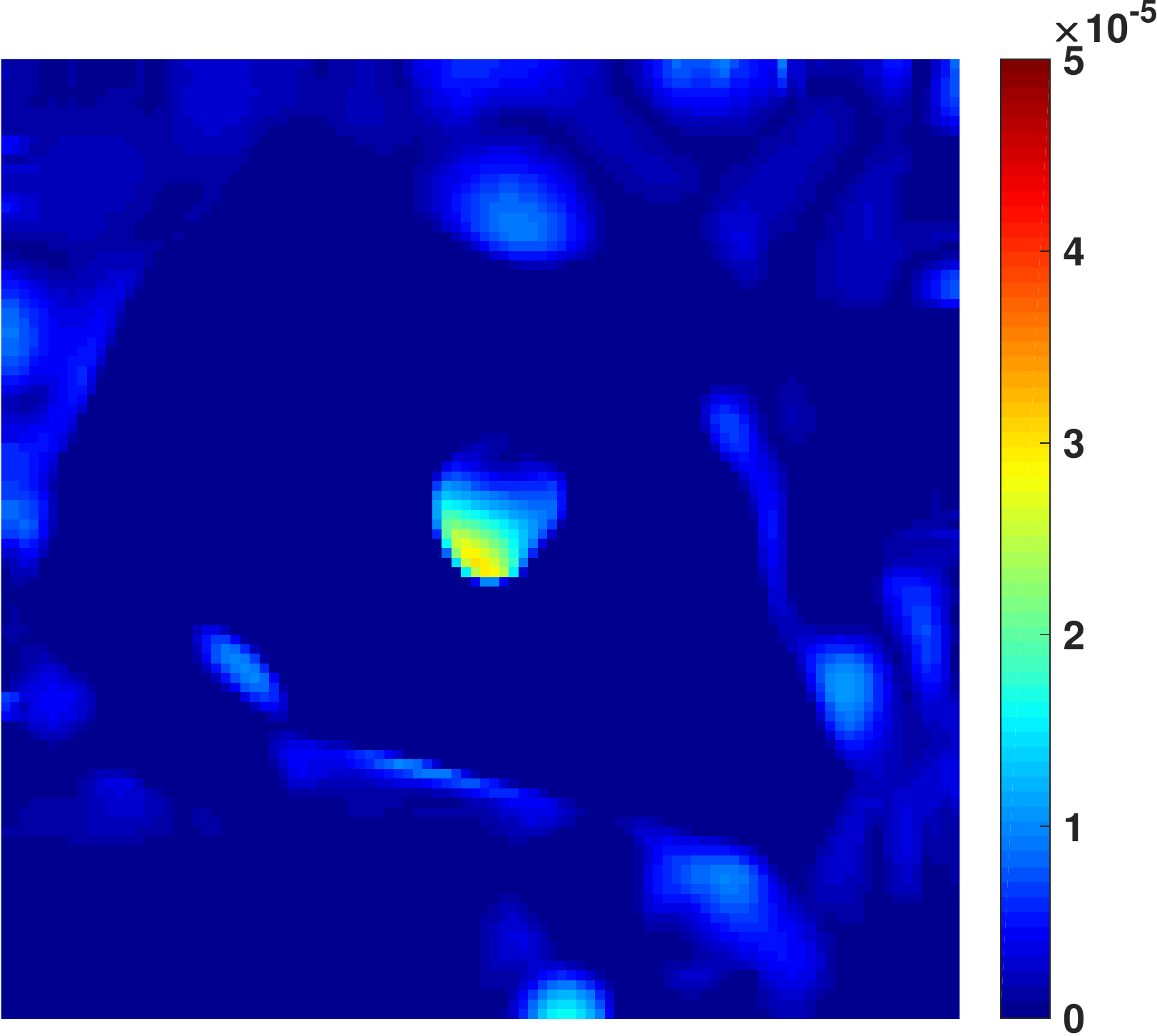} 
\end{tabular}
\caption{The polarization error images for the forward-jet model (first column) and the counter-jet model (second column) showing the pixels where the polarization constraint is not satisfied. These images are computed from the reconstructions obtained from the TV problem without constraint (first row), $\ell_1$+ TV problem without constraint (second row) and Polarized SARA without constraint (third row). All the images are shown in linear scale. Note that in the case of imposing this constraint, the corresponding polarization error images only have around 0.5~$\%$ non-zero pixels, as ensured in the stopping criterion.}
\label{fig:flux_err_im}
\end{figure}  

For visual comparison of the results obtained from these tests, we show the reconstructed images and the respective error images. The latter are computed by taking the absolute difference between the true and the reconstructed images. Out of the 5 simulations, the displayed images correspond to the simulation results with the least NRMSE.
The results for the forward-jet model images are shown in Figs~\ref{fig:rec_images_avery_I} and \ref{fig:rec_images_avery_P}, respectively for the intensity image $I$ and the linear polarization image $P$. In both the figures, the first row shows the ground truth image, whereas the second row shows the \textsc{cs-clean} reconstruction followed by its error image. For the \textsc{cs-clean} reconstruction, the shown image corresponds to the model image convolved with the restoring beam of FWHM size giving the minimum error for this method. Third and fourth row show the results for the TV and $\ell_1$+ TV  problems, respectively. In these rows, while the first two columns show the reconstructed and the error images obtained by without imposing the polarization constraint in the reconstruction process, the respective images obtained in the presence of the constraint are shown in the last two columns. Similarly, for the last row, column-wise, the following are displayed: reconstructed image for Polarized SARA without constraint and its error image; reconstructed image for Polarized SARA and its error image.
In the same manner, the results for the intensity image and the linear polarization image for the counter-jet model are shown in Figs~\ref{fig:rec_images_jason_I} and \ref{fig:rec_images_jason_P}, respectively.

Comparing the different regularizations from these figures, we can observe that the reconstructions obtained using the TV and $\ell_1$~+~TV regularizations are similar, while employing the SARA regularization leads to a better reconstruction quality. Firstly, in the case of the intensity image, for both forward and counter-jet models, the central region is much more resolved for the SARA regularization. It is in contrast with the reconstructions obtained by the TV and $\ell_1$~+~TV regularizations, where only the sharp edges are retained, leading to the staircase effect. One can recall that this effect arises due to the definition of the TV regularization, which tends to promote piece-wise constant images.
Secondly, for the linear polarization images, while all the regularizations produce diffused emissions in the background, these artefacts in the background are lower in the case of SARA regularization.  
In particular for the counter-jet model, the SARA regularization performs significantly better than the other two. Note that with the same noise variance, the low intensity values of this model provides lesser signal-to-noise ratio than the forward-jet model images. Thus, the image reconstruction is much more challenging in this case. The superiority of the SARA regularization over other regularizations in reconstructing these images is also supported by the error images. All these images shown in the linear scale, one can notice that for the TV and $\ell_1$~+~TV regularizations, these images have more residual, especially in the background. Furthermore, for the \textsc{cs-clean} reconstructions, it can be observed that the reconstruction quality is worse than that obtained by using any other sparsifying regularization, especially for the linear polarization image, validating the high errors observed in Fig.~\ref{fig:I_nrmse}. These observations are consistent with those obtained in other studies \citep{Chael2016, Akiyama2017b, Akiyama2017a}.

Regarding the comparison between the cases with and without polarization constraint, reduction in the artificial diffused background emissions, especially for the linear polarization images, by enforcing the constraint can be noticed from the presented results. This is supported by the visual inspection of the results as well as by the lesser residual in the error images. Note that the improvement in reconstruction quality is remarkable for the TV and $\ell_1$~+~TV regularizations, especially for the counter-jet model.

Overall, we can deduce that the results highlight the suitability of the SARA regularization for EHT imaging. Moreover, the use of the polarization constraint not only imposes the physical coherency between the reconstructed images, but it also tends to improve the reconstruction quality, independently of the choice of the sparsifying regularization. The latter is even more evident in the reconstruction of the the images with low signal-to-noise ratio, as observed for the linear polarization images for the counter-jet model. Furthermore, note that the non-physical reconstructions obtained in the absence of the constraint are more likely to appear in the background where the total intensity image has smaller values. 
To illustrate this assertion, the corresponding polarization error images are presented in Fig.~\ref{fig:flux_err_im} for the forward and counter-jet models, respectively in first and second columns. As previously mentioned, these images basically show the pixels where the polarization constraint is not satisfied by the reconstructed Stokes images. Note that having only 0.5~$\%$ (corresponding to the chosen stopping criterion) of such undesirable pixels, we do not show the polarization error images obtained in the presence of the constraint. In Fig.~\ref{fig:flux_err_im}, the images are shown row-wise for the following: TV problem without constraint (first row), $\ell_1$+ TV problem without constraint (second row) and Polarized SARA without constraint (third row). 
It can be clearly seen from these images that not imposing the constraint leads to the reconstruction of many pixels with physically unacceptable values. Another observation is regarding the SARA regularization, which performs better in suppressing these pixels than the other two regularizations, coherent with the values in Table~\ref{tab:flux}.




\vspace*{-0.0cm}
\section{Conclusion and Discussion} \label{sec:conc}
We have presented a new method, named Polarized SARA, for joint estimation of sparse Stokes images in the context of radio interferometry (RI), considering explicitly the polarization constraint. The latter is used to exploit the correlation between the Stokes images, imposing the polarization intensity as a lower bound on the total intensity image. This constraint cannot be managed by the classical optimization tools, and hence, we have proposed to deal with it using the techniques of epigraphical projection. In addition, our method leverages the sparsity of the underlying images using SARA regularization, consisting in promoting average sparsity of each Stokes parameter using the weighted $\ell_1$ norm encompassed in a reweighted scheme. Thanks to this weighting, the proposed method does not require the tuning of any regularization parameter and only the noise bound needs to be specified. To solve the resultant image reconstruction problem, we have designed an iterative proximal primal-dual algorithm. In this respect, the proposed approach presents the first application of sparsity based optimization techniques for the reconstruction of Stokes images, taking into account the polarization constraint. Moreover, our algorithm presents a highly versatile structure. This allows the incorporation of different sparsifying regularizations in the algorithm.

We have applied the proposed Polarized SARA method on the simulated EHT datasets. For the choice of sparsifying regularization, apart from the SARA regularization, we have also considered the TV and $\ell_1$~+~TV regularizations, the latter two being suggested in \cite{Akiyama2017b} for full-polarization EHT imaging. To judge the effect of the polarization constraint on the reconstruction qaulity, we have also generalized the problem considered in \cite{Akiyama2017b} to take into account this constraint. It is solved using a modified version of the proposed algorithm. On the one hand, comparing between the cases with and without the polarization constraint, the results indicate the importance of imposing it in reconstructing physically acceptable images. Additionally, irrespective of the considered sparsifying regularization, the enforcement of this constraint tends to enhance the reconstruction quality, particularly for the linear polarization images. This enhancement is significant for the results obtained by solving the TV and $\ell_1$~+~TV problems with constraint. Thus, we can conclude that the polarization
constraint is highly effective in producing images not only with physical meaning, but also with lesser artefacts. On the other hand, regarding the choice of sparsifying regularization, the results demonstrate the superiority of the SARA regularization in producing images with lesser artefacts and higher resolution, without requiring the convolution of the reconstructed images with any restoring beam. The reconstructions obtained by the SARA regularization also prevail over those obtained by the standard \textsc{cs-clean} algorithm. 
Moreover, Polarized SARA yields better results than solving the TV-based problems, both with and without the constraint, for polarimetric imaging. Note that SARA regularization has been initially proposed for total intensity (Stokes $I$) imaging in RI. Our results not only confirm the good performance of this regularization on Stokes $I$, but also highlight the fact that this regularization is well adapted for polarimetric imaging. Finally, the proposed Polarized SARA method stands out as a promising candidate for polarimetric imaging in RI. 

It is worth emphasizing again that the proposed approach has been developed to solve for the general RIME formalism~\eqref{eq:RIME}. However, in the current simulation settings, we have dealt with the case when the calibration terms, DIEs and DDEs, are absent. For the future work, we plan to consider the general setting of RIME, which essentially consists in adapting the measurement operator to take into account these terms. In such a case, with known DIEs and DDEs, the problem to reconstruct the Stokes images can be solved employing a similar approach as the one described in this article. Furthermore, the more challenging case of unknown calibration terms can be tackled using the ideas from our previous work on joint calibration and imaging for Stokes $I$ \citep{Repetti2017}.

Additionally, in the context of linear polarimetric imaging, a potential future prospect is to reconstruct Stokes $I$ image with the linear polarization image $P$ directly, instead of obtaining the image $P$ from the estimations of Stokes $Q$ and $U$. This stems from the fact that the magnitudes of the latter two can vary depending on the orientation of the chosen coordinate system, whereas the magnitude of the linear polarization image is independent of the choice of the coordinate system. Moreover, the problem can be solved for the rotationally invariant electric and magnetic components, defined from Stokes $Q$ and $U$ parameters \citep{Wiaux2007}. 

Our work can also be directly adapted to hyperspectral imaging as the polarization constraint acts at each wavelength independently. More complex priors can obviously be incorporated into the associated minimization problem, in line of recent results by one of the authors \citep{Abdulaziz2016}. Such developments are of critical interest for Faraday synthesis \citep{Brentjens2005, Bell2012}.

We also note that an interesting extension of the Polarized SARA method would be to apply it on VLBI observations, where the visibility phase cannot be measured. In this case, we can either apply our technique on the self-calibrated Stokes $I$ data or work with phase-insensitive observables, as in the case of optical interferometry (OI) \citep{Thiebaut2010}. With this in mind, we can combine the current proposed method with other sparse modelling techniques in OI \citep{Birdi2016, Akiyama2017a}. This will be advantageous for polarimetric imaging from VLBI observations.


\section*{Acknowledgements}

This work was supported by the UK Engineering and Physical Sciences Research Council (EPSRC, grant EP/M011089/1). 



\bibliographystyle{mnras}




%


\bsp	
\label{lastpage}
\end{document}